\documentclass[useAMS,usenatbib]{mn2e}

\usepackage{times}  
\usepackage{listings}
\usepackage{graphics}
\usepackage{subfigure}
\usepackage{graphicx}
\usepackage{amssymb}
\usepackage{hyperref}
\usepackage{aas_macros}

\def\simprop{ \lower .75ex \hbox{$\sim$} \llap{\raise .27ex \hbox{$\propto$}} }

\title[I. Gas in early-type galaxies]{
The origin of the atomic and molecular gas contents of early-type galaxies. I. A new test of galaxy formation physics}
\author[Claudia del P. Lagos et al.]{
\parbox[t]{\textwidth}{
\vspace{-1.0cm}
Claudia del P. Lagos$^{1}$,
Timothy A. Davis$^{1}$,
Cedric G. Lacey$^{2}$, 
Martin A. Zwaan$^{1}$,
Carlton M. Baugh$^{2}$,
Violeta Gonzalez-Perez$^{1}$,
Nelson D. Padilla$^{3}$
}
\vspace*{6pt} \\
$^{1}$European Southern Observatory, Karl-Schwarzschild-Strasse 2, 85748, Garching, Germany.\\
$^{2}$Institute for Computational Cosmology, Department of Physics,
University of Durham, South Road, Durham, DH1 3LE, UK.\\
$^{3}$Departamento Astronom\'ia y Astrof\'isica, Pontificia Universidad
Cat\'olica de Chile, Av. Vicu\~na Mackenna 4860, Stgo., Chile
\vspace*{-0.5cm}}

\begin{document}


\pagerange{\pageref{firstpage}--\pageref{lastpage}} \pubyear{2011}

\maketitle

\label{firstpage}

\begin{abstract}
{We study the atomic (HI) and molecular hydrogen (H$_2$) contents of early-type galaxies (ETGs) and their gas sources using the
GALFORM model of galaxy formation. This model uses a self-consistent
calculation of the star formation rate (SFR), which depends on the H$_2$ content of galaxies. 
We first present a new analysis of HIPASS and ATLAS$^{\rm 3D}$ surveys, with special emphasis on ETGs.
The model predicts HI and H$_2$ contents of ETGs in agreement with the observations 
from these surveys only if partial ram pressure stripping of the hot gas is included, 
showing that observations of neutral gas in `quenched' galaxies place stringent constraints on 
the treatment of the hot gas in satellites.  
We find that $\approx 90$\% of ETGs at z=0 have neutral gas contents supplied by radiative cooling from their hot halos, 
8\% were supplied by gas accretion from minor mergers that took place in the last 
1Gyr, while 2\% were supplied by mass loss from old stars. 
The model predicts neutral gas fractions 
strongly decreasing with increasing bulge fraction. This is due to the impeded disk regeneration in ETGs, resulting 
from both active galactic nuclei feedback and environmental quenching by partial ram pressure 
stripping of the hot gas.}
\end{abstract}

\begin{keywords}
galaxies: formation - galaxies : evolution - galaxies: ISM - stars: formation
\end{keywords}

\section{Introduction}

The classic picture of early-type galaxies (ETGs), 
which include elliptical and lenticular galaxies, is that they are 
`red and dead', without any significant star formation, 
and contain mainly old stellar populations \citep{Bower92}. 
ETGs have also been long connected to the red sequence in the color-magnitude relation, establishing a 
strong connection between quenching of star formation and morphological transformation 
(e.g. \citealt{Bower92}; \citealt{Strateva01}; \citealt{Baldry04}; \citealt{Balogh04}; \citealt{Bernardi05}; \citealt{Schiminovich07}). 
By analysing the S\'ersic index of galaxies 
in the star formation rate-stellar mass plane, \citet{Wuyts11} 
showed that galaxies in the star forming sequence (the so called `main sequence' of galaxies) are typically 
disk-like galaxies (with  S\'ersic indices close to $1$), while passive galaxies tend 
to have higher S\'ersic indices (typically $>3$). Wuyts et al. also showed that these trends are 
observed in galaxies from $z=0$ to $\approx 2$, suggesting that the relation between morphology and quenching is 
fundamental and that is present over most of the star formation history of galaxies. 

Although this simple paradigm of `red and dead' ETGs is qualitatively sufficient to explain their location 
on the red sequence of galaxies, it is far from being quantitatively correct. 
High quality, resolved observations of the different components of ETGs, 
mainly from the ATLAS$^{\rm 3D}$\footnote{{\tt http://www-astro.physics.ox.ac.uk/atlas3d/}.} multi-wavelength survey \citep{Cappellari11}, 
showed that this paradigm is too simplistic.
This survey showed that at least $20$\% of ETGs 
have molecular and atomic hydrogen contents large enough to be detected 
(\citealt{Young11}; \citealt{Serra12}; see also \citealt{Welch10} for similar results from an independent survey). 
The approximate detection limits for molecular hydrogen (H$_2$) and atomic hydrogen (HI) masses in the 
ATLAS$^{\rm 3D}$ are $\approx 10^7-10^8\,M_{\odot}$. 
Large amounts of cold gas in ETGs are frequently found when star formation is observed (e.g. \citealt{Davis14}).
Some of these galaxies with ongoing star formation lie on the red sequence of galaxies in the color-magnitude 
relation (\citealt{Kaviraj07}; \citealt{Smith12}, \citealt{Young13}). 
All of this evidence points to a large fraction of ETGs, which were before seen as `passive' in terms of their colours, having 
star formation rates and cold interstellar medium (ISM) 
contents that can be rather large. From this it is reasonable to conclude that the quenching of galaxies 
is indeed more complex than the simple picture of `passive, red and dead' ETGs.

This shift of paradigm in ETGs poses new questions regarding how we understand the formation of this galaxy population. 
For instance, how do we understand the presence of a non-negligible cold ISM in ETGs 
and their location on the red sequence of galaxies in the color-magnitude diagram?
 Was the cold gas accreted recently or does it come from internal processes, such as 
recycling of old stars? These questions are at the core of the understanding of the quenching of galaxies 
and the decline of the star formation activity with time. 

Simulations of galaxy formation have long explored the origin of galaxy morphologies 
in the context of the hierarchical growth of structures. Pioneering ideas about the formation of galaxy disks and bulges 
were presented by \citet{Toomre77} and \citet{White78}. Toomre proposed for the first time that galaxy mergers could lead 
to the formation of spheroids, which was implemented in early semi-analytic models of galaxy 
formation (e.g. \citealt{Baugh96}; \citealt{Kauffmann96}; \citealt{Cole00}). 
However, with the advent of large area surveys, and more sophisticated 
cosmological $N$-body simulations, it became clear that 
major mergers (mergers between galaxies with mass ratios $\gtrsim 0.3$) could not be the only 
formation mechanism of spheroids (e.g. see \citealt{LeFevre00} for an observational example 
 and \citealt{Naab03} for a theoretical work) 
 because of their expected rareness, which is incompatible with the large numbers of ETGs observed 
(\citealt{Bernardi03}; \citealt{Lintott08}).
Theoretical work on the formation mechanisms of spheroids led to the conclusion 
that minor mergers (e.g. \citealt{Malbon07}; \citealt{Parry09}; 
\citealt{Hopkins10}; \citealt{Bournaud11}; \citealt{Naab13}) 
and disk instabilities (e.g. \citealt{Mo98}; \citealt{Gammie01}; \citealt{Bournaud09}; \citealt{Krumholz10}; \citealt{Elmegreen10}) 
can also play a major role. Many studies exploited numerical simulations and semi-analytic models 
to study the formation of ETGs and their mass assembly with interesting predictions, for example 
that massive ETGs assemble their mass relatively late but have stellar populations that are very old 
(e.g. \citealt{Baugh96}; \citealt{Kauffmann96}; \citealt{DeLucia06}; \citealt{Parry09}), and that the formation paths 
for ETGs can be many, going from 
having had one or more major mergers, to having had no mergers at all (e.g. \citealt{Naab13}). 

Despite all this progress, little 
attention has been paid to the study of the neutral gas content of the ETG population. The 
atomic and molecular gas contents of ETGs may provide 
 strong constraints on the recent accretion history. \citet{Lagos10}, for example, show that the neutral gas content 
of galaxies is very sensitive to short term variations in the accretion history, while the stellar mass and optical colours are not.
 Similarly, \citet{Serra14} show that although simulations can reproduce the nature of slow and fast rotators 
in the early-type population, the atomic hydrogen content predicted by the same simulations is too low.  
Another reason to believe that 
the neutral gas content of ETGs will provide strong constraints on galaxy formation models, is that they show 
different correlations between their gas and stellar contents. 
For example, for normal star-forming galaxies, there is a good correlation between 
the HI mass and the stellar mass (e.g. \citealt{Catinella10}; \citealt{Cortese11}; \citealt{Huang12}; \citealt{Wang14}), while 
ETGs show no correlation between these two quantities (e.g. \citealt{Welch10}; \citealt{Serra12}). 
Different physical mechanisms are then driving the HI content of ETGs. Similar conclusions were reached 
for molecular hydrogen (e.g. \citealt{Saintonge11}; \citealt{Lisenfeld11}; \citealt{Young11}; \citealt{Boselli14}). 

The motivation behind this paper is to investigate the neutral gas content of ETGs in hierarchical galaxy formation models 
and relate them to the formation and quenching mechanisms of ETGs.
We explore the question of the origin of the atomic and molecular gas contents of ETGs and attempt to connect this to 
their observed HI and H$_2$ contents and their stellar mass content. In paper II (Lagos et al. 
in prep.), we will explore the question of the alignments between the angular momenta of the 
gas disk and the stellar contents of ETGs.
For the current study, we use three flavours of the semi-analytical model {\texttt{GALFORM}}
in a $\Lambda$CDM cosmology (\citealt{Cole00}),
namely those of \citet{Lagos12}, \cite{Gonzalez-Perez13}, and Lacey et al. (2014, in prep.).
The three models include the improved treatment of SF implemented by \citet{Lagos10}. This extension
 splits the hydrogen content of the ISM
 into HI and H$_2$. In addition, these three models allow bulges to grow through minor and major galaxy mergers 
and through global disk instabilities. The advantage of using three different flavours of {\texttt{GALFORM}}
 is the ability to characterise the robustness of the trends found. The outputs of the three models
shown in this paper will be made publicly available through the Millennium
database\footnote{\tt http://gavo.mpa-garching.mpg.de/Millennium}.   

This paper is organised as follows. In $\S 2$ we present the observations of the HI and H$_2$ content 
of the entire galaxy population and of ETGs, along with the gas mass functions and gas fraction distribution functions 
for these two populations. In $\S 3$, we describe the galaxy formation model
and the main aspects which relate to the growth of bulges: star formation, disk and bulge build-up, recycling of 
intermediate and low mass stars and the treatment of the partial ram pressure stripping of the hot gas. We also describe the main 
differences between the three
{\tt GALFORM} flavours and the dark matter simulations used.
 In $\S 4$ we compare the model predictions with observations of the neutral gas content of ETGs and show the impact of 
including partial ram pressure stripping of the hot gas.
In $\S 5$ we analyse the connection between the neutral gas content of ETGs, their bulge fraction and quenching and explain 
the physical processes behind this connection. In $\S 6$ we analyse all the sources that contribute to the neutral gas 
content of ETGs and their environmental dependence. Finally, our main conclusions are presented in  $\S 7$. 

\section{The observed neutral hydrogen content of local galaxies}\label{obssec}

\begin{figure}
\begin{center}
\includegraphics[trim = 0.9mm 0.3mm 1mm 0.5mm,clip,width=0.5\textwidth]{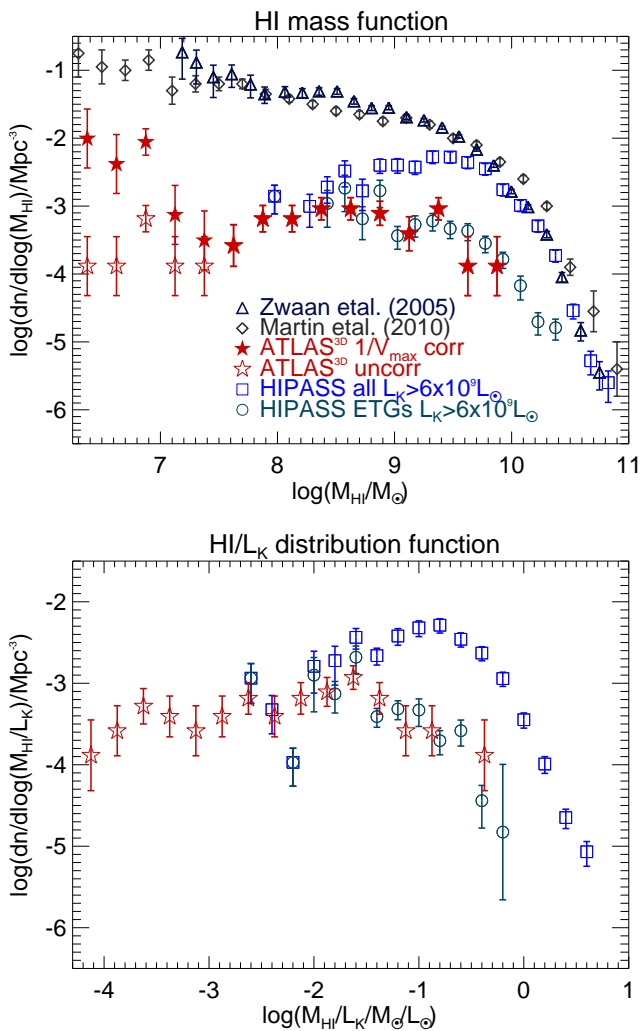}
\caption{{\it Top panel:} The HI mass function from \citet{Zwaan05} and \citet{Martin10}
at $z=0$, which include both early and late-type galaxies; the HI mass function of ETGs from HIPASS and from 
the ATLAS$^{\rm 3D}$ survey (with and without $1/V_{\rm max}$ correction), 
calculated in this work. {\it Bottom panel:} Distribution 
function of the ratio between the HI mass to the $K$-band luminosity for all and ETGs for the HIPASS and 
the ATLAS$^{\rm 3D}$ surveys, calculated in this work.
Note that we express densities and masses in physical units.}
\label{HIobs}
\end{center}
\end{figure}

Our aim is to study the neutral gas content of ETGs and show how this compares 
to the overall galaxy population. We first need to define the observational datasets that we will use 
as the main constraints on the galaxy formation simulations\footnote{The units in the observations 
are expressed in physical units. Note that 
we normalised all the observational datasets to the choice of $h=0.73$.}. 
We focus on the HI and H$_2$ gas contents 
of galaxies and how these compare to the stellar content. We do this through comparisons 
with the $K$-band luminosity, which is closely related to the stellar mass in galaxies but 
directly measured by observations. We extensively use the HI Parkes
All-Sky Survey (HIPASS; \citealt{Meyer04}), the 
Arecibo Legacy Fast ALFA Survey (ALFALFA; \citealt{Giovanelli05}), the 
Five College Radio Astronomy Observatory CO$(1-0)$ survey of \citet{Keres03} and 
the ATLAS$^{\rm 3D}$ survey \citep{Cappellari11}. 

The ATLAS$^{\rm 3D}$ survey is of particular interest as it 
is a volume limited sample of ETGs, where all ETGs within a volume 
of $1.16\times 10^{5}\,\rm Mpc^{3}$ and with $K$-band rest-frame luminosities of 
$L_{\rm K}>6\times 10^{9}\,L_{\odot}$ were studied in detail. The total number of ETGs in the 
ATLAS$^{\rm 3D}$ catalogue is $260$. The 
morphological classification was obtained through careful visual inspection.
The ATLAS$^{\rm 3D}$ survey contains multi-wavelength information, such as 
broad-band photometry in the $B$-, $r$- and $K$-bands \citep{Cappellari11}, as well as $21$~cm interferometry (presented in \citealt{Serra12}), 
from which the HI mass is derived, and 
CO$(1-0)$ single dish observations, presented in \citet{Young11}, from which the H$_2$ mass is derived, along with 
detailed stellar kinematic information. This survey 
provides an in depth view of ETGs in the local Universe and shows how the neutral gas
 content correlates with other galaxy properties. 

\begin{figure}
\begin{center}
\includegraphics[trim = 0.9mm 0.3mm 1mm 0.5mm,clip,width=0.5\textwidth]{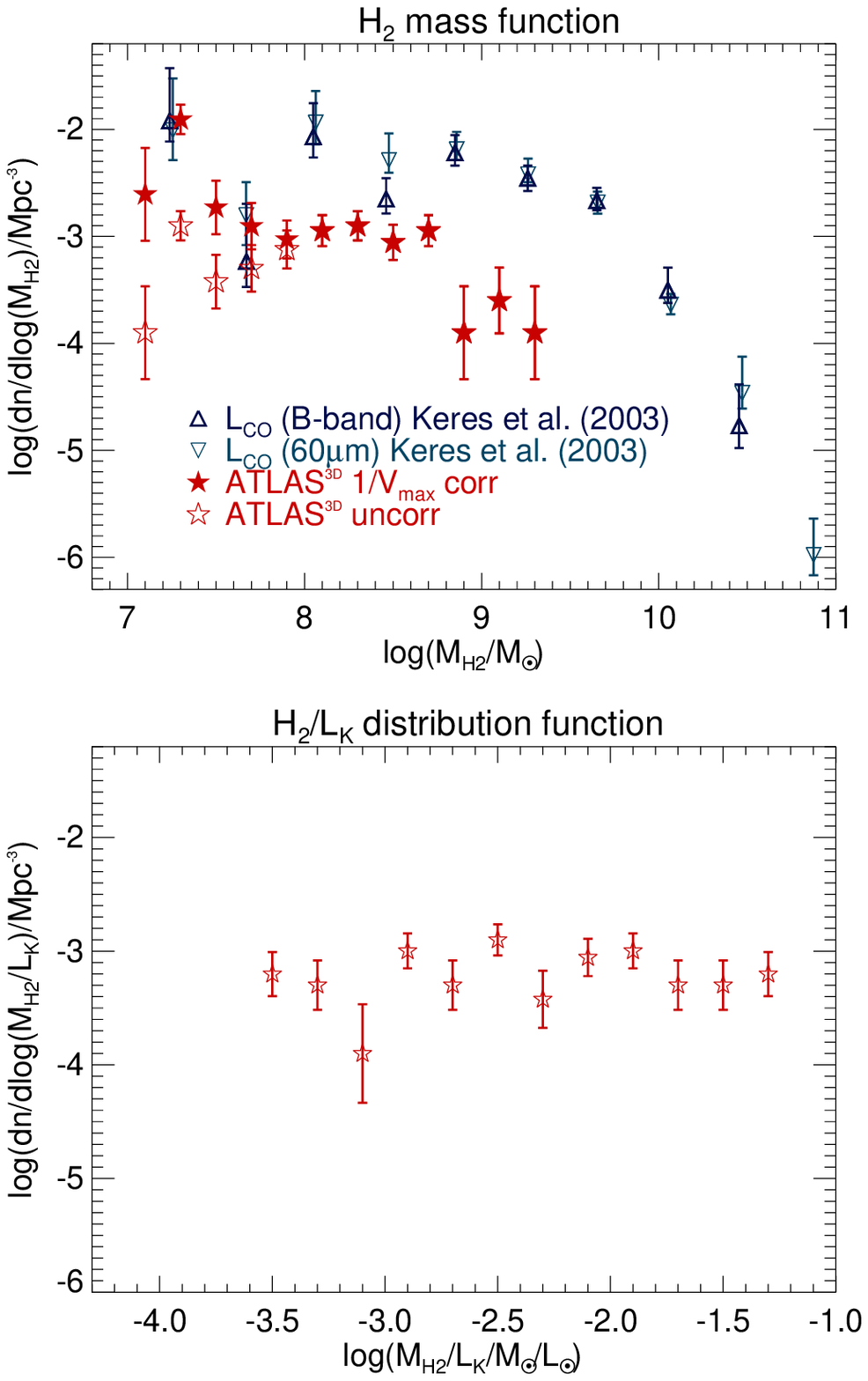}
\caption{{\it Top panel:} The H$_2$ mass function from \citet{Keres03} 
 using CO$(1-0)$ observations of parent samples selected in 
 $B$-band and $60\mu$m, as labelled, and adopting 
a Milky Way $\rm H_2$-to-CO conversion factor,
$N_{\rm H_2}/\rm cm^{-2}=2\times10^{-20}\, I_{\rm CO}/\rm K\,km\,s^{-1}$.
 Here $N_{\rm H_2}$ is the column density of H$_2$ and
$I_{\rm CO}$ is the integrated CO$(1-0)$ line intensity per unit
surface area. Also shown are the H$_2$ mass function of ETGs from the ATLAS$^{\rm 3D}$ survey (stars), with and without 
the $1/V_{\rm max}$ correction, also for a Milky Way $\rm H_2$-to-CO conversion factor. 
 {\it Bottom panel:} Distribution 
function of the H$_2$ gas fraction for the ATLAS$^{\rm 3D}$ survey.}
\label{H2obs}
\end{center}
\end{figure}

We constructed HI and H$_2$ mass functions for the ATLAS$^{\rm 3D}$ objects. The HI survey presented in 
\cite{Serra12} includes every ATLAS$^{\rm 3D}$ ETG visible with the Westerbork Synthesis Radio Telescope, 
a total of $166$ objects.
We took HI masses from \cite{Serra12} to construct a HI mass function, 
which we plot in Fig.~\ref{HIobs}. This HI survey is incomplete in depth due to 
observations being performed using a fixed $12$ hour integration. This means that 
HI masses $<10^{7.5}\,M_{\odot}$ could only be detected 
in nearby objects. Thus, we correct the lower HI mass bins for incompleteness using the standard $1/V_{\rm max}$ method \citep{Schmidt68}.
Note that since ATLAS$^{\rm 3D}$ is complete above a $K$-band luminosity of $L_{\rm K}>6\times 10^{9}\,L_{\odot}$, 
we use a V$_{\rm max}$ calculated for the HI mass only.
Total H$_2$ masses are available from \citet{Young11} for all galaxies in the full ATLAS$^{\rm 3D}$ survey volume.
We constructed H$_2$ mass functions from this data in the same way (see Fig.~\ref{H2obs}). 
The IRAM-30m telescope observations have a fixed noise limit, and Fig.~\ref{H2obs} 
shows both the original and a corrected mass function (where the lower H$_2$ mass bins have been 
corrected using the $1/V_{\rm max}$ method). 

Blind HI surveys, such as ALFALFA and HIPASS provide information on the HI content of 
galaxies in larger volume. HI mass functions were derived from these two surveys and presented in 
\citet{Martin10} and \citet{Zwaan05}, respectively. These HI mass functions are shown in the top-panel of Fig.~\ref{HIobs}.
However, because we are interested in how ETGs 
differ from the overall galaxy population and in isolating the physical processes 
leading to such differences, we need a proxy for the stellar mass of these HI-selected galaxies.
To this end, we use the HIPASS survey cross-matched with the 
Two Micron All-Sky Survey ($2$MASS; \citealt{Jarrett00}) to obtain $K$-band luminosities 
for the HIPASS galaxies (see \citealt{Meyer08a}). We limit the analysis to the southern HIPASS 
sample \citep{Meyer04}, because the completeness function for this sample is well-described and the 
HI mass function is determined accurately 
\citep{Zwaan05}. We find that $86$\% of the southern HIPASS galaxies have $K$-band counterparts, and 
all of these have morphological classifications which are described in \citet{Doyle05}.  
Galaxy morphologies are taken from the SuperCOSMOS Sky Survey \citep{Hambly01}, and are obtained by 
visual inspection, predominantly in the $b_J$-band.

With the subsample of HIPASS galaxies with $K$-band luminosities and 
assigned morphological types, we calculate the HI mass function for ETGs, take to be 
 those identified as `E-Sa'. For this, we use the maximum likelihood
equivalents of the $1/V_{\rm max}$ values determined by \citet{Zwaan05}. These values represent
the maximum volume over which each of the HIPASS galaxies could have been detected, 
but they are corrected so as to remove the effects of large scale inhomogeneities in the HIPASS survey (see \citealt{Zwaan05} for 
details). 
We add an additional selection in the $K$-band luminosity to match the selection used in ATLAS$^{\rm 3D}$ 
and re-calculate the HI mass function of the overall galaxy population using the same $V_{\rm max}$ 
values determined by Zwaan et al. Note that we do not need to recalculate $V_{\rm max}$ due to the $K$-band survey 
being much deeper than the HI survey (i.e. it is able to detect small galaxies further out than the HI survey). 
 The results of this exercise are shown in the top-panel of Fig.~\ref{HIobs}. We also show the 
estimated HI mass function of the ATLAS$^{\rm 3D}$ ETGs. There is very good agreement between the 
HI mass function of ETGs from the HIPASS and ATLAS$^{\rm 3D}$ surveys in the range where they overlap 
(despite the inclusion of Sa galaxies in the HIPASS sample, which are absent in ATLAS$^{\rm 3D}$). 

ATLAS$^{\rm 3D}$ provides an important insight into the HI mass function of ETGs in the regime of 
low HI masses that are not present in HIPASS. In the case of the HI mass function of all galaxies that have 
$K$-band luminosities above $6\times 10^9\,L_{\odot}$, we find that the HI mass function is fully recovered down to 
$M_{\rm HI}\approx 5\times 10^9\,M_{\odot}$, with a drop in the number density of lower HI masses due to 
the $K$-band luminosity limit. Note that this drop is not because of incompleteness but instead is a real feature 
connected to the minimum HI mass that normal star-forming galaxies with $L_{\rm K}>6\times 10^9\,L_{\odot}$ have. 
By normal star-forming galaxy we mean those that lie on the main sequence of galaxies in the 
plane of star formation rate (SFR) vs. stellar mass. 
The HIPASS matched sample is complete for $K$-band luminosities $L_{\rm K}>6\times 10^9\,L_{\odot}$. 
The turn-over observed at $M_{\rm HI}\approx 5\times 10^9\,M_{\odot}$ is simply the HI mass expected for a
normal star-forming galaxy with $L_{\rm K}\approx 6\times 10^9\,L_{\odot}$. 

Similarly, we calculate the distribution function of the ratio of HI mass to the $K$-band 
luminosity, which we refer to as the HI gas fraction. 
We do this for all HIPASS galaxies with $L_{\rm K}>6\times 10^9\,L_{\odot}$ 
in the cross-matched catalogue and for the subsample of ETGs. 
This is shown in the bottom-panel of Fig.~\ref{HIobs}. Also shown is the distribution of $M_{\rm HI}/L_{\rm K}$ 
for the ETGs in the ATLAS$^{\rm 3D}$ survey. Note that these distribution functions provide 
higher order constraints on galaxy formation simulations than the more commonly used scaling relations between the 
HI mass or H$_2$ mass and stellar mass (e.g. \citealt{Catinella10}; 
\citealt{Saintonge11}). This is because this distribution allows us to test not only if the amount of neutral gas in model galaxies 
is in the expected proportion to their stellar mass, but also that the number density of galaxies with different 
gas fractions is correct. \citet{Kauffmann12} show that this is a stronger constraint on semi-analytic models of galaxy formation, 
as the distribution of the neutral gas fraction depends on the quenching mechanisms included in the models, and the 
way they interact. So far, no such comparison has been presented for cosmological hydrodynamical simulations. 

We perform the same exercise we did for HI masses but now for H$_2$ masses. In this case there are no 
blind surveys of carbon monoxide or any other H$_2$ tracer, and therefore the data for large samples of 
galaxies is scarce. \citet{Keres03} reported the first and only attempt to derive the
local luminosity function (LF) of $\rm CO(1-0)$. This was done using
 $B$-band and a $60\,\mu$m selected samples, and with follow up using the 
Five College Radio Astronomy Observatory. This is shown in the top-panel of Fig.~\ref{H2obs}. 
Also shown is the H$_2$ mass function of ETGs from ATLAS$^{\rm 3D}$. 
Here, we adopt a Milky Way $\rm H_2$-to-CO conversion factor,
$N_{\rm H_2}/\rm cm^{-2}=2\times10^{-20}\, I_{\rm CO}/\rm K\,km\,s^{-1}$, for both 
the Keres et al. sample and the ATLAS$^{\rm 3D}$ (but see \citet{Lagos12} for more on conversions).
 Here $N_{\rm H_2}$ is the column density of H$_2$ and
$I_{\rm CO}$ is the integrated CO$(1-0)$ line intensity per unit
surface area.
In the bottom-panel of Fig.~\ref{H2obs} we show the distribution function of 
the $M_{\rm H_2}/L_{\rm K}$ ratios (we which refer to as H$_2$ gas fractions) for the ATLAS$^{\rm 3D}$ sources.

Throughout this paper we compare different flavours of {\texttt{GALFORM}} to the set of observations 
presented in Figs.~\ref{HIobs} and \ref{H2obs}.

\section{Modelling the evolution of the morphology, neutral gas content and star formation in galaxies}\label{modelssec}

Here we briefly describe the {\texttt{GALFORM}} 
semi-analytical model of galaxy formation and evolution (introduced by \citealt{Cole00}), 
focusing on the aspects that are relevant to the build-up of ETGs. 

The {\tt GALFORM} model 
takes into account the main physical processes
that shape the formation and evolution of galaxies. These are: (i) the collapse
and
merging of dark matter (DM) halos, (ii) the shock-heating and radiative cooling
of gas inside
DM halos, leading to the formation of galactic disks, (iii) quiescent star
formation in galaxy disks, (iv) feedback
from supernovae (SNe), from heating by active galactic nuclei (AGN) 
 and from photo-ionization of the
inter-galactic medium (IGM), (v) chemical 
enrichment of stars and gas, and (vi) galaxy mergers driven by
dynamical friction within common DM halos which can trigger bursts of star formation,
and lead to the formation of spheroids (for a review of these
ingredients see \citealt{Baugh06} and \citealt{Benson10b}). 
Galaxy luminosities are computed from the predicted star formation and
chemical enrichment histories using a stellar population synthesis model (see \citealt{Gonzalez-Perez13} to 
see the impact of using different models). 

In the rest of this section we describe the star formation (SF) law used and how this connects            
to the two-phase interstellar medium (ISM; $\S$~\ref{SFlaw}), the recycled fraction and yield from newly formed stars and how
that gas fuels the ISM ($\S$~\ref{Sec:Randp}), the physical processes that give rise
to discs and bulges in {\tt GALFORM} ($\S$\ref{Sec:MorphoTrans}) and 
 the modelling of the ram-pressure
stripping of the hot gas (\S~\ref{Sec:rampressure}) in satellite galaxies. 
We put our focus into these processes because we aim to distinguish the contribution of each of them to 
the neutral gas content of ETGs: galaxy mergers, hydrostatic cooling and recycling from stars. 
These processes are the same in the three variants of {\tt GALFORM} 
we use in this paper: the models of  Lagos et al. (2012; Lagos12), 
Gonzalez-Perez et al. (2014; Gonzalez-Perez14) and Lacey et al. (2014, in prep.; Lacey14), albeit with different parameters.
In $\S$~\ref{Models} we describe the differences between these variants. 
 We finish the section with a short description of the
$N$-body cosmological simulation used and the parameters adopted ($\S$~\ref{Cosmos}).

\subsection{Interstellar medium gas phases and the star formation law}\label{SFlaw}

In {\tt GALFORM} the SF law developed in Lagos et al. (2011b, hereafter `L11') is adopted. 
In this SF law the atomic and molecular 
phases of the neutral hydrogen in the ISM are explicitly distinguished.
L11 found that the SF law that gives the best agreement with 
the observations without the need for further calibration is the 
empirical SF law of \citet{Blitz06}. 
Given that the SF law has been well constrained in spiral and dwarf galaxies in the 
local Universe, L11 decided to implement this molecular-based SF law only in the quiescent SF mode (SF following gas 
accretion onto the disk), keeping the original prescription of \citet{Cole00} 
for starbursts (driven by galaxy mergers and 
global disk instabilities). 

{\it Quiescent Star Formation.} The empirical SF law of Blitz \& Rosolowsky has the form,
\begin{equation}
\Sigma_{\rm SFR} = \nu_{\rm SF} \Sigma_{\rm mol},
\label{Eq.SFR}
\end{equation}
\noindent where $\Sigma_{\rm SFR}$ and $\Sigma_{\rm mol}$ are the surface
densities of SFR and molecular gas, respectively, 
and $\nu_{\rm SF}$ is the inverse of the SF 
timescale for the molecular gas.
The molecular gas mass includes the contribution from helium.
The ratio between the molecular and total gas mass, $\rm f_{\rm mol}$, depends on
the internal hydrostatic pressure through $\Sigma_{\rm H_2}/\Sigma_{\rm HI}=\rm f_{\rm mol}/(f_{\rm mol}-1)=
(P_{\rm
ext}/P_{0})^{\alpha}$. Here HI and H$_2$ only include
hydrogen (which in total corresponds
to a fraction $X_{\rm H}=0.74$ of the overall cold gas mass).
To calculate $\rm P_{\rm
ext}$, we use the approximation from \citet{Elmegreen89}, in which the pressure 
depends on the surface density of gas and stars. The parameters 
$\nu_{\rm SF}$, $P_{0}$ and $\alpha$ are given in \S~\ref{Models} for each of the three {\tt GALFORM} variants.

{\it Starbusts.} {For starbursts the situation is less clear than in star formation in disks mainly due to 
observational uncertainties, such as 
the conversion between CO and H$_2$ in starbursts, and the 
intrinsic compactness of star-forming regions, which have prevented 
a reliable characterisation of the SF law (e.g. \citealt{Genzel10}).
For this reason we choose to apply the BR 
law only during quiescent SF (fuelled by accretion of cooled
gas onto galactic disks) and retain 
the original SF prescription for
starbursts (see \citealt{Cole00} and L11 for details).} 
In the latter, the SF timescale is proportional to the bulge dynamical
timescale above a minimum floor value 
and involves the whole cold gas content of the galaxy, $\rm SFR={\it M}_{\rm cold}/\tau_{\rm SF}$ 
(see \citealt{Granato00} and \citealt{Lacey08} for details). The SF timescale is defined as

\begin{equation}
\tau_{\rm SF}=\rm max(\tau_{\rm min},f_{\rm dyn}\tau_{\rm dyn}), 
\label{SFlawSB}
\end{equation}

\noindent where $\tau_{\rm dyn}$ is the bulge dynamical timescale,  $\tau_{\rm min}$ is a minimum duration 
adopted for starbursts and $f_{\rm dyn}$ is a free parameter. The values of the parameters  
$\tau_{\rm min}$ and $f_{\rm dyn}$ are given in \S~\ref{Models} for each of the three {\tt GALFORM} variants.

\subsection{Recycled fraction and yield}\label{Sec:Randp}

In {\tt GALFORM} we
 adopt the instantaneous mixing approximation
for the metals in the ISM. This implies that the metallicity
of the cold gas mass instantaneously absorbs the fraction of recycled mass and newly synthesised metals
 in recently formed stars, neglecting the time delay for the ejection of gas and metals from stars.

The recycled mass injected back to the ISM by newly born stars
is calculated from the initial mass function (IMF) as,

\begin{equation}
R=\int_{m_{\rm min}}^{m_{\rm max}}\, (m-m_{\rm rem})\phi(m)\, {\rm d} m, 
\label{Eq:ejec}
\end{equation}

\noindent where $m_{\rm rem}$ is the remnant mass and the IMF is defined
as $\phi(m)\propto dN(m)/dm$. Similarly, we define the yield as

\begin{equation}
p =\int_{m_{\rm min}}^{m_{\rm max}}\, m_{\rm i}(m)\phi(m) {\rm d} m, 
\label{Eq:yield}
\end{equation}

\noindent where $m_{\rm i}(m)$ is the mass of newly synthesised metals ejected by stars of initial mass $m$.
 The integrations limits are taken to be $m_{\rm min}=1\, M_{\odot}$ and
$m_{\rm max}=120\, M_{\odot}$. Stars with masses
$m<1\, M_{\odot}$ have lifetimes longer than the age of the Universe, and therefore they do not contribute
to the recycled fraction and yield.
{The quantities $m_{\rm rem}(m)$ and $m_{\rm i}(m)$ depend on the initial mass of a star and are calculated 
by stellar evolution theory. The stellar evolution model we use for intermediate stars ($1M_{\odot}<m\lesssim 8M_{\odot}$) is 
\citet{Marigo01} (i.e. which provides $m_{\rm rem}(m)$ and $m_{\rm i}(m)$ for those types of stars), while 
for massive stars, $m\gtrsim 8M_{\odot}$,  we use \citet{Portinari98}.}

We describe in \S~\ref{Models} the IMF adopted in the three 
variants of {\tt GALFORM}.

\subsection{Morphological transformation of galaxies}\label{Sec:MorphoTrans}

\subsubsection{The build-up of discs}\label{Sec:Cooling}

Galaxies form from gas which cools from the hot halo observing conservation of angular momentum.
As the temperature decreases,
thermal pressure stops supporting the gas which then settles in a 
rotating disks \citep{Fall80}.

We model the gas profile of the hot gas with a $\beta$ profile \citep{Cavaliere76},

\begin{equation}
\rho_{\rm hot}(r)\propto (r^2+r^2_{\rm core})^{-3\beta_{\rm fit}/2},
\label{beta-prof}
\end{equation}

\noindent The simulations of \citet{Eke98} show that $\beta_{\rm fit}\approx 2/3$ and that $r_{\rm core}/R_{\rm  NFW}\approx1/3$.
In {\tt GALFORM}, we adopt the above values and use $\rho_{\rm hot}(r)\propto (r^2+R^2_{\rm NFW}/9)^{-1}$, 
where $R_{\rm NFW}$ is the scale radius of the \citet{Navarro97} profile of the dark matter halos.

During a timestep $\delta t$ in the integration of galaxy properties, we calculate the amount of gas that cools and estimate
the radius at which $\tau_{\rm cool}(r_{\rm cool})=t-t_{\rm form}$, where $t_{\rm form}$ corresponds to the
time at which the halo was formed and $t$ is the current time. The gas inside $r_{\rm cool}$ is cool enough to be accreted
onto the disk. However, in order to be accreted onto the disk, the cooled gas should have had enough time to
fall onto the disk. Thus, the gas that has enough time to cool and be accreted onto the disk is that within the radius 
in which the free-fall time and the cooling time are smaller than $(t-t_{\rm form})$, defined as
$r_{\rm ff}$ and $r_{\rm cool}$, respectively.
The mass accreted onto the disk simply corresponds to the
hot gas mass enclosed within $r={\rm min}[r_{\rm cool},r_{\rm ff}]$.

We calculate $r_{\rm cool}$ from the cooling time, which is defined as 

\begin{equation}
\tau_{\rm cool}(r)=\frac{3}{2}\,\frac{\mu\, m_{\rm H}\,k_{\rm B}\, T_{\rm hot}}{\rho_{\rm hot}(r)\, \Lambda(T_{\rm hot},Z_{\rm hot})}.
\end{equation}

\noindent Here, $\Lambda(T_{\rm hot},Z_{\rm hot})$ is the cooling function that depends on the gas temperature, $T_{\rm hot}$, 
which corresponds to the virial temperature of the halo ($T_{\rm hot}= T_{\rm V}$) and
the metallicity $Z_{\rm hot}$
(i.e. the ratio between the mass in metals heavier than Helium and total gas mass).
 The cooling rate per unit volume is
$\epsilon_{\rm cool}\propto \rho^2_{\rm hot}\, \Lambda(T_{\rm hot},Z_{\rm hot})$.
In {\tt GALFORM} we adopt the cooling function tabulated of \citet{Sutherland93}.

\subsubsection{The formation of spheroids}\label{BuildUpBulges} 

Galaxy mergers and disk instabilities give rise to the formation of spheroids and
 elliptical galaxies. Below we describe both physical processes.

{\it Galaxy mergers.} When DM halos merge, we assume that the galaxy hosted by the most massive progenitor halo becomes
the central galaxy, while all the other galaxies become satellites orbiting
the central galaxy. These orbits gradually decay towards the centre due to energy and angular
momentum losses driven by dynamical friction with the halo material.

Eventually, given sufficient time satellites spiral in and merge with the central galaxy. Depending on the amount of gas and baryonic mass
 involved in the galaxy merger, a starburst can result. The time for the satellite to hit the central galaxy
is called the orbital timescale, $\tau_{\rm merge}$, which is calculated following \citet{Lacey93} as

\begin{equation}
\tau_{\rm merge}=f_{\rm df}\, \Theta_{\rm orbit}\, \tau_{\rm dyn}\, \left[\frac{0.3722}{{\rm ln}(\Lambda_{\rm Coulomb})}\right]\, \frac{M}{M_{\rm sat}}.
\end{equation}

\noindent Here, $f_{\rm df}$ is a dimensionless adjustable
parameter which is $f_{\rm df}>1$ if the satellite's halo is efficiently
stripped early on during the infall, $\Theta_{\rm orbit}$ is a function of the orbital parameters,
$\tau_{\rm dyn}\equiv \pi\,R_{\rm v}/V_{\rm v}$ is the dynamical timescale of the halo,
${\rm ln}(\Lambda_{\rm Coulomb})={\rm ln}(M/M_{\rm sat})$ is the Coulomb logarithm,
$M$ is the halo mass of the central
galaxy and $M_{\rm sat}$ is the mass of the satellite, including the mass of the DM halo in which the galaxy
was formed.

\citet{Lagos12} and \citet{Gonzalez-Perez13} used the $\Theta_{\rm orbit}$ function calculated in \citet{Lacey93},

\begin{equation}
\Theta_{\rm orbit}=\left[\frac{J}{J_{\rm c}(E)}\right]^{0.78}\, \left[\frac{r_{\rm c}(E)}{R_{\rm v}}\right]^{2},
\label{Eq:merger:Lacey}
\end{equation}

\noindent where $J$ is the initial angular momentum and $E$ is the energy of the satellite's orbit, and
$J_{\rm c}(E)$ and $r_{\rm c}(E)$ are, respectively, the angular momentum and radius of a circular orbit with the same energy as
that of the satellite. Thus, the circularity of the orbit corresponds to $J/J_{\rm c}(E)$. The function
$\Theta_{\rm orbit}$ is well described by a log normal distribution with
median value $\langle{\rm log}_{10} \Theta_{\rm orbit} \rangle=-0.14$ and
dispersion $\langle({\rm log}_{10} \Theta_{\rm orbit}-\langle{\rm log}_{10} \Theta_{\rm orbit} \rangle)^2 \rangle^{1/2}=0.26$. 
These values are not correlated with satellite galaxy properties. Therefore, for each satellite, the value of $\Theta_{\rm orbit}$
is randomly chosen from the above distribution. Note that the dependence of $\Theta_{\rm orbit}$ on $J$ in
Eq~\ref{Eq:merger:Lacey} is a fit to numerical estimates. Lacey et al. (2014) use the updated dynamical friction function 
of \citet{Jiang07}, which slightly changes the dependence on the mass ratio of the satellite to the central galaxy. The net 
effect of such a change is that minor mergers occur faster, while major mergers occur slower when compared to the 
\citet{Lacey93} prescription.

If the merger timescale is less than the time that has elapsed since the formation of the halo, i.e. if 
$\tau_{\rm merge}<t-t_{\rm form}$, we proceed to merge the satellite with the central galaxy at $t$.
If the total mass of gas plus stars of the primary (largest) and secondary galaxies involved in a merger are
$M_{\rm p}=M_{\rm cold,p}+M_{\star,p}$ and
$M_{\rm s}=M_{\rm cold,s}+M_{\star,s}$, the outcome of the galaxy merger depends on the galaxy mass ratio,
$M_{\rm s}/M_{\rm p}$,  and
the fraction of gas in the primary galaxy, $M_{\rm cold,p}/M_{\rm p}$:

\begin{itemize}
\item $M_{\rm s}/M_{\rm p}>f_{\rm ellip}$ drives a major merger. In this case all the stars
present are rearranged into a spheroid. In addition, any cold gas in the merging system is assumed to undergo
a burst of SF and the stars formed are added to the spheroid component. We typically take
$f_{\rm ellip}=0.3$, which is within the range found in simulations (e.g. see \citealt{Baugh96} for a discussion).
\item $f_{\rm burst}<M_{\rm s}/M_{\rm p}\le f_{\rm ellip}$ drives minor mergers. In this case all the stars
in the secondary galaxy are accreted onto the primary galaxy spheroid, leaving the stellar disk
of the primary intact. In minor mergers the presence of a starburst depends on the cold gas content of the
primary galaxy, as set out in the next bullet point.
\item $f_{\rm burst}<M_{\rm s}/M_{\rm p}\le f_{\rm ellip}$ and $M_{\rm cold,p}/M_{\rm p}>f_{\rm gas,burst}$ drives a starburst
in a minor merger. The perturbations introduced by the secondary galaxy are assumed to drive all the cold gas from
both galaxies to the new spheroid, producing a starburst. There is no starburst if $M_{\rm cold,p}/M_{\rm p}<f_{\rm gas,burst}$.
The Bau05 and Bow06 models adopt $f_{\rm gas,burst}=0.75$ and $f_{\rm gas,burst}=0.1$, respectively.
\item $M_{\rm s}/M_{\rm p}\le f_{\rm burst}$ results in the primary disk remaining unchanged. As before, the stars accreted
from the secondary galaxy are added to the spheroid, but the overall gas component (from both galaxies) stays in the disk,
and the stellar disk of the primary is preserved. The Bau05 and Bow06 models adopt $f_{\rm burst}=0.05$ and
$f_{\rm burst}=0.1$, respectively.
\end{itemize}

{\it Disk instabilities.} If the disk becomes sufficiently massive that its self-gravity is dominant, then it is unstable to small perturbations
by satellites or DM substructures. The criterion for instability was described by \citet{Efstathiou82} and \citet{Mo98}
and introduced into {\tt GALFORM} by \citet{Cole00},

\begin{equation}
\epsilon=\frac{V_{\rm circ}(r_{\rm d})}{\sqrt{G\, M_{\rm d}/r_{\rm s}}}.
\label{DisKins}
\end{equation}

\noindent Here, $V_{\rm circ}(r_{\rm d})$ is the circular velocity of the disk at the half-mass radius, $r_{\rm d}$,
$r_{\rm s}$ is the scale radius of the disk and
$M_{\rm d}$ is the disk mass (gas plus stars). If $\epsilon<\epsilon_{\rm disk}$, where $\epsilon_{\rm disk}$ is a parameter, then 
the disk is considered to be unstable.
In the case of unstable disks, stars and gas in the disk are accreted onto the spheroid and the gas inflow drives a starburst.
Lagos12 and Gonzalez-Perez14 adopt $\epsilon_{\rm disk}=0.8$, while Lacey14 adopt a slightly higher value, 
$\epsilon_{\rm disk}=0.9$.

\subsection{Gradual Ram pressure stripping of the hot gas}\label{Sec:rampressure}

The standard treatment of the hot gas in accreted satellites in {\tt GALFORM} is usually referred to as `strangulation'\footnote{Another 
way this is referred to in the literature is `starvation', but both terms refer to the same process: complete removal 
of the hot gas reservoiro f galaxies when they become satellites.} of the hot gas.
In this extreme case, the ram pressure stripping of the satellite's hot gas reservoir by the hot gas in the main halo is 
completely efficient and is assumed to occur as soon as a galaxy becomes a satellite.
This treatment has shown to drive redder colour of satellite galaxies \citep{Font08}. As 
we are studying ETGs, which tend to be found more frequently in denser environments, 
we test the impact of a more physical and gradual process, the partial ram pressure stripping of the hot gas, 
on the neutral gas content of ETGs. Simulations show that the amount of gas removed from the satellite's hot reservoir 
depends upon the ram pressure experienced which is turn is determined by the peri-centre of the orbit 
\citep{McCarthy08}.

Here we briefly describe the more physical partial ram-pressure stripping model
introduced by \citet{Font08}.
The partial ram-pressure stripping of the hot gas is applied to a spherical distribution 
of hot gas. The model considers that all the host gas outside the stripping radius, $r_{\rm str}$, is removed from the 
host gas reservoir and transferred to the central galaxy halo. The stripping radius is defined 
as the radius where the ram pressure, $P_{\rm ram}$, equals the gravitational restoring force per unit area of the 
satellite galaxy, $P_{\rm grav}$. The ram pressure is defined as,

\begin{equation}
P_{\rm ram} \equiv \rho_{\rm gas,p}\,v^{2}_{\rm sat},
\label{RPdefinition}
\end{equation}

\noindent and the gravitational pressure as 

\begin{equation}
P_{\rm grav}\equiv \alpha_{\rm rp}\,\frac{G\,M_{\rm tot,sat}(r_{\rm str})\,\rho_{\rm gas,s}(r_{\rm str})}{r_{\rm str}}. 
\label{Pgravdefinition}
\end{equation}

\noindent Here, $\rho_{\rm gas,p}$ is the gas density of the parent halo, $v_{\rm sat}$ is the velocity of the satellite with 
respect to the parent halo gas medium, $M_{\rm tot,sat}(r_{\rm str})$ is the total mass of the satellite galaxy (stellar, gas and 
dark matter components) enclosed within $r_{\rm str}$ and $\rho_{\rm gas,s}(r_{\rm str})$ is the hot gas density of the satellite galaxy 
at $r_{\rm str}$. In this model, $r_{\rm str}$ is measured from the centre of the satellite galaxy sub-halo. The coefficient 
$\alpha_{\rm rp}$ is a geometric constant of order unity. In this paper we use $\alpha_{\rm rp}=2$ which is the value found by 
\citet{McCarthy08} in their hydrodynamical simulations. The hot gas of the parent halo follows the density profile 
of Eq.~\ref{beta-prof}. 

This model assumes that the hot gas of the satellite galaxy inside $r_{\rm str}$ remains intact while the hot gas outside 
is stripped on approximately a sound crossing time. In {\tt GALFORM}, $r_{\rm str}$ is calculated at the time a galaxy becomes 
satellite solving Eq.~\ref{RPdefinition} and setting the ram pressure to its maximum value, which occurs at the peri-centre 
of the orbit of the satellite galaxy. The hot gas outside $r_{\rm str}$ is instantaneously stripped once the galaxy crosses the virial 
radius of the parent halo. This simplified modelling overestimates the hot gas stripped between the time the satellite galaxy crosses 
the virial radius and the first passage. \citet{Font08} argue that this is not a bad approximation as the timescale for the latter 
is only a small fraction of the time a satellite galaxy spends orbiting in the parent halo. Font et al. also argue that in terms of 
hot gas removal, ram pressure is the major physical mechanism, while tidal heating and stripping are secondary effects. 

The infall velocity of the satellite galaxy is randomly sampled from the $2$-dimensional distribution of infalling velocity 
 of the dark matter substructures, measured by \citet{Benson05} from a large suite of cosmological simulations. Then, the peri-centre 
radius and velocity at the peri-centre are computed by assuming that the orbital energy and angular momentum are conserved and 
by treating the satellite as a point mass orbiting within a Navarro-Frenk-White gravitational potential with the same 
total mass and concentration as the parent halo. 

The remaining hot gas in the satellite galaxy halo can cool down and feed the satellite's disc. The cooling of this remaining hot gas 
is calculated by assuming that the mean density of the hot gas of the satellite is not altered by the stripping process, 
and using a nominal hot halo mass that includes both the current hot gas mass and the hot gas mass that has been stripped. The difference 
with the standard calculation described in $\S$~\ref{Sec:Cooling} is that the cooling radius cannot be larger than $r_{\rm str}$. 
As star formation continues to take place in satellite galaxies, there will be an additional source of hot gas which corresponds 
to the winds escaping the galaxy disk that mix or evaporate to become part of the hot halo gas. Most of this star formation takes place 
when the satellite galaxy is on the outer parts of its orbit, where the ram pressure is small. \citet{Font08} then suggested that a fraction, 
$\epsilon_{\rm strip}$ (less than unity) of this gas is actually stripped from the hot halo of the satellite. 
Font et al. discussed the effects of different values for $\epsilon_{\rm strip}$ and adopted $\epsilon_{\rm strip}=0.1$ 
to reproduce the colours of satellite galaxies. Throughout this paper we adopt the same value for $\epsilon_{\rm strip}$, but 
we discuss in Appendix~\ref{RPpars} the effect of varying it.

Finally, in order to account for the growth of the parent halo and the effect this has on the ram pressure, 
 the ram pressure is recalculated for each satellite galaxy every time the parent halo doubles its mass 
compared to the halo mass at the instant of the initial stripping event. 

We test the effect of partial ram pressure stripping of the hot gas by including the above modelling into the three 
variants of {\tt GALFORM}, and we refer to 
the variants with partial ram pressure stripping of the hot gas as 
Lagos12+RP, Gonzalez-Perez14+RP and Lacey14+RP.

\subsection{Differences between the Lagos12, Gonzalez-Perez14 and Lacey14 models}\label{Models} 

The Lagos12 model is a development of the model originally described in \citet{Bower06}, which was the 
first variant of {\tt GALFORM} to include AGN feedback as the mechanism suppressing gas cooling 
in massive halos. The Lagos12 model assumes a universal initial mass function (IMF), 
the \citet{Kennicutt83} IMF\footnote{The distribution of the masses of stars 
formed follows ${\rm d}N(m)/{\rm d\, ln}\,m \propto m^{-x}$, where $N$ is the number of stars of mass $m$ formed, 
 and $x$ is the IMF slope. For a \citet{Kennicutt83} IMF, $x=1.5$ for masses in the range $1\,M_{\odot}\le m\le 100\,M_{\odot}$ and 
$x=0.4$ for masses $m< 1\,M_{\odot}$.}.
Lagos12 extend the model of Bower et al. by including the self-consistent SF law 
described in $\S$~\ref{SFlaw}, and adopting $\nu_{\rm SF}=0.5\,\rm Gyr^{-1}$, 
$\rm log(P_{0}/k_{\rm B} [\rm cm^{-3} K])=4.23$, where 
$\rm k_{\rm B}$ is is Boltzmann's constant, and $\alpha=0.8$, 
which correspond to the values of the parameters reported by \citet{Leroy08} for local spiral 
and dwarf galaxies. 
This choice of SF law greatly reduces the parameter space of the model and also 
extends its predictive power by directly modelling the atomic and molecular hydrogen content of galaxies. 
All of the subsequent models that use the same SF law have also the ability to predict
the HI and H$_2$ gas contents of galaxies. 
Lagos12 adopt longer duration starbursts (i.e. larger $f_{\rm dyn}$) 
compared to Bower et al. to improve the agreement with the observed luminosity 
function in the rest-frame ultraviolet (UV) at high redshifts. Lagos12 adopts $\tau_{\rm min}=100\, \rm Myr$ and 
$f_{\rm dyn}=50$ in Eq.~\ref{SFlawSB}. The Lagos12 model was developed in the Millennium simulation, which 
assumed a WMAP1 cosmology \citep{Spergel03}.  

The Gonzalez-Perez14 model updated the Lagos12 model to the WMAP7 cosmology \citep{Komatsu11}. A small number 
of parameters were recalibrated to recover 
 the agreement between the model predictions and the observed evolution 
of both the UV and $K$-band luminosity functions. These changes include 
a slightly shorter starburst duration, i.e.  $\tau_{\rm min}=50\, \rm Myr$ and $f_{\rm dyn}=10$, and 
 weaker supernovae feedback. See \citet{Gonzalez-Perez13} for more details.

The Lacey14 model is also developed in the WMAP7 cosmology but it differs 
from the other two flavours in that it adopts a bimodal IMF. The IMF describing 
SF in disks (i.e. the quiescent mode) is the same as the universal IMF in the other two models, but 
a top-heavy IMF is adopted for starbursts (i.e. with an IMF slope $x=1$). This choice motivated by \citet{Baugh05} who used a bimodal 
IMF to recover the agreement between the model predictions and observations of 
the number counts and redshift distribution of submillimeter galaxies.
 We note, however, that Baugh et al. adopted a more top-heavy IMF for starbursts with 
 $x=0$. 
The stellar population synthesis model used for Lacey14 is also different. 
While both Lagos12 and Gonzalez-Perez14 use \citet{Bruzual03}, the Lacey14 model uses 
\citet{Maraston05}.
Another key difference between the Lacey14 model and the other two {\tt GALFORM} flavours considered here, is that 
 Lacey14 adopt a slightly larger value of the SF efficiency rate, $\nu_{\rm SF}=0.74\,\rm Gyr^{-1}$, still within 
the range allowed {by the most recent observation compilation of \citet{Bigiel11}}, making SF more efficient. 

\subsection{The $N$-body simulations and cosmological parameters}\label{Cosmos}

We use halo merger trees extracted from the Millennium cosmological N-body
simulation (adopting WMAP1 cosmology; \citealt{Springel05}) and its WMAP7 counterpart.
The Millennium simulation\footnote{Data from the Millennium simulation is available on a relational 
database accessible from {\tt http://galaxy-catalogue.dur.ac.uk:8080/Millennium}.} has the following cosmological parameters:
$\Omega_{\rm m}=\Omega_{\rm DM}+\Omega_{\rm baryons}=0.25$ (giving a
baryon fraction of $0.18$), $\Omega_{\Lambda}=0.75$, $\sigma_{8}=0.9$
and $h=0.73$. The resolution of the $N$-body
simulation is fixed at a halo mass of $1.72 \times 10^{10} h^{-1} M_{\odot}$. 
\citet{Lagos14} show that much higher resolution merger trees are needed to fully resolve the 
HI content of galaxies from $z=0$ to $z=10$. However, in this work we are concerned about 
 galaxies with $L_{\rm K}\gtrsim 10^9\,L_{\odot}$, which are well resolved in the Millennium simulations. 
The Lacey14 and Gonzalez-Perez14 were developed in the WMAP7 version of the Millenniun simulation, where the cosmological parameters are 
 $\Omega_{\rm m}=\Omega_{\rm DM}+\Omega_{\rm baryons}=0.272$ (with a
baryon fraction of $0.167$), $\Omega_{\Lambda}=0.728$, $\sigma_{8}=0.81$
and $h=0.704$ (WMAP7 results were presented in \citealt{Komatsu11}). 

Throughout this work we show gas masses in units of $M_{\odot}$, luminosities 
in units of $L_{\odot}$ and number densities in units of $\rm Mpc^{-3}\,dex^{-1}$. This implies 
that we have evaluated the $h$ factors.
The largest difference driven by the different cosmologies is in the number density, but this is only $0.05$~dex, which 
is much smaller than the differences between the models or between model and observations.

\section{The neutral gas content of local early-type galaxies: models vs. observations}\label{Sec:ModelComparison}

\begin{figure}
\begin{center}
\includegraphics[trim = 0.9mm 0.3mm 1mm 3.45mm,clip,width=0.47\textwidth]{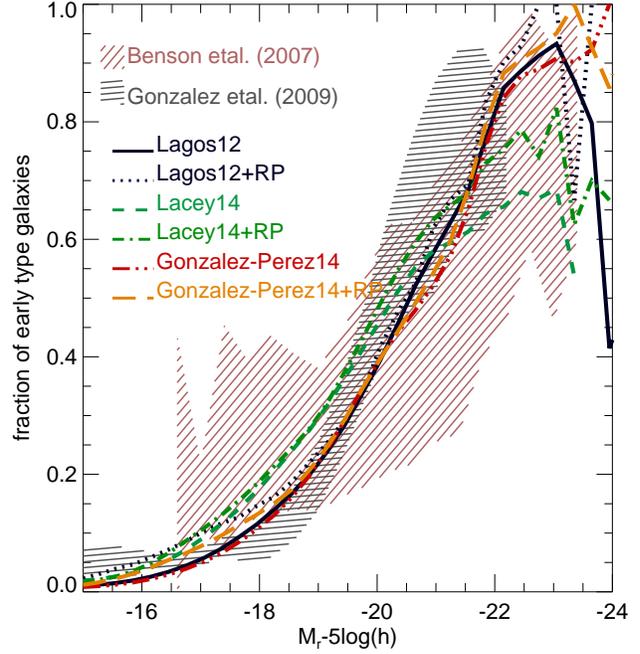}
\caption{Fraction of ETGs as a function of the rest-frame $r$-band absolute magnitude, 
for the Lagos12, Lacey14, Gonzalez-Perez14 and the variants including partial ram pressure of the hot gas (`+RP'), as labelled.
 Early-type galaxies in the models are those with a bulge-to-total stellar mass ratio 
$B/T>0.5$. The shaded regions correspond to the observational estimates of \citet{Benson07} and 
\citet{Gonzalez09} using the SDSS, as labelled. In the case of \citet{Gonzalez09} the upper limit of the shaded region is given by 
the S\'ersic index selection, while the lower limit is given by the concentration selection (see text for details).}
\label{ETratios}
\end{center}
\end{figure}

In this section we compare the predicted properties of ETGs in the three variants of 
{\tt GALFORM} with and without the inclusion of partial ram pressure stripping of the hot gas.
This modification has little effect on the
local $b_J$-band and $K$-band luminosity functions in the three variants 
of {\tt GALFORM}. These observables are usually 
considered to be the main constraints for finding the best set of model 
parameters (see for example \citealt{Bower10} and \citealt{Ruiz13}). 
Other $z=0$ properties, such as half-mass radii, gas and stellar metallicity, are also 
insensitive to the inclusion of partial ram pressure stripping. We therefore conclude that the partial 
ram pressure stripping versions of the 
three {\tt GALFORM} models provide a representation of the local Universe as good as the standard models. 

The first comparison we perform is the fraction of galaxies that are ETGs as a function of galaxy luminosity. This is a 
crucial step in our analysis, as we aim to characterise the neutral gas content of the ETG population. 
Throughout we will refer to ETGs in the model as those having bulge-to-total stellar mass 
ratios, $B/T$, $>0.5$. Although this selection criterion is very sharp and has been analysed in detail in the literature (see for example 
 \citealt{Weinzirl09} and \citealt{Khochfar11}), we find that our results are not sensitive to 
 the threshold $B/T$ selecting ETGs, as long at this threshold is $B/T>0.3$. We analyse this selection criterion  
in more detail in $\S$~\ref{Robustness}. 

Fig.~\ref{ETratios} shows the fraction of ETGs, $f_{\rm early}$, as a function of the $r$-band
absolute magnitude at $z=0$ for 
the three {\tt GALFORM} models described in \S~\ref{Models} and their variants including partial ram pressure of the hot gas.
Observational estimates of $f_{\rm early}$ are also shown in Fig.~\ref{ETratios}, for 
three different ways of selecting ETGs. The first one corresponds to \citet{Benson07}, in which 
a disc and a bulge component were fitted to $r$-band 
images of $8,839$ bright galaxies selected from the SDSS Early Data Release. The free parameters of the fitting of the disk and bulge
components of each galaxy are the bulge ellipticity and disc
inclination angle, $i$. The second corresponds to \citet{Gonzalez09}, where the $r$-band concentration, $c$, and S\'ersic index, $n$, 
of the SDSS were used to select ETGs: $c>2.86$ or $n>2.5$. The upper and lower limits in the shaded region for the Gonzalez et al. 
measurements correspond to the two early-type selection criteria. 

All the models predict a trend 
of increasing $f_{\rm early}$ with increasing $r$-band luminosity in good agreement with the observations within the errorbars. 
Note that the inclusion of partial ram pressure stripping of the hot gas leads to a 
slightly larger $f_{\rm early}$ in galaxies with $M_{r}-5\,\rm log(\it h\rm)>-18$ in the three variants 
of {\tt GALFORM}. The same happens for the brightest galaxies,  $M_{r}-5\,\rm log(\it h\rm)<-22$.
Both trends are due to the higher frequency of disk instabilities in the models when partial ram pressure stripping is included; 
the continuous fueling of 
neutral gas in satellite galaxies due to cooling from their hot halos (which in the case of partial ram pressure stripping 
is preserved to some extent) drives more star formation in discs, lowering the 
stability parameter of Eq.~\ref{DisKins}. These lower stability parameters result in more disk instabilities, driving the formation 
of spheroids. The space allowed by the observations is large enough so that we cannot 
discriminate between models.

\subsection{The H$_2$-to-HI mass ratio dependence on morphology}

\begin{figure}
\begin{center}
\includegraphics[trim = 0.0mm 0.3mm 1mm 3.45mm,clip,width=0.49\textwidth]{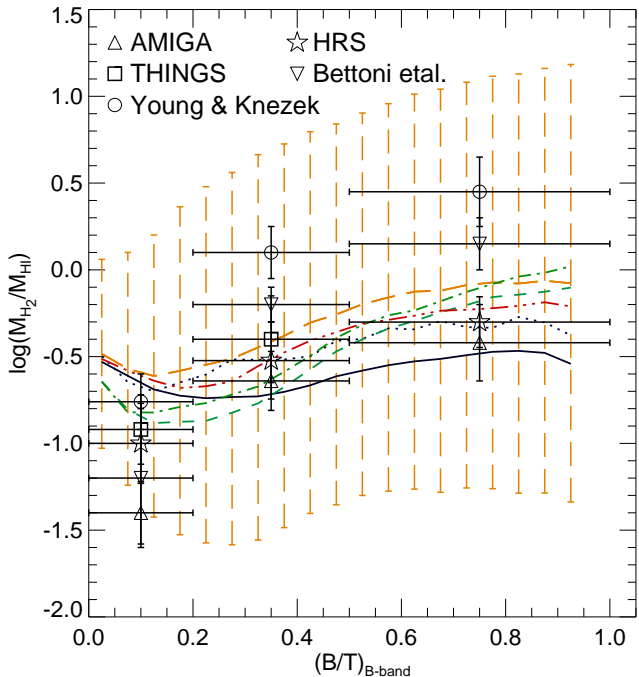}
\caption{{Molecular-to-atomic hydrogen mass ratio, $M_{\rm H_2}/M_{\rm HI}$, as a function
of the bulge-to-total luminosity in the $B$-band, $(B/T)_{B}$, in the same models of Fig.~\ref{ETratios} (lines as labelled in 
Fig.~\ref{ETratios}), 
for galaxies with
absolute $B$-band magnitudes, $M_B-5\rm \, log(h)<-19$.
Lines show the median of the relations for each model and the 
$10$ and $90$ percentiles are only shown for the Gonzalez-Perez14+RP model, as an illustration of the 
dispersion. The other model show very similar $10$ and $90$ percentiles.
Observational results from \citet{Young89}, \citet{Bettoni03},
\citet{Leroy08} (THINGS sample), \citet{Lisenfeld11} (AMIGA sample) and \citet{Boselli14b} (HRS sample) 
are shown as symbols, as labelled, and we combine them so that
$B/T<0.2$ corresponds to Irr, Sm, Sd galaxies; $0.2<B/T<0.5$ corresponds
to Sc, Sb, Sa galaxies; $B/T>0.5$ corresponds to E and S0 galaxies (see
\citealt{deVaucouleurs91} for a description of each morphological type).}}
\label{ScalingMstar2}
\end{center}
\end{figure}

It has been shown observationally that the ratio between the  
H$_2$ and HI masses correlates strongly with morphological type, with ETGs 
 having higher H$_2$/HI mass ratios than late-type galaxies
(e.g. \citealt{Young89}; \citealt{Bettoni03}; \citealt{Lisenfeld11}; \citealt{Boselli14b}).
Fig.~\ref{ScalingMstar2} shows the H$_2$/HI mass ratio in the models as a function of the
bulge-to-total luminosity ratio in the $B$-band, $B/T_{\rm B}$, for all galaxies 
with $B$-band
absolute magnitude of $M_B-5\rm \, log(h)<-19$.
This magnitude limit is chosen as it roughly corresponds to the selection criteria
applied to the observational data shown in Fig.~\ref{ScalingMstar2}.
The observational data have morphological types
derived from a visual classification of $B$-band images \citep{deVaucouleurs91}, and have
also been selected in blue bands
(e.g. \citealt{Simien86}; \citealt{Weinzirl09}).

The models predict a relation between the H$_2$/HI mass ratio and $B/T_{\rm B}$
that is in good agreement with the observations.
Note that, for $B/T_{\rm B}<0.2$, the models slightly overpredict the median 
H$_2$/HI mass ratio. The latter has been also observed in the recent Herschel 
Reference Survey (HRS, \citealt{Boselli14b}; stars in Fig.~\ref{ScalingMstar2}) for 
galaxies of morphological types later than Sd (including irregular galaxies).
{In {\tt GALFORM} there is a monotonic relation between H$_2$/HI mass ratio and 
stellar mass in a way that H$_2$/HI decreases with decreasing stellar mass (see \citealt{Lagos11} for a detailed discussion).
On the other hand the relation between stellar mass and $B/T$ is not monotonic, in a way that 
the median stellar mass in the lowest $B/T$ bins ($B/T<0.1$) is higher than 
at $B/T\sim 0.2$. This drives the slight increase of H$_2$/HI at the lowest $B/T$. 
The physical reason why $B/T$ is not monotonically correlated to 
stellar mass is because environment plays an important role in the morphology
(for example in the number of galaxy mergers, and disk instabilities), which makes it a more 
transient property of galaxies, while stellar mass is not necessarily correlated to environment
but to halo mass. This will be discussed in more detail in paper II.}
 
Of the three {\tt GALFORM} variants, the model that predicts the steepest slope for the relation between H$_2$/HI mass ratio and $B/T$ 
is the Lacey14 model. However, when including partial 
ram pressure stripping of the hot gas, all the models show a slight increase in this slope, with 
ETGs having higher H$_2$/HI mass ratios. This comes from the higher gas surface densities that ETGs 
have on average when including partial ram pressure, which drive higher hydrostatic pressure. 
The strangulation scenario, which removes the hot gas instantaneously, drives a quick depletion of the cold gas reservoir in galaxies as 
star formation continues, while in the partial ram pressure scenario the cold gas reservoir is still replenished due 
to continuous inflow of gas from the satellite's remaining hot halo.

\subsection{The HI content of early-type galaxies}

\begin{figure}
\begin{center}
\includegraphics[trim = 0.9mm 0.3mm 1mm 0mm,clip,width=0.47\textwidth]{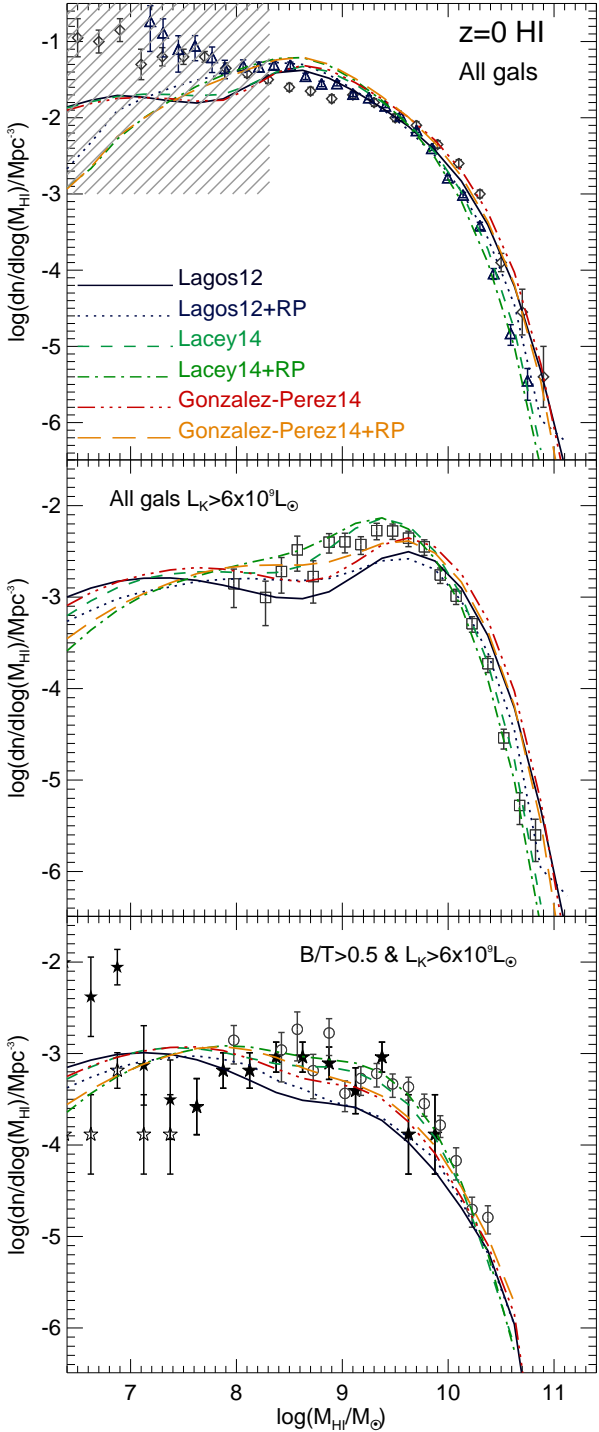}
\caption{{\it Top panel:} The HI mass function of all galaxies at $z=0$ for the models Lagos12, Lacey14 and Gonzalez-Perez14 
and its variants including partial ram pressure stripping of the hot gas (`+RP'), as labelled. Observations correspond to 
\citet{Zwaan05} (triangles) and \citet{Martin10} (diamonds). The shaded region shows the range where 
the number density of galaxies decline due to halo mass resolution effects.
{\it Middle panel:} as in the top panel, but here the HI mass function is shown for galaxies with $K$-band luminosities 
$L_{\rm K}>6\times 10^9\,L_{\odot}$. Observations correspond to the analysis of HIPASS presented in $\S$~\ref{obssec}.
{\it Bottom panel:} as in the top panel but for ETGs (those with a bulge-to-total stellar mass 
ratio $>0.5$) and $K$-band luminosities  
$L_{\rm K}>6\times 10^9\,L_{\odot}$. Observations correspond to the analysis of HIPASS (circles) and the ATLAS$^{\rm 3D}$ 
(with and without volume correction as filled and empty stars, respectively)
surveys containing galaxies with $L_{\rm K}>6\times 10^9\,L_{\odot}$ and described 
in $\S$~\ref{obssec}.}
\label{HIRPcomp}
\end{center}
\end{figure}

\begin{figure}
\begin{center}
\includegraphics[trim = 0.9mm 0.3mm 1mm 0mm,clip,width=0.49\textwidth]{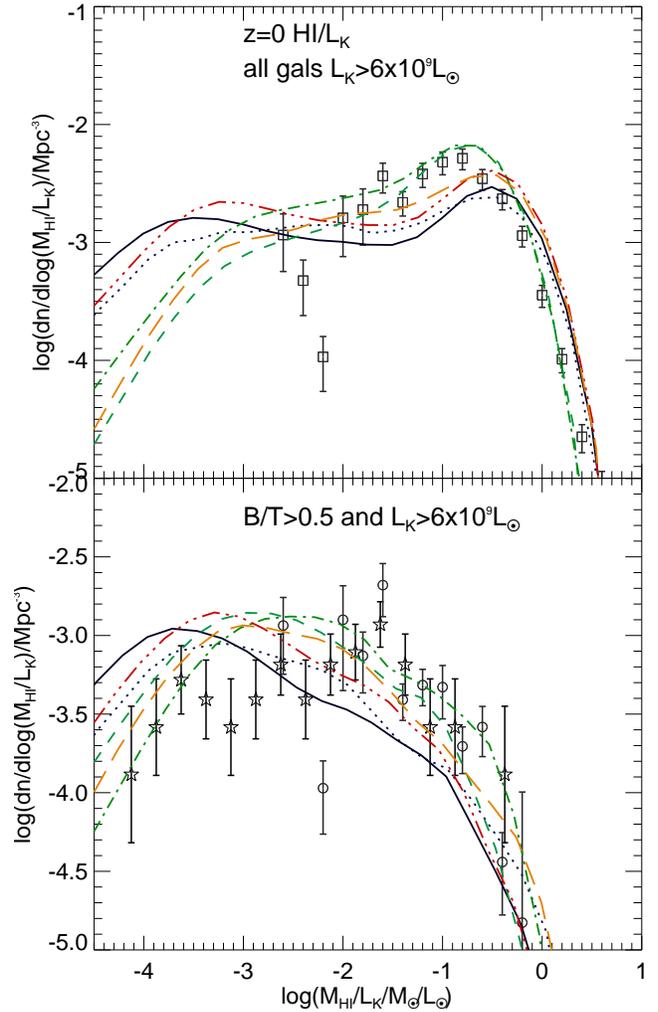}
\caption{{\it Top panel:} the distribution function of the ratio between the HI mass and the $K$-band luminosity 
for the same models shown in Fig.~\ref{HIRPcomp} for galaxies with $L_{\rm K}>6\times 10^9\,L_{\odot}$. Observations 
correspond to the analysis of HIPASS presented in $\S$~\ref{obssec}.
{\it Bottom panel:} As in the top panel but for ETGs ($B/T>0.5$) with $L_{\rm K}>6\times 10^9\,L_{\odot}$.
Observations correspond to the analysis of HIPASS (circles) and ATLAS$^{3D}$ (stars) presented in $\S$~\ref{obssec}.}
\label{HIRPcompv2}
\end{center}
\end{figure}

Fig.~\ref{HIRPcomp} shows the model predictions for the HI mass function for all galaxies (top panel), for the subsample of 
galaxies with $K$-band luminosities $L_{\rm K}>6\times 10^9\,L_{\odot}$ (middle panel), and for ETGs with the same 
 $K$-band luminosity cut (bottom panel).
The observational results in Fig.~\ref{HIRPcomp} are described in \S~\ref{obssec}. 
For the overall galaxy population (top panel of Fig.~\ref{HIRPcomp}), the six models 
provide a good description of the HI mass function, and the inclusion of partial 
ram pressure stripping of the hot gas has little effect. 

For the galaxy population with $L_{\rm K}>6\times 10^9\,L_{\odot}$ (middle panel of Fig.~\ref{HIRPcomp}), the Lagos12 and Gonzalez-Perez14 
models predict a slightly lower number density at the peak of the HI mass distribution compared to 
the observations, while the Lacey14 model 
predicts a HI mass distribution in good agreement with the observations throughout the full HI mass range. 
Note that the inclusion of partial ram pressure stripping of the hot gas has the effect of increasing the number density 
of galaxies with $10^8\,M_{\odot}<M_{\rm HI}<3\times 10^9\,M_{\odot}$, and decreasing the number density of 
galaxies with $M_{\rm HI}<10^8\,M_{\odot}$. The reason for this is that many of the galaxies with low 
HI masses ($M_{\rm HI}<10^8\,M_{\odot}$) become more gas rich when partial ram pressure stripping 
is included {compared to the case of strangulation of the hot gas}, and move to higher HI masses 
($10^8\,M_{\odot}<M_{\rm HI}<3\times 10^9\,M_{\odot}$). 
This slight change improves the agreement with the observations in the three {\tt GALFORM} 
models, particularly around the turnover at $M_{\rm HI}\approx 5\times 10^9\,M_{\odot}$ in the 
mass function shown in the middle panel of Fig.~\ref{HIRPcomp}.

In the case of ETGs with $L_{\rm K}>6\times 10^9\,L_{\odot}$ (bottom panel of Fig.~\ref{HIRPcomp}), 
the Lagos12 model predicts a number density of 
galaxies with $10^{9}\,M_{\odot}<M_{\rm HI}<10^{10}\,M_{\odot}$ lower than observed, while 
the predictions from the Gonzalez-Perez14 and Lacey14 models agree well with the observations.
The inclusion of 
partial ram pressure stripping of the hot gas in the three models has the effect of increasing the 
number density of ETGs with HI masses $M_{\rm HI}>10^8\,M_{\odot}$. This increase 
allows the models to get closer to the observations, and particularly the Lacey14+RP model predicts 
a HI mass function of ETGs in very good agreement with the observations.

The HI mass function has become a standard constraint on the {\tt GALFORM} model since 
\citet{Lagos11}. However, \citet{Lemonias13} show that a stronger constraint on simulations of galaxy formation is 
provided by the  
conditional mass function of gas, or similarly, the gas fraction distribution. Here, we compare 
the predictions for the HI gas fraction distribution function with observations in Fig.~\ref{HIRPcompv2}. 
The HI gas fraction is taken with respect to the $K$-band luminosity to 
allow direct comparisons with the observations without the need of having to convert between different adopted IMFs, which 
would be the case if stellar mass was used.
 \citet{Mitchell13} show that when simulations and observations adopt different IMF, the 
comparison between the stellar masses predicted by the models 
and observations is misleading. Instead, a full SED fitting needs to be performed 
to make a fair comparison in such a case. Note that the same applies when the star formation histories 
adopted in the observations differ significantly from the simulated galaxies.

The top panel of 
Fig.~\ref{HIRPcompv2} shows the predicted HI gas fraction for galaxies with $L_{\rm K}>6\times 10^9\,L_{\odot}$.
The observations are described in \S~\ref{obssec}.
The Lagos12 and Gonzalez-Perez14 
models predict a peak of the gas fraction distribution at higher gas fractions than observed, while the Lacey14 predicts a 
peak closer to the observed one (i.e. $M_{\rm HI}/L_{\rm K}\approx 0.15\,M_{\odot}/L_{\odot}$). 
The galaxies at the peak of the HI gas fraction 
distribution also lie in the main sequence of galaxies in the SFR-stellar mass plane \citep{Lagos10}. 
The inclusion of partial ram pressure of the hot gas 
 increases the number density of galaxies with HI gas fractions $M_{\rm HI}/L_{\rm K}>10^{-3}\,M_{\odot}/L_{\odot}$ 
and reduces the number density 
of galaxies with lower HI gas fractions. The physical reason behind these trends is that 
the inclusion of partial ram pressure stripping increases the 
HI gas fraction of gas poor galaxies, compared to the strangulation scenario, due to replenishment of their cold gas 
reservoir. The bottom panel of Fig.~\ref{HIRPcompv2} shows the HI gas fraction distribution of ETGs ($B/T>0.5$) 
with $L_{\rm K}>6\times 10^9\,L_{\odot}$. The HI gas fraction distribution of 
ETGs is very different from that of all galaxies, showing a much broader distribution with 
a tail to very low HI gas fractions in all six models. The Lagos12 and Gonzalez-Perez14 models predict lower HI gas fractions 
for ETGs than observed, while the predictions from Lacey14 are closer to the observations. 
 By including partial ram pressure stripping of the hot gas, the HI gas fractions increase in all the models due to the 
replenishment of the cold gas reservoirs in satellite galaxies. The predictions of the Lacey14+RP model are a very good description 
of the observations, with a peak in the number density of ETGs in the range 
 $M_{\rm HI}/L_{\rm K}\approx 0.002-0.02\,M_{\odot}/L_{\odot}$.
Overall, the HI content of ETGs is moderately sensitive to the treatment of the hot gas content of satellite galaxies,  
while the overall galaxy population does not show the same sensitivity to this physical process due to the dominance of 
central galaxies.

\subsection{The H$_2$ content of early-type galaxies}

\begin{figure}
\begin{center}
\includegraphics[trim = 0.9mm 0.3mm 1mm 0mm,clip,width=0.49\textwidth]{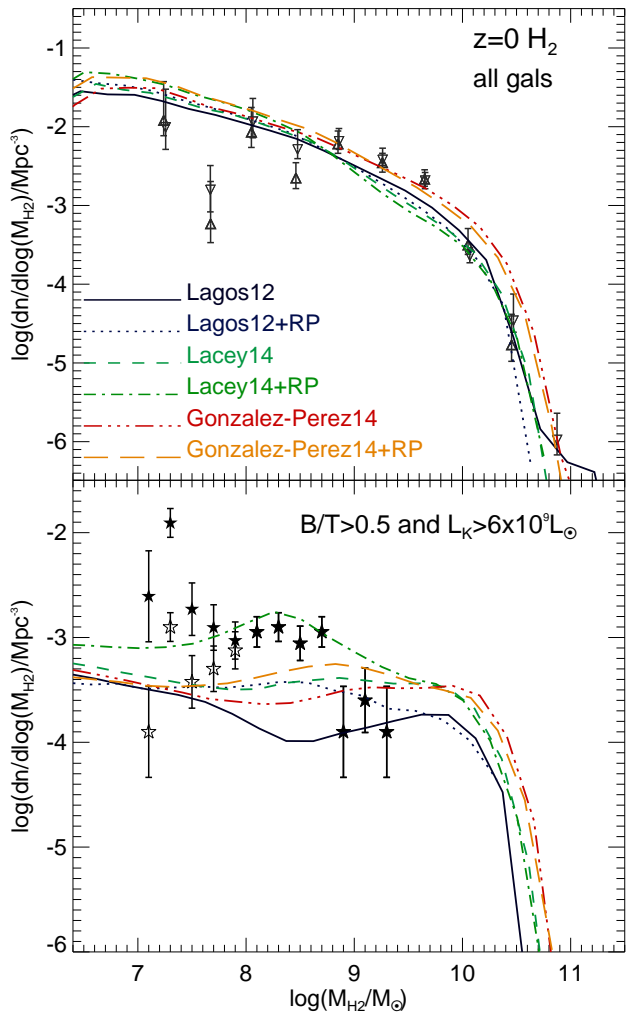}
\caption{{\it Top panel:} the H$_2$ mass function at $z=0$ for all galaxies in the Lagos12, Lacey14 and Gonzalez-Perez14 models 
and their variants 
including partial ram pressure stripping of the hot gas (`+RP').  
Observations 
from the $60\mu$m (downwards pointing triangles) and $B$-band (triangles) samples of \citet{Keres03} are also shown (see $\S$~\ref{obssec}). 
{\it Bottom panel:} as in the top panel but for ETGs ($B/T>0.5$) 
with $L_{\rm K}>6\times 10^9\,L_{\odot}$. Observations correspond to 
ETGs from the ATLAS$^{3D}$ survey with (filled stars) and without (open stars) volume 
correction (see \S~\ref{obssec} for details).}
\label{H2RPcomp}
\end{center}
\end{figure}

\begin{figure}
\begin{center}
\includegraphics[trim = 0.9mm 0.3mm 1mm 0mm,clip,width=0.47\textwidth]{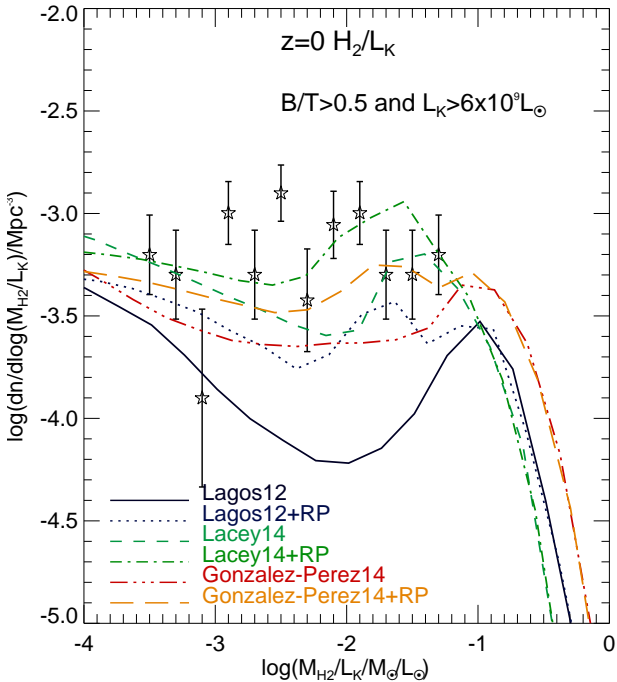}
\caption{The distribution function of the ratio between the H$_2$ mass and the $K$-band luminosity for 
ETGs ($B/T>0.5$)
with $L_{\rm K}>6\times 10^9\,L_{\odot}$, for the models as labelled. Observations correspond to 
ETGs from the ATLAS$^{3D}$ survey (see \S~\ref{obssec} for details of the 
observational dataset).}
\label{H2RPcompv2}
\end{center}
\end{figure}

The top panel of Fig.~\ref{H2RPcomp} shows the predicted H$_2$ mass functions for all galaxies.
The observations are from \citet{Keres03} and are described in \S~\ref{obssec}.   
All the models provide a good description of the observations. The largest differences between the models 
are at the high-mass end. The Gonzalez-Perez14 model predicts the highest number densities of galaxies 
with $M_{\rm H_2}>5\times 10^{9}\,M_{\odot}$, 
although it is still in agreement with the observations within the errorbars. Note that the inclusion of partial ram pressure 
stripping of the hot gas leads to very little change.
This is due to the dominance of central galaxies in the H$_2$ mass function, which are 
only indirectly affected by the treatment of partial ram pressure stripping.
In the bottom panel of Fig.~\ref{H2RPcomp} we show the H$_2$ mass function of ETGs ($B/T>0.5$) with 
$K$-band luminosities $L_{\rm K}>6\times 10^9\,L_{\odot}$. The differences between the models are significant: the Lagos12, 
Gonzalez-Perez14 and Lacey14 models predict a number density of ETGs with H$_2$ masses 
$M_{\rm H_2}>10^7\,M_{\odot}$ much lower than is observed, by factors of $10$, $6$ and $4$, respectively. It is only when partial 
ram pressure stripping of the 
hot gas is included that their predictions get closer to the observations. In particular, the Lacey14+RP model predicts 
a number density of ETGs with $M_{\rm H_2}>10^7\,M_{\odot}$ that is very close to the observations. The Lagos12+RP and 
Gonzalez-Perez14+RP models are still 
a factor of $3-4$ lower than the observations. 
At the high-mass end of the H$_2$ mass function for all galaxies (top panel of Fig.~\ref{H2RPcomp}), 
the contribution from ETGs is significant (although not dominant), while for the HI,
ETGs are only a minor contribution. The reason for this is simply the higher H$_2$-to-HI mass ratios in ETGs compared to
late-type galaxies (see Fig.~\ref{ScalingMstar2}).

Fig.~\ref{H2RPcompv2} shows the H$_2$ gas fraction distribution for ETGs with  $L_{\rm K}>6\times 10^9\,L_{\odot}$ 
for the same six models of Fig.~\ref{H2RPcomp}. 
Similarly to the H$_2$ mass function, the Lagos12 and 
Gonzalez-Perez14 models predict a number density of ETGs with 
$M_{\rm H_2}/L_{\rm K}>10^{-3}\,M_{\odot}/L_{\odot}$ lower than observed, while their variants including 
 partial ram pressure stripping of the hot gas
 predict higher number densities, in better agreement with the observations. 
The Lacey14+RP model provides the best description of the observed H$_2$ gas fractions. The physical reason for the 
higher number density of H$_2$ `rich' ETGs in the models including 
partial ram pressure stripping of the hot gas is that the replenishment of the cold gas reservoir leads to an increase in the 
surface density of gas. Since the HI saturates at $\Sigma_{\rm H_2}\approx 10\,M_{\odot}\,\rm pc^{-2}$, due to H$_2$ self-shielding at higher 
densities, 
the effect of cold gas replenishment in the ISM has a stronger effect on the H$_2$ reservoir than on the HI. 

The incorporation of partial ram pressure stripping of the hot gas brings the models into 
 better agreement with the observed gas fractions of galaxies, and particularly of ETGs. 
This indicates that partial ram pressure of the hot gas is relevant in a wide range of environments.
Note that do not include any 
description of the ram pressure stripping of the cold gas, which has been shown to take place in clusters 
through observations of the HI and H$_2$ contents of 
galaxies in the Virgo cluster (e.g. \citealt{Cortese11}; \citealt{Boselli14}). 
However, no deficiency of HI or H$_2$ has been observed in galaxies in environments other than clusters. 
\citet{Tecce10} using galaxy formation models show that ram pressure 
stripping of the cold gas is relevant only in halos with $M_{\rm halo}>3\times 10^{14}\,h^{-1}\,M_{\odot}$.
Since most of the galaxies in the ATLAS$^{\rm 3D}$ and 
HIPASS surveys are not cluster galaxies, we expect the effect of the ram pressure stripping of the cold gas to be insignificant 
in our analysis. 

The study of the neutral gas content of ETGs offers independent constraints on the 
modelling of the ram pressure stripping of the hot gas. \citet{Font08} used the fraction of passive to active 
galaxies as the main 
constraint on the satellite's hot gas treatment.
Here we propose that the exact levels of activity or cold gas content of galaxies that are 
classified as passive offer new, independent constraints.

The Lacey14+RP model agrees the best with the
observations of HI and H$_2$ in different galaxy populations. 
This is the first time such a successful model is presented. 
\citet{Serra14} use a sample of hydrodynamical simulations of galaxies and find that 
the simulations have difficulties reproducing the HI masses of ETGs.
 This may partially be due to the small sample of simulated ETGs
analysed by Serra et al. ($50$ in total). Here, by taking the full simulated galaxy population, we can make a statistically robust comparison 
with the observed ETG population.

\subsection{Expectations for the evolution of the HI and H$_2$ mass functions}
\begin{figure}
\begin{center}
\includegraphics[trim = 0.9mm 0.3mm 1mm 0.45mm,clip,width=0.46\textwidth]{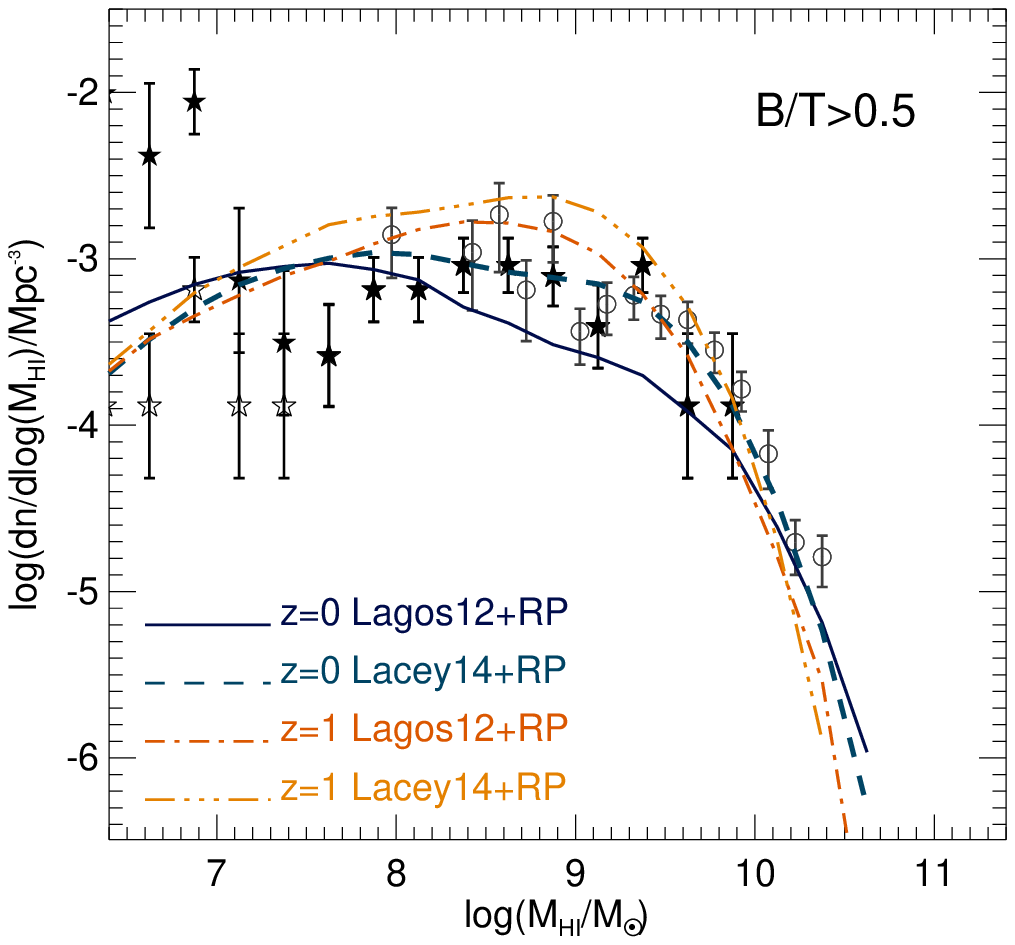}
\includegraphics[trim = 0.9mm 0.3mm 1mm 0.45mm,clip,width=0.46\textwidth]{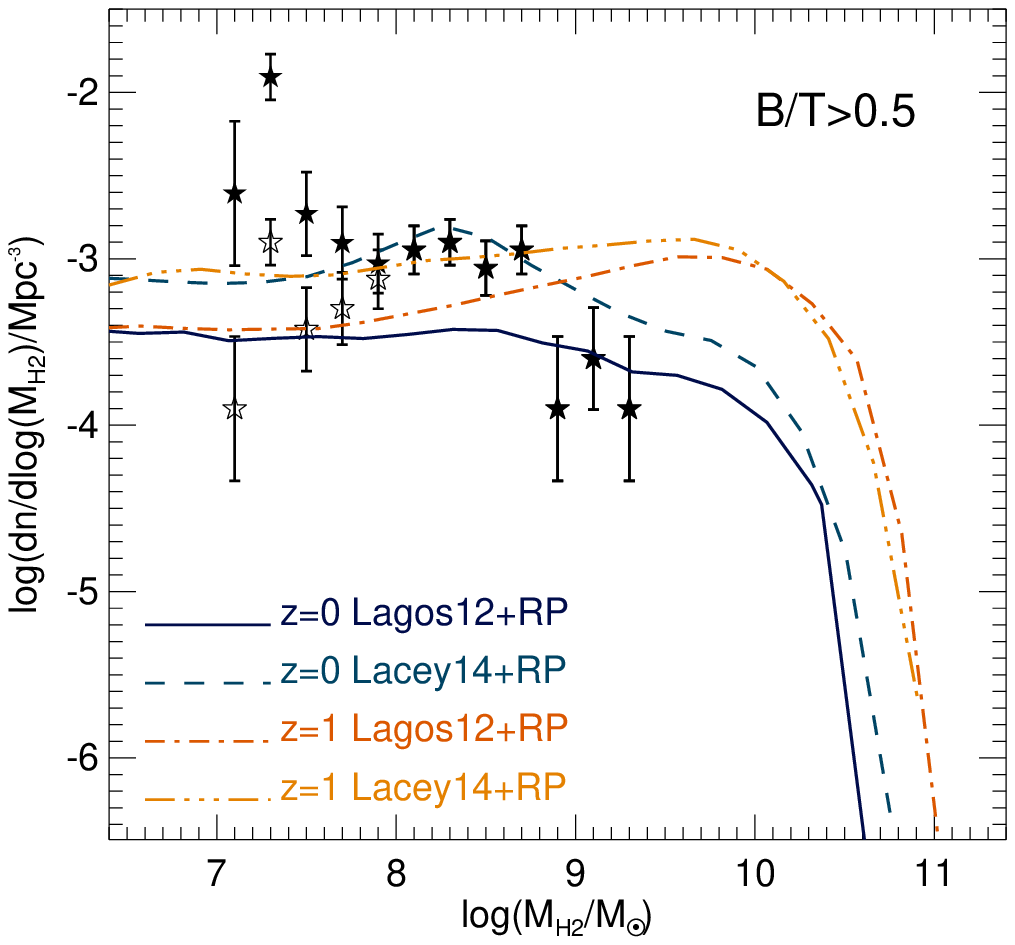}
\caption{{\it Top panel:} The HI mass function of ETGs ($B/T>0.5$) at $z=0$ and 
$z=1$ for the Lagos12+RP and Lacey14+RP models. Observations are as in the bottom panel of Fig.~\ref{HIRPcomp}.
{\it Bottom panel:} As in the top panel, but for H$_2$. Observations are as in the bottom panel of Fig.~\ref{H2RPcomp}.}
\label{HIH2Evolution}
\end{center}
\end{figure}

In the near future, the Australian Square Kilometer Array (ASKAP) and 
the South African Karoo Array Telescope (MeerKAT) will be able to trace the evolution of the HI gas content of ETGs 
towards redshifts higher than $z=0.1$, while current millimeter telescopes, such as 
the Plateau de Bure Interferometer and the Atacama Large Millimeter Array (ALMA), can already trace H$_2$ in ETGs. 
To provide insights into the expected redshift evolution of the HI and H$_2$ mass functions 
of ETGs, we show in Fig.~\ref{HIH2Evolution} the predictions for the mass functions at $z=0$ and $z=1$ 
for the Lagos12+RP and Lacey14+RP models, which give the lowest and highest number densities of ETGs (see 
Figs.~\ref{HIRPcomp} and \ref{H2RPcomp}). 

The top panel of Fig.~\ref{HIH2Evolution} shows that 
both models predict an ETG HI mass function weekly evolving with redshift. On the contrary, a large 
increase in the number density of ETGs with large H$_2$ masses, $M_{\rm H_2}>10^{10}\,M_{\odot}$, 
from $z=0$ to $z=1$ is predicted by both models (bottom panel of Fig.~\ref{HIH2Evolution}). This is driven by 
the predicted increase of the H$_2$/HI mass ratio as well as an increase in the overall gas content of ETGs with increasing redshift. 
\citet{Lagos11} and \citet{Lagos14} present detailed studies which unveil the physics behind this evolution.
In short this is due to a combination of higher gas contents and more compact galaxies at high redshift, 
which increase the hydrostatic pressure in galaxies, and therefore the H$_2$/HI mass ratio.

\section{The morphological transformation and quenching of galaxies}\label{Robustness}

\begin{figure}
\begin{center}
\includegraphics[trim = 0.1mm 0.3mm 1mm 0.45mm,clip,width=0.49\textwidth]{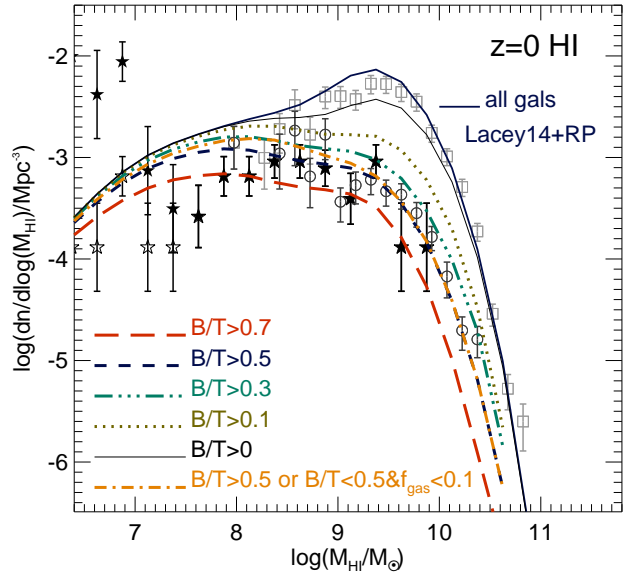}
\caption{The HI mass function 
for galaxies with  
$L_{\rm K}>6\times 10^9\,L_{\odot}$ in the Lacey14+RP model. Galaxy populations with different 
bulge-to-total mass ratios are shown: all galaxies (solid thick line), $B/T>0$ (solid thin line), 
$B/T>0.1$ (dotted line), $B/T>0.3$ (triple dot-dashed line), $B/T>0.5$ (dashed line) and 
 $B/T>0.7$ (long dashed line).
We also show the HI mass function for ETGs selected in an alternative way: 
in addition to those with $B/T>0.5$, galaxies with $B/T<0.5$
that are gas poor, $M_{\rm gas}/M_{\rm stellar}=f_{\rm gas}<0.1$, can also appear as early-types.
The latter values are consistent  
with those presented in \citet{Khochfar11}).}
\label{MorphoSel}
\end{center}
\end{figure}

\begin{figure}
\begin{center}
\includegraphics[trim = 0.1mm 0.3mm 1mm 0.45mm,clip,width=0.49\textwidth]{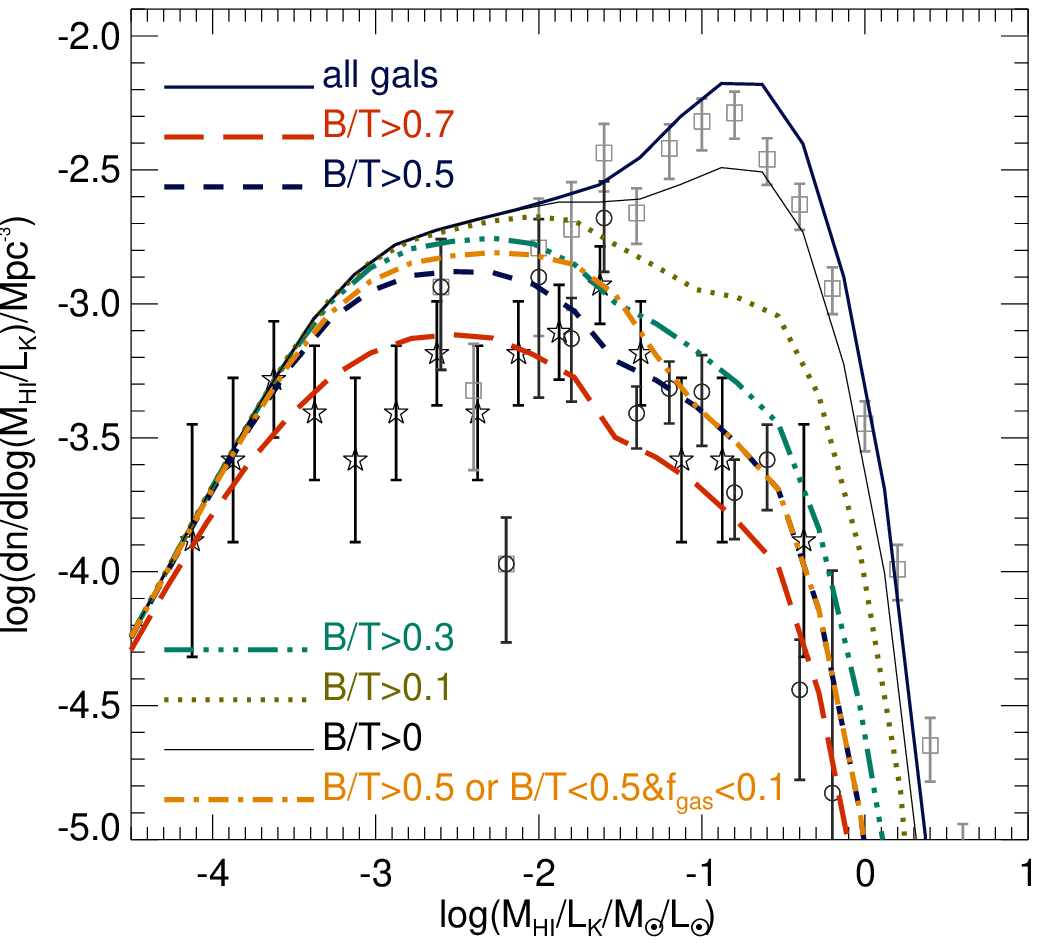}
\includegraphics[trim = 0.1mm 0.3mm 1mm 0.45mm,clip,width=0.49\textwidth]{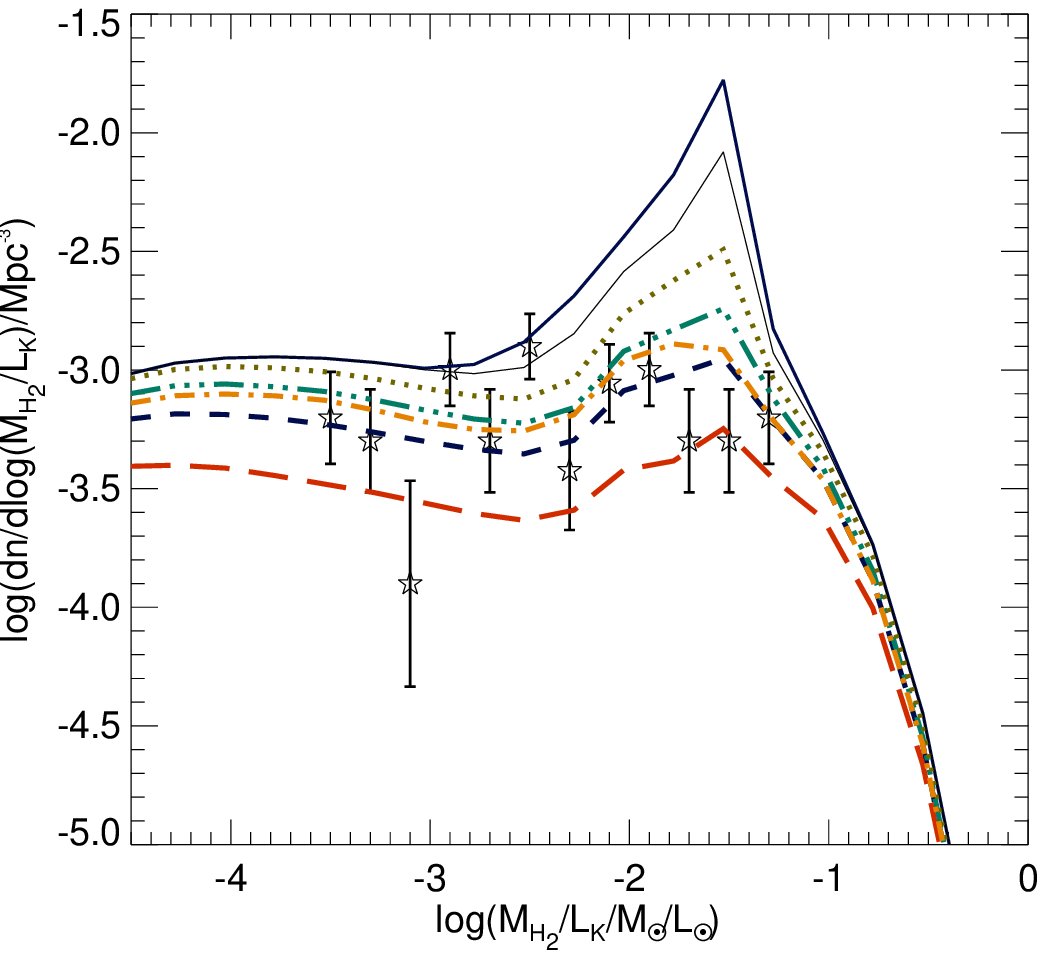}
\caption{Same as Fig.~\ref{MorphoSel} but for the HI gas fraction (top panel) and H$_2$ gas fraction (bottom panel) 
distribution functions.}
\label{MorphoSel3}
\end{center}
\end{figure}

A key aspect in the analysis performed here is the morphological selection of ETGs. Observationally, 
morphologies are derived from visual inspection of optical images of galaxies 
(see for example \citealt{Cappellari11} for the ATLAS$^{\rm 3D}$ survey sample selection). 
This means that among the galaxies selected as early-type there 
could be contamination by edge-on spirals that are gas poor, and that therefore would appear red with no spiral arms. 
Following this argument, \citet{Khochfar11} suggested that to select ETGs in the models that are comparable to the observed ones 
 one needs to include truly bulge dominated objects, which are selected by their bulge-to-total mass ratio, and 
late-type galaxies, which are gas poor. Khochfar et al. suggest the following selection to select early-type looking galaxies: $B/T>0.5$ and 
$B/T<0.5$ with gas fractions $<0.1$. Note that of the latter subpopulation, only the fraction that are edge-on oriented 
will be confused as early-types, and therefore only a small fraction will contribute to the ETG population.
We assume random inclinations for late-type galaxies, 
select those that have inclination angles $>45^{\circ}$ and add them to the sample of galaxies with $B/T>0.5$. 
The HI mass function and gas fractions of ETGs obtained using this selection criterion are shown in Fig.~\ref{MorphoSel} and 
Fig.~\ref{MorphoSel3}, respectively, for the Lacey14+RP model. The focus on the latter model as it predicts HI and H$_2$ masses 
of ETGs in best agreement with the observations.
The effect of including the contamination 
from gas poor late-type galaxies is very small and therefore does not change the results presented earlier.

Another interesting question is how much the $B/T$ threshold to select ETGs in the model affects our results. 
In order to answer this question we show in Fig.~\ref{MorphoSel} and Fig.~\ref{MorphoSel3} different $B/T$ thresholds to select ETGs in the 
Lacey14+RP model. $B/T$ thresholds lower than $0.5$ have the expected impact of increasing the number density of galaxies compared 
to the canonical value of $0.5$, particularly at 
$M_{\rm HI}>10^7\,M_{\odot}$, $M_{\rm HI}/L_{\rm K}>5\times 10^{-3}M_{\odot}/L_{\odot}$ and 
throughout the whole H$_2$ gas fraction range. This increase is of a factor of 
$2$ for $B/T=0.3$ and $7$ for $B/T=0.1$. In addition, about $40$\% of the galaxies with 
$L_{\rm K}>6\times 10^9\,L_{\odot}$ are pure disks (no bulges; see difference between thick and thin solid lines in 
Fig.~\ref{MorphoSel} and Fig.~\ref{MorphoSel3}), which are mainly located around the peaks of the 
HI and H$_2$ gas fraction distributions of all galaxies with $L_{\rm  K}>6\times 10^9\,L_{\odot}$ (see Fig.~\ref{MorphoSel3}). 
This shows that the development of a small bulge 
is connected with an important gas depletion in galaxies. Spheroids in the model are formed when galaxies undergo a starburst, 
either driven by a galaxy merger 
or a global disk instability. The fact that these galaxies remain bulge dominated is because large disks fail to regrow after 
the formation of the spheroid. To adopt a higher $B/T$ threshold has the expected effect of lowering the number density of ETGs.

The number density of galaxies with low HI content is only weekly dependent on the $B/T$ threshold. This is due to 
a connection 
between low gas fractions and large $B/T$; i.e. if a galaxy has a large bulge 
fraction it will also be gas poor. This has been observed in the ATLAS$^{\rm 3D}$ \citep{Cappellari13}, 
where ETGs with the largest velocity dispersions (a tracer of bulge fraction) have the lowest gas fractions. 
This implies that the modelling of the morphological transformation in {\tt GALFORM} is able to capture the processes 
that lead to the relation between bulge fraction and gas depletion.
Thresholds of $B/T$ used to select ETGs which are in the range $0.3-0.6$ produce similar results 
 and therefore does not
affect the analysis presented here.

\begin{figure}
\begin{center}
\includegraphics[trim = 0.1mm 0.3mm 1mm 0.45mm,clip,width=0.45\textwidth]{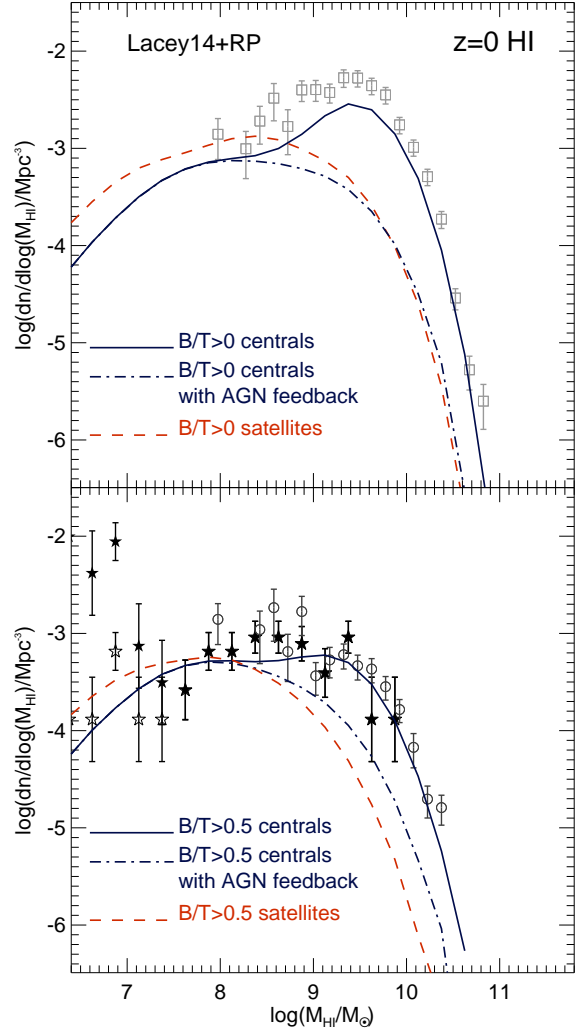}
\caption{{\it Top panel:} The HI mass function for galaxies with 
$B/T>0$ and $L_{\rm K}>6\times 10^9\,L_{\odot}$ separated into central (solid line) and satellite galaxies (dashed line) 
in the Lacey14+RP model. The subpopulation of central galaxies with AGN heating the hot halo of galaxies 
is also shown as dot-dashed line. Observations correspond to the analysis of HIPASS galaxies with $L_{\rm K}>6\times 10^9\,L_{\odot}$. 
{\it Bottom panel:} Same as in the top panel but for galaxies with 
$B/T>0.5$. Observations here correspond to the ATLAS$^{\rm 3D}$ and HIPASS, described in $\S$~\ref{obssec}.}
\label{MorphoSel2}
\end{center}
\end{figure}

The key question is why do galaxies fail to rebuild large disks and remain spheroid dominated with low 
gas fractions? First of all, ETGs have relatively old stellar populations as we will see in $\S$~\ref{CompetingSources}, 
and therefore we need to understand why after the formation of bulges, galaxy disks fail to reaccrete significant quantities of gas 
to keep forming stars at the level of the main sequence of galaxies in the SFR-M$_{\rm stellar}$ plane.  
This question is then related to 
the physical mechanisms quenching star formation in ETGs.
In order to understand why large disks fail to regrow, we first look into the nature of galaxies 
with $B/T>0$. The top panel of 
Fig.~\ref{MorphoSel2} shows the HI mass function of galaxies with $B/T>0$ and 
$L_{\rm K}>6\times 10^9\,L_{\odot}$ separated into centrals and satellites. The satellite galaxy population makes up 
most of the tail of low HI masses, $M_{\rm HI}<5\times 10^8\,M_{\odot}$. These galaxies 
have little cold gas replenishment after they become satellites as they continuously lose part of their hot gas reservoir due to 
the continuous action of ram pressure stripping. This has the consequence of 
lowering the cooling rates. However, satellite galaxies do preserve some neutral gas reservoir; i.e. they hardly completely deplete their 
gas content. The mechanism for this is connected to the dependence of the star formation timescale with the gas surface density. 
Low gas surface densities produce low H$_2$/HI ratios and low SFRs. 
As the gas is being depleted, the star formation timescale becomes longer and longer, allowing satellite galaxies to retain 
their gas reservoir. This mechanism drives the population of satellite galaxies with low gas fractions. This is consistent with  
observations, where there is a non negligible fraction of ETGs with H$_2$ and/or HI 
contents detected in high mass groups or clusters (e.g. \citealt{Young11}). 

In the case of central galaxies, there is a population with HI masses $M_{\rm HI}>10^9\,M_{\odot}$ and HI gas fractions of
$\approx 0.15\,M_{\odot}/L_{\odot}$ that have cooling rates large enough to replenish their cold gas contents.
The tail of central galaxies with low HI masses, i.e. $M_{\rm HI}<10^9\,M_{\odot}$, fail to replenish their gas reservoirs 
and rebuild a new disk due to the effect of AGN heating their hot halo (see dashed line in the top panel of Fig.~\ref{MorphoSel2}). 
In {\tt GALFORM}, 
AGN feedback acts in halos where the cooling time is larger than the free fall time at the cooling radius 
(`hot accretion' mode; \citealt{Fanidakis10b}). 
In these halos, the AGN power is examined and if 
it is greater than the cooling luminosity, the cooling flows are switched off (see \citealt{Bower06}). 
This means that in central galaxies under the action of AGN feedback, there will be no further gas accretion onto the galaxy, 
driving a low HI and H$_2$ gas contents. Note that, in the model, 
the close connection between bulge fraction and gas depletion naturally arises 
in galaxies where AGN feedback operates. This is because the black hole grows together with the bulge 
(\citealt{Fanidakis10b}). The consequence of this is that large black hole masses, hosted by large bulges, 
are capable of large mechanical luminosities which can more easily affect their hot halo.
These large bulges are also connected to large bulge-to-total stellar mass ratios 
due to the impeded disk regrowth. 

\begin{figure}
\begin{center}
\includegraphics[trim = 0.1mm 0.3mm 1mm 0.45mm,clip,width=0.49\textwidth]{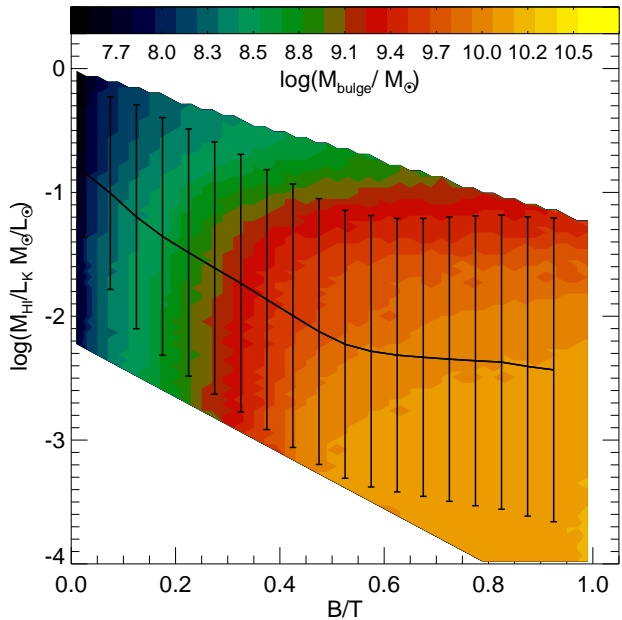}
\caption{{HI gas fraction as a function of the bulge-to-total stellar mass for galaxies in the Lacey14+RP model 
with $L_{\rm K}>6\times 10^9L_{\odot}$. The solid lines and errorbars represent the median and 10 and 90 percentiles 
of the distribution. In colours we show the mean bulge mass in $2$-dimensional bins of HI gas fraction and $B/T$, in an arbitrary 
 region encompassing the errorbars.
The mean bulge masses are as labelled in the bar at the top of the figure. The wiggle features in the coloured region
 are an artifact of the binning.}} 
\label{BTMbulge}
\end{center}
\end{figure}

{We show the HI gas fraction as a function of $B/T$ in Fig.~\ref{BTMbulge} 
with the background colour scheme showing the mean bulge mass in $2$-dimensional bins of HI gas fraction and $B/T$. 
There is a clear anti-correlation between 
the HI gas fraction and $B/T$. In addition, at a fixed $B/T$ there is a trend of increasing HI gas fraction with 
decreasing bulge mass. The latter is related to the stronger AGN feedback in higher mass bulges that lead to gas depletion. 
Recent high redshift observations show evidence for the strong connection between quenching and bulge mass \citep{Lang14}. 
Lang et al. point to the bulge mass as a fundamental property related to star formation quenching, rather than the bulge fraction, which 
in our model is also understood due to the effect of AGN feedback.}

\begin{figure}
\begin{center}
\includegraphics[trim = 0.1mm 0.3mm 1mm 0.45mm,clip,width=0.49\textwidth]{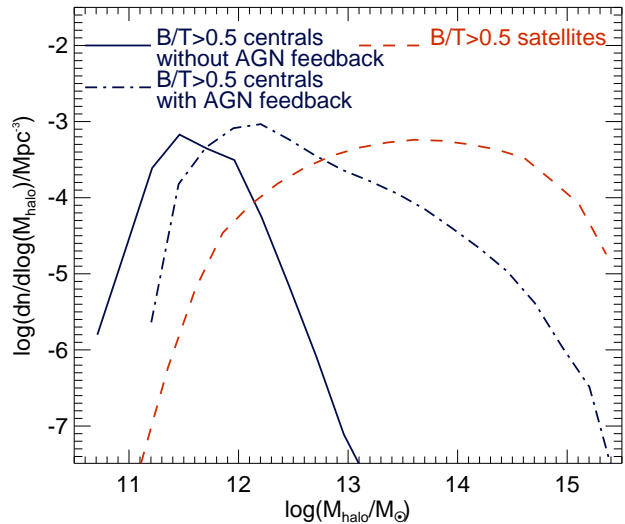}
\caption{Host halo mass distribution for ETGs ($B/T>0.5$) in the Lacey14+RP model 
that are satellites (dashed line), centrals under the action of AGN feedback (dot-dashed line) and 
centrals without AGN feedback (solid line).}
\label{Mhalos}
\end{center}
\end{figure}

Focusing on the population of galaxies with $B/T>0.5$ (bottom panel of Fig.~\ref{MorphoSel2}), one can find similar trends. Most of 
the galaxies with $M_{\rm HI}<10^8\,M_{\odot}$ correspond to satellite galaxies, while the central galaxy population is 
responsible for the massive end of the HI mass function. The tail of central galaxies with 
$M_{\rm HI}<5\times 10^8\,M_{\odot}$ is under the influence of AGN feedback which explains the 
low HI contents. There is a subpopulation of ETGs with large HI masses and gas fractions, 
 $M_{\rm HI}>10^9\,M_{\odot}$ and 
$M_{\rm HI}/L_{\rm K}\gtrsim 0.1\,M_{\odot}/L_{\odot}$, respectively. The latter population is also the one living in low mass haloes.
 Fig.~\ref{Mhalos} shows the distribution of masses of the halos hosting ETGs in the 
Lacey14+RP model. ETGs with the highest {HI gas contents, which correspond to 
centrals without AGN feedback (see Fig.~\ref{MorphoSel2}),} are hosted by low mass halos, 
with a median host halo mass of $3\times 10^{11}\,M_{\odot}$; centrals with AGN feedback on are hosted by higher mass halos, 
with median masses of $2\times 10^{12}\,M_{\odot}$, while satellites are distributed throughout a wider range of halo masses, with 
a median mass of $3\times 10^{13}\,M_{\odot}$. In the observations similar trends are seen. \citet{Young11} show that 
the ETGs with the highest H$_2$ masses also reside in the lowest density environments, {which in our model correspond 
to ETGs that are central galaxies and are not undergoing AGN feedback}. Similarly, 
\citet{Serra12} reported that the HI mass as well as the ratio $M_{\rm HI}/L_{\rm K}$ decrease with increasing 
density of the environment. We find that these trends are simply a reflection of the halo masses in which either AGN feedback 
switches on or environmental quenching acts more effectively (where ram pressure stripping of the hot gas
 is effective enough to remove significant 
amounts of hot gas).

Our conclusion is that most ETGs with low gas fractions correspond to satellite galaxies, 
which due to environmental quenching (partial ram pressure stripping of the hot gas), are unable to regrow a significant disk. 
The rest of the ETGs with low gas fractions are central galaxies under the action of AGN feedback. The break at the high end of the 
 HI gas fraction distribution, traced by HIPASS, is due to a population of ETGs with large HI masses that are 
still forming stars at the same level of spiral galaxies.  

\section{The competing sources of the neutral gas content of early-type galaxies}\label{CompetingSources}

One the aims of this paper is to answer the question: what is the source of the HI and H$_2$ gas contents in ETGs? 
We follow all the gas sources throughout the star formation history of galaxies identified as ETGs today: 
radiative cooling from hot halos (which 
we refer to as `cooling'), mass loss from old stars (which
we refer to as `recycling') and galaxy mergers (see Appendix~\ref{GasSourcesContribution} for details of how 
we do this). 
Note that in our model the recycled mass from old stars does not incorporate into the hot halo and therefore 
it can be distinguished from the gas coming from cooling. 

We study the sources of the gas in ETGs in two of the models shown in $\S$~\ref{Sec:ModelComparison} 
to find general 
trends present in the different models as well as variations between them. We focus here on the 
Lagos12+RP and Lacey14+RP models as, after including partial ram pressure stripping of the hot gas, they give the lowest and highest number 
densities of ETGs, respectively. 

\begin{table*}
\begin{center}
\caption{The percentage of ETGs with $L_{\rm K}>6\times 10^9\,L_{\odot}$ in the Lacey14+RP and Lagos12+RP models 
under different selection criteria, which we group in four categories: neutral gas content, neutral gas sources, mergers 
and disk instability.}\label{Contributions}
\begin{tabular}{l c c}
\\[3pt]
\hline
Selection & ETGs ($L_{\rm K}>6\times 10^9\,L_{\odot}$) &  ETGs ($L_{\rm K}>6\times 10^9\,L_{\odot}$) \\
 & Lacey14+RP & Lagos12+RP\\ 
\hline
Neutral gas content \\
\hline
ETGs with neutral gas content $M_{\rm HI}+M_{\rm H_2}>10^7\,M_{\odot}$ & $58$\% & $65$\%\\
\hline
Neutral gas sources of the sample of ETGs with $M_{\rm HI}+M_{\rm H_2}>10^7\,M_{\odot}$.\\
\hline
ETGs with current gas content dominated by mergers & $7.5$\%  & $17$\%\\
ETGs with current gas content dominated by recycling & $1.5$\% & $0.8$\%\\
ETGs with current gas content dominated by cooling & $91$\%  & $82$\%\\
\hline
Mergers of ETGs with $M_{\rm HI}+M_{\rm H_2}>10^7\,M_{\odot}$.\\
\hline
ETGs that had a merger in the last $1$~Gyr & $11$\% & $25$\%\\
ETGs that had a merger-driven starburst in the last $1$~Gyr & $1$\% & $1$\%\\
Mergers in ETGs that took place in $M_{\rm halo}<10^{14}\,M_{\odot}\,h^{-1}$ & $95$\%  & $94$\% \\
Mergers in ETGs that increased the neutral gas content by a factor of $>2$ &  $69$\%  & $66$\%\\
\hline
Disk instabilities on ETGs with $M_{\rm HI}+M_{\rm H_2}>10^7\,M_{\odot}$.\\
\hline
ETGs that had a disk instability in the last $1$~Gyr & $4$\% & $2$\%\\
\hline
\end{tabular}
\end{center}
\end{table*}

We first estimate the fraction of ETGs
 that have neutral gas masses ($M_{\rm HI}+M_{\rm H_2}$) $>10^7\,M_{\odot}$. We find that $58$\% of ETGs with 
$K$-band luminosities $L_{\rm K}>6\times 10^9\,L_{\odot}$ in the Lacey14+RP 
model and $65$\% in the Lagos12+RP model have $M_{\rm HI}+M_{\rm H_2}>10^7\,M_{\odot}$. 
We analyse these sub-samples of ETGs 
and estimate the fraction of the ETGs with neutral gas 
contents supplied mainly by mergers, recycling or 
cooling (summarised in Table~\ref{Contributions}).
Most ETGs have neutral gas contents supplied predominantly by cooling. A smaller percentage 
have neutral gas contents supplied mainly by mergers ($\approx 8$\% for the Lacey14+RP and 
$17$\% for the Lagos12+RP model) or by recycling ($\approx 1.5$\% for the Lacey14+RP and $0.8$\% 
for the Lagos12+RP model).
The latter percentages are not sensitive to the $K$-band luminosity or stellar mass of ETGs. However, they 
are sensitive to the current neutral gas content and halo mass. 
In order to get an insight into the properties of ETGs that have different gas suppliers, 
we show in Table~\ref{Contributions} the fraction of ETGs that had a minor merger in the last $1$~Gyr, 
the fraction of these that increased the neutral gas content by at least a factor of $2$, 
and the fraction of ETGs that had a starburst driven by either a galaxy merger or a disk instability in the last $1$~Gyr. 
The main conclusions we draw from Table~\ref{Contributions} are:
\begin{itemize}
\item Minor mergers took place in a tenth of the ETG population with $L_{\rm K}>6\times 10^9\,L_{\odot}$ and 
$M_{\rm HI}+M_{\rm H_2}>10^7\,M_{\odot}$ in the last 
$1$~Gyr in the Lacey14+RP model. 
Only $10$\% of these resulted in a starburst, although none of these starbursts made a significant contribution 
to the stellar mass build-up (mass weighted stellar ages are usually $>7$~Gyr). 
The large percentage of minor mergers that did not drive starbursts in the last $1$~Gyr 
is due to the very low mass ratios between the accreted galaxy and 
the ETG, which is on average $\approx 0.05$. Such small mass ratios are not considered to drive starbursts in {\tt GALFORM} 
unless they are very gas rich (see $\S$~\ref{BuildUpBulges}). For the Lagos12+RP model the fraction of galaxies 
that had a minor merger in the last $1$~Gyr is higher, $25$\%, with a smaller fraction ($0.05$) of these driving 
starbursts. The mass ratios of these minor merger events are also very small, which explains the small percentage 
of merger driven starbursts.
\item $\approx 68$\% of minor mergers in ETGs 
in both the Lacey14+RP model Lagos12+RP models
increased the neutral gas content significantly (at least by a factor of $2$).
The frequency of minor mergers times the percentage of those which increased the gas content significantly 
explains the percentages of ETGs with neutral gas contents supplied by 
minor merger accretion in the models.
\item Of these minor merger accretion episodes, $\approx 95$\% in both the Lacey14+RP and Lagos12+RP models
took place in halos with masses $<10^{14}\,M_{\odot}$, which implies
that this source of neutral gas accretion is negligible in cluster environments. This agrees with  
the observations of \citet{Davis11}.
\item There is only a small percentage, $~\approx 3$\%, of ETGs with $L_{\rm K}>6\times 10^9\,L_{\odot}$ and
$M_{\rm HI}+M_{\rm H_2}>10^7\,M_{\odot}$,   
 that had a starburst driven by disk instabilities in the last $1$~Gyr. These galaxies have very small disks (usually
 the bulge half-mass radius is larger than the disk half-mass radius) and 
 $B/T\gtrsim 0.9$. In the model we use the properties of galaxy disks to 
determine whether they are unstable under small perturbations (see Eq.~\ref{DisKins}). In reality, one would expect that 
such large bulges dominate over the gravity of the disk, stabilizing it. \citet{Martig13} show this to happen in 
hydrodynamical simulations of individual galaxies: self-gravity of the disk is reduced when it is embedded in a bulge, 
preventing gas fragmentation. This results in an overall
lower efficiency of star formation in ETGs. Our model does not capture this physics showing that it 
needs improvement to account for these cases. 

\end{itemize}
\begin{figure}
\begin{center}
\includegraphics[trim = 0.9mm 0.3mm 1mm 0.45mm,clip,width=0.43\textwidth]{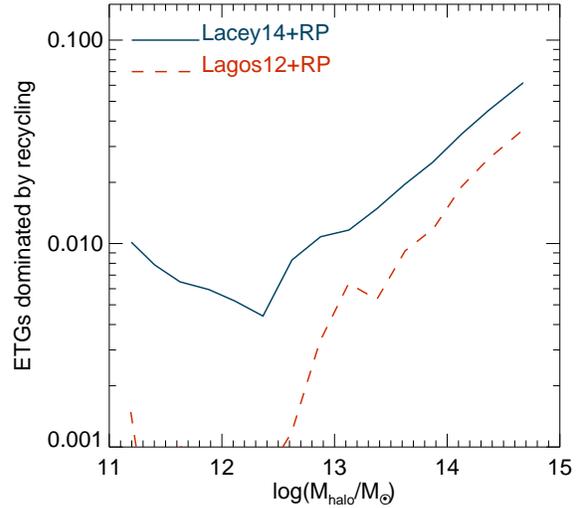}
\caption{Fraction of ETGs ($B/T>0.5$) with $L_{\rm K}>6\times 10^9\,L_{\odot}$ and 
$M_{\rm HI}+M_{\rm H_2}>10^7\,M_{\odot}$ at $z=0$ in the Lacey14+RP and Lagos12+RP models 
that have most of their cold gas supplied by mass loss from intermediate- and low-mass stars, 
as a function of the host halo mass.}
\label{FracGasContributions}
\end{center}
\end{figure}

The largest differences found between the two models is in the fraction of ETGs with 
$M_{\rm HI}+M_{\rm H_2}>10^7\,M_{\odot}$ and the percentage of those with current neutral gas contents
dominated by merger accretion. These differences are due to a combination of the different threshold values 
to examine disk instabilities and the different dynamical friction prescriptions used by the models ($\S$~\ref{BuildUpBulges}).  
In the Lacey14+RP model more disk instabilities take place due to the higher $\epsilon_{\rm disk}$, which drives 
a more rapid gas exhaustion in the galaxies that are prone to disk instabilities. Many of the galaxies that 
go through disk instabilities in the Lacey14+RP model do not do so in the Lagos12+RP model due to 
lower value of $\epsilon_{\rm disk}$. In the case of the dynamical friction, the prescription used by the 
Lagos12+RP model produces more minor mergers at lower redshifts than the prescription used in the Lacey13+RP model.

\begin{figure}
\begin{center}
\includegraphics[trim = 0.9mm 0mm 1mm 0.45mm,clip,width=0.43\textwidth]{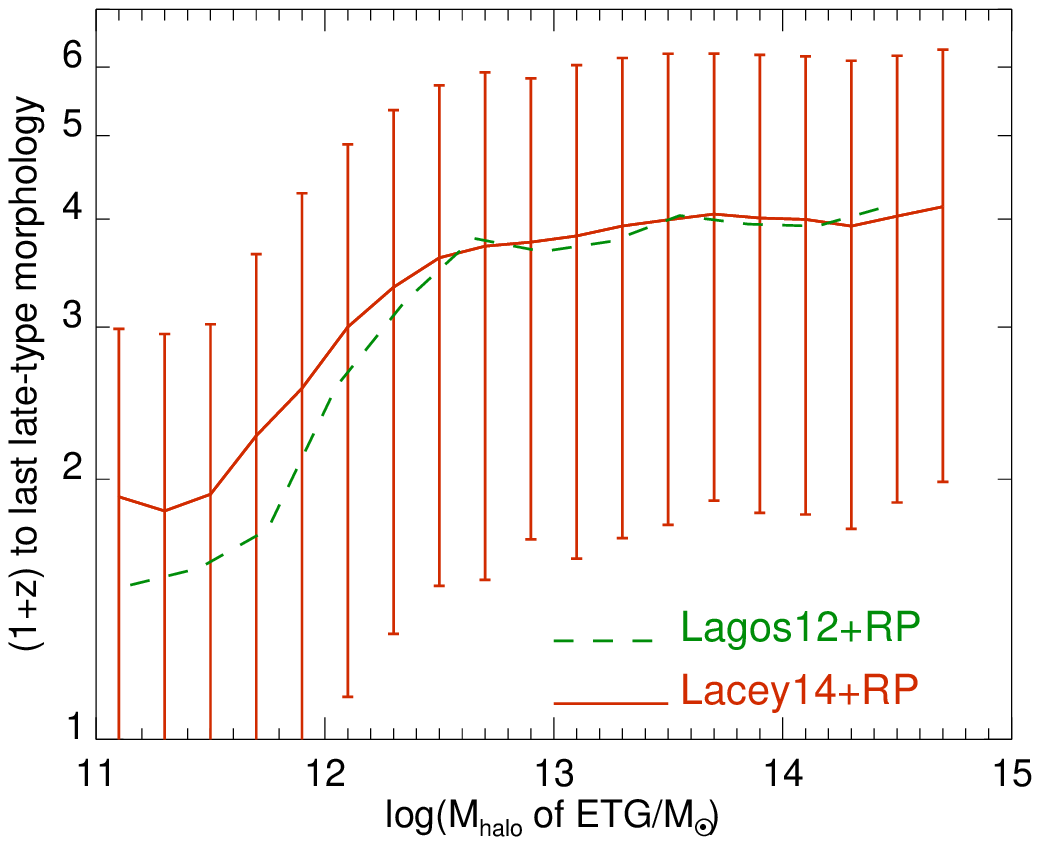}
\includegraphics[trim = 0.9mm 0mm 1mm 0.45mm,clip,width=0.43\textwidth]{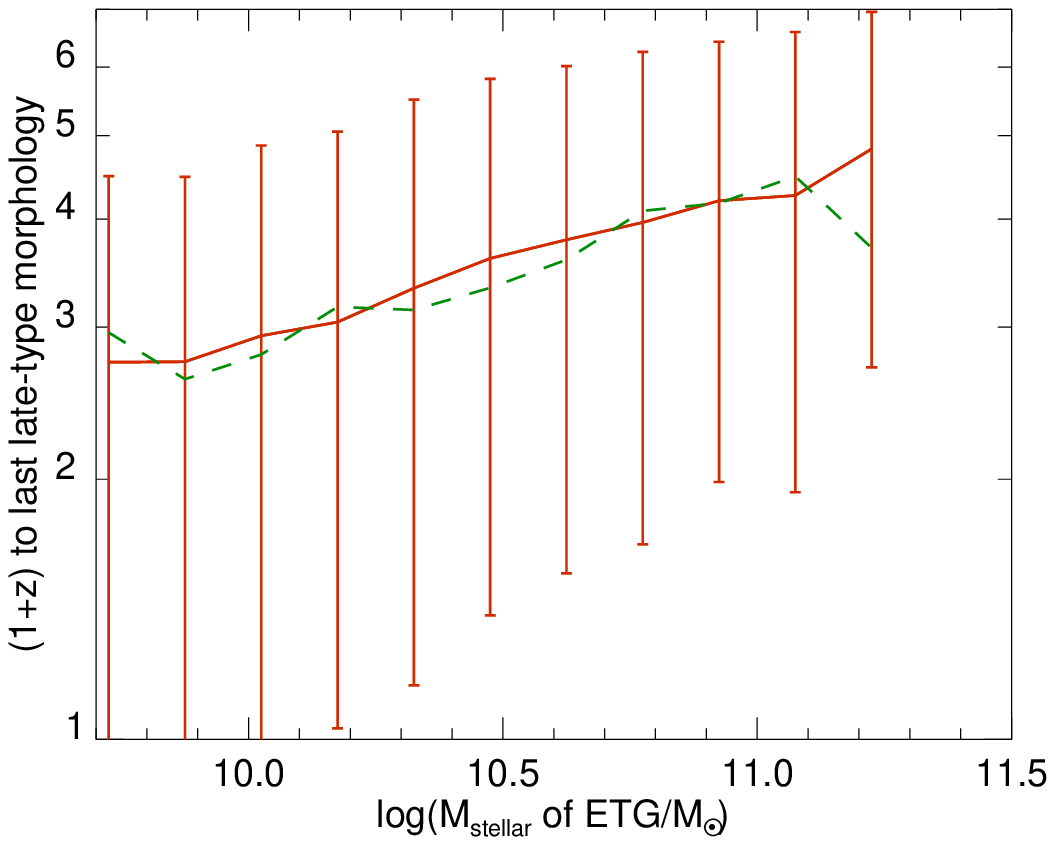}
\caption{{\it Top Panel:} Look-back time to the last time ETGs ($B/T>0.5$) selected as $L_{\rm K}>6\times 10^9\,L_{\odot}$,  
$M_{\rm HI}+M_{\rm H_2}>10^7\,M_{\odot}$ at $z=0$ 
had a $B/T<0.5$ (late-type morphology), expressed in $1+z$, as a function of the current host halo mass, for the 
Lacey14+RP and Lagos12+RP models. Lines with errorbars show the median and 10 and 90 percentiles of the distributions, respectively. 
For clarity, errorbars are only shown for the Lacey14+RP model, but the ones in the Lagos12+RP model are of similar magnitude. {\it Bottom panel:} 
{As in the top panel but here the look-back time to the last time ETGs had a $B/T<0.5$ is shown as a function of stellar mass.}}
\label{Misalignment2}
\end{center}
\end{figure}

Another interesting dichotomy between cluster environments and lower mass halos is that the amount of ETGs that have neutral 
gas contents supplied mainly by mass loss from intermediate and low-mass stars increases with increasing 
halo mass. This is shown in Fig.~\ref{FracGasContributions} for the models 
Lagos12+RP and Lacey14+RP. 
In cluster environments, we expect $~7$\% of ETGs to have neutral gas contents mainly supplied by recycling 
 in the Lacey14+RP model and $3$\% in the Lagos12+RP model, 
while that percentage drops dramatically at halo masses $M_{\rm halo}<10^{14}\,M_{\odot}$.
Note that both models show a minimum contribution from recycling at 
$M_{\rm halo}\approx 2\times 10^{12}\,M_{\odot}$, which is connected to the halo mass in which feedback, either by 
stellar feedback in the lower halo masses or AGN in the higher halo masses, is the 
least effective. At this halo mass we have the most efficient accretion of newly cooled gas, which minimises the 
contribution from mass loss from old stars.

The higher frequency of ETGs with neutral gas contents dominated by internal origin in higher mass halos 
takes place together with the aging of 
the bulge. In Fig.~\ref{Misalignment2}, we show the look-back time to the last time ETGs had a bulge-to-total stellar mass ratio 
$<0.5$ (a late-type morphology), {as a function of the current halo and stellar mass. 
There is a positive correlation between the current halo and stellar mass and the last time these ETGs were late-type:
ETGs residing in high mass halo have had an early-type morphology for longer time 
than those residing in lower mass halos, and similarly, the most massive ETGs have a tendency of having had an early-type morphology for longer 
than lower mass ETGs}.
Note, however, that the dispersion around 
these relations is very high, showing that there is no single path for the formation of spheroids and that the star formation 
history of ETGs can be quite complex (see also \citealt{Naab13}). 
The Lagos12+RP model predicts systematically lower times at halo masses $M_{\rm halo}<10^{13}\, M_{\odot}$ 
than the Lacey14+RP model, which is due to the 
higher disk instability threshold in the latter model, which drove ETGs to undergo disk instabilities 
on average earlier than in the Lagos12+RP model.
 
\section{Conclusions}\label{conclusion} 

We have studied the current neutral gas content of ETGs 
and its origin in the context of hierarchical galaxy formation.
We first use the HIPASS and ATLAS$^{\rm 3D}$ surveys to quantify the HI and H$_2$ gas fraction distribution functions for the overall galaxy 
population and ETGs observationally.
We then explored the predictions for the neutral gas content of galaxies in three flavours of 
the {\tt GALFORM} semi-analytic model of galaxy formation, the Lagos12, Gonzalez-Perez14 and 
Lacey14 models and performed a thorough comparison with observations. 

For quiescent star formation,
the three models use the pressure-based SF law of \citet{Blitz06}, in which the ratio between the
 surface density of H$_2$ and HI is derived from the radial profile of the hydrostatic pressure of the disk. 
The SFR is then calculated from the surface density of H$_2$. The advantage of this SF law is that the atomic and molecular gas phases
of the ISM of galaxies are explicitly distinguished, which allows us to compare the predictions for the HI and H$_2$ contents 
of ETGs directly with observations.
 Other physical processes in the three models are different, such
as the IMF adopted and the strength of both the SNe and the AGN feedback, as well as the cosmological parameters.
We also tested the importance of the modelling of the processing of the hot gas of galaxies once they become satellites.
The original {\tt GALFORM} flavours include a strangulation treatment of the hot gas: once 
galaxies become satellites they immediately lose all of their hot gas reservoir. We run the three {\tt GALFORM} flavours with a different 
hot gas processing: the 
partial ram pressure stripping of the hot gas of satellite galaxies, which depends upon the orbit followed by 
the satellite galaxy, with cooling continuing onto the satellite galaxies.

Our conclusions are:

(i) The three flavours of {\tt GALFORM} predict overall HI and H$_2$ mass functions in good agreement with the 
observations, regardless of the treatment of the hot gas of galaxies once they become satellites. However, 
when focusing exclusively on the ETG population, the inclusion of partial ram pressure stripping of the hot gas (as opposed 
to the strangulation scenario) 
results in the models predicting ETGs with 
 higher contents of HI and H$_2$, improving the agreement with the observations.
 This shows that the HI and H$_2$ gas contents of ETGs are a great test for the modelling of the hot gas stripping in 
 simulations of galaxy formation. Moreover, the gas fraction distribution is a statistical measurement which is particularly 
good for placing constraints on models.

(ii) The presence of a bulge in galaxies is strongly correlated with depleted HI and H$_2$ gas contents in the three 
{\tt GALFORM} models tested. This close correspondence between the bulge fraction and the depleted neutral gas contents 
in ETGs has been observed 
and here we provide a physical framework to understand it.
We show that this is due to AGN feedback in central galaxies, and the environmental quenching 
due to partial ram pressure stripping of the hot gas in satellite galaxies. In the former, 
the black hole mass is correlated with the bulge mass, 
which implies that feedback can be stronger in larger bulges (as the Eddington luminosity increases with back hole mass). 
Galaxies experiencing AGN feedback do not accrete significant amounts of newly cooled gas, which impedes the regeneration of 
a prominent disk. In the case of satellites, the lower accretion rates due to depletion of the hot gas prevent the 
regrowth of a substantial disk, or drive the exhaustion of the gas in the disk, leaving a (close to) gas-free disk. 
There is a fraction of ETGs though that are neither experiencing AGN feedback, nor environmental quenching, that 
have normal HI and H$_2$ contents which are comparable to those obtained for late-type galaxies of the same mass. 

(iii) We find that about $\approx 90$\% of ETGs accreted most of their neutral gas from 
the hot halo through radiative cooling. A lower fraction have current HI and H$_2$ contents supplied by accretion from minor galaxy mergers 
(ranging from $8$\% to $17$\%, depending on the model). An even smaller fraction ($0.5-2$\%) have their neutral gas content 
supplied mass loss from intermediate and low mass stars. Interestingly, most of those galaxies are 
hosted by large mass halos ($M_{\rm halo}>10^{14}\,M_{\odot}$; clusters of galaxies), while most 
of those dominated by minor merger accretion are in non-cluster environments ($M_{\rm halo}<10^{14}\,M_{\odot}$). 
We find that the source of the HI and H$_2$ gas in ETGs has strong consequences for the expected alignment between the gas disk and 
the stellar component, which we discuss in depth in paper II (Lagos et al. in prep.). 

(iv) We find a general trend of increasing look-back time to the last time ETGs were late-types ($B/T<0.5$) 
with increasing {host halo mass and stellar mass}.
 However, these trends are characterised by 
a very large dispersion around the median, suggesting that the paths for the formation of ETGs of a given 
stellar mass are variable and non self-similar. The latter is due to the stochastic nature of galaxy mergers and 
disk instabilities. 

Our analysis shows the power of studying the gas contents of galaxies, and          
how sub-samples of them are affected by different physical processes. In particular, 
our work points to the need for improved modelling of ram pressure stripping of the hot gas, which has 
an important effect in a wide range of environments.
Although we show how the model we include, originally 
developed by \citet{Font08}, works well with the observational constraints we currently have, 
it may be too simplistic. For example 
this model does not explicitly take into account the three dimensional position and velocity 
 of galaxies in the simulation, which means that we do not consider 
their specific position in the halo to calculate the ram pressure stripping throughout its transit. This will be possible 
with high resolution simulations, given that for such a detailed analysis it is necessary to resolve all halos with few hundreds 
 particles at least.
In the future, we suggest that the study of the HI and H$_2$ gas contents 
of galaxies classified as ``passive'' will provide stringent constraints on the details of the ram pressure 
stripping modelling.

\section*{Acknowledgements}

We thank Martin Meyer, Diederik Kruijssen, Andreas Schruba and Paolo Serra for very motivating discussions.
The research leading to these results has received funding from 
the European Community's Seventh
Framework Programme ($/$FP7$/$2007-2013$/$) under grant agreement no 229517 and 
the Science and Technology Facilities Council grant number ST/F001166/1.
This work used the DiRAC Data Centric system at Durham University, operated by the Institute for Computational Cosmology on behalf of the STFC DiRAC HPC Facility ({\tt www.dirac.ac.uk}). This equipment was funded by BIS National E-infrastructure capital grant ST/K00042X/1, STFC capital grant ST/H008519/1, and STFC DiRAC Operations grant ST/K003267/1 and Durham University. DiRAC is part of the National E-Infrastructure.
 VGP acknowledges support from a European Research
Council Starting Grant (DEGAS-259586).

\bibliographystyle{mn2e_trunc8}
\bibliography{OriginGasContentElls}

\begin{thebibliography}{109}
\expandafter\ifx\csname natexlab\endcsname\relax\def\natexlab#1{#1}\fi

\bibitem[{{Baldry} {et~al.}(2004){Baldry}, {Glazebrook}, {Brinkmann},
  {Ivezi{\'c}}, {Lupton}, {Nichol}, \& {Szalay}}]{Baldry04}
{Baldry} I.~K., {Glazebrook} K., {Brinkmann} J., {Ivezi{\'c}} {\v Z}., {Lupton}
  R.~H., {Nichol} R.~C., {Szalay} A.~S., 2004, \apj, 600, 681

\bibitem[{{Balogh} {et~al.}(2004){Balogh}, {Baldry}, {Nichol}, {Miller},
  {Bower}, \& {Glazebrook}}]{Balogh04}
{Balogh} M.~L., {Baldry} I.~K., {Nichol} R., {Miller} C., {Bower} R.,
  {Glazebrook} K., 2004, \apjl, 615, L101

\bibitem[{{Baugh}(2006)}]{Baugh06}
{Baugh} C.~M., 2006, Reports on Progress in Physics, 69, 3101

\bibitem[{{Baugh} {et~al.}(1996){Baugh}, {Cole}, \& {Frenk}}]{Baugh96}
{Baugh} C.~M., {Cole} S., {Frenk} C.~S., 1996, \mnras, 283, 1361

\bibitem[{{Baugh} {et~al.}(2005){Baugh}, {Lacey}, {Frenk}, {Granato}, {Silva},
  {Bressan}, {Benson}, \& {Cole}}]{Baugh05}
{Baugh} C.~M., {Lacey} C.~G., {Frenk} C.~S., {Granato} G.~L., {Silva} L.,
  {Bressan} A., {Benson} A.~J., {Cole} S., 2005, \mnras, 356, 1191

\bibitem[{{Benson}(2005)}]{Benson05}
{Benson} A.~J., 2005, \mnras, 358, 551

\bibitem[{{Benson}(2010)}]{Benson10b}
---, 2010, \physrep, 495, 33

\bibitem[{{Benson} {et~al.}(2007){Benson}, {D{\v z}anovi{\'c}}, {Frenk}, \&
  {Sharples}}]{Benson07}
{Benson} A.~J., {D{\v z}anovi{\'c}} D., {Frenk} C.~S., {Sharples} R., 2007,
  \mnras, 379, 841

\bibitem[{{Bernardi} {et~al.}(2003){Bernardi}, {Sheth}, {Annis}, {Burles},
  {Eisenstein}, {Finkbeiner}, {Hogg}, {Lupton}, {Schlegel}, {SubbaRao},
  {Bahcall}, {Blakeslee}, {Brinkmann}, {Castander}, {Connolly}, {Csabai},
  {Doi}, {Fukugita}, {Frieman}, {Heckman}, {Hennessy}, {Ivezi{\'c}}, {Knapp},
  {Lamb}, {McKay}, {Munn}, {Nichol}, {Okamura}, {Schneider}, {Thakar}, \&
  {York}}]{Bernardi03}
{Bernardi} M., {Sheth} R.~K., {Annis} J., {Burles} S., {Eisenstein} D.~J.,
  {Finkbeiner} D.~P., {Hogg} D.~W., {Lupton} R.~H. {et~al}, 2003, \aj, 125,
  1817

\bibitem[{{Bernardi} {et~al.}(2005){Bernardi}, {Sheth}, {Nichol}, {Schneider},
  \& {Brinkmann}}]{Bernardi05}
{Bernardi} M., {Sheth} R.~K., {Nichol} R.~C., {Schneider} D.~P., {Brinkmann}
  J., 2005, \aj, 129, 61

\bibitem[{{Bettoni} {et~al.}(2003){Bettoni}, {Galletta}, \&
  {Garc{\'{\i}}a-Burillo}}]{Bettoni03}
{Bettoni} D., {Galletta} G., {Garc{\'{\i}}a-Burillo} S., 2003, \aap, 405, 5

\bibitem[{{Bigiel} {et~al.}(2011){Bigiel}, {Leroy}, {Walter}, {Brinks}, {de
  Blok}, {Kramer}, {Rix}, {Schruba}, {Schuster}, {Usero}, \&
  {Wiesemeyer}}]{Bigiel11}
{Bigiel} F., {Leroy} A.~K., {Walter} F., {Brinks} E., {de Blok} W.~J.~G.,
  {Kramer} C., {Rix} H.~W., {Schruba} A. {et~al}, 2011, \apjl, 730, L13+

\bibitem[{{Blitz} \& {Rosolowsky}(2006)}]{Blitz06}
{Blitz} L., {Rosolowsky} E., 2006, \apj, 650, 933

\bibitem[{{Boselli} {et~al.}(2014{\natexlab{a}}){Boselli}, {Cortese},
  {Boquien}, {Boissier}, {Catinella}, {Gavazzi}, {Lagos}, \&
  {Saintonge}}]{Boselli14}
{Boselli} A., {Cortese} L., {Boquien} M., {Boissier} S., {Catinella} B.,
  {Gavazzi} G., {Lagos} C., {Saintonge} A., 2014{\natexlab{a}}, \aap, 564, A67

\bibitem[{{Boselli} {et~al.}(2014{\natexlab{b}}){Boselli}, {Cortese},
  {Boquien}, {Boissier}, {Catinella}, {Lagos}, \& {Saintonge}}]{Boselli14b}
{Boselli} A., {Cortese} L., {Boquien} M., {Boissier} S., {Catinella} B.,
  {Lagos} C., {Saintonge} A., 2014{\natexlab{b}}, \aap, 564, A66

\bibitem[{{Bournaud} {et~al.}(2011){Bournaud}, {Chapon}, {Teyssier}, {Powell},
  {Elmegreen}, {Elmegreen}, {Duc}, {Contini}, {Epinat}, \&
  {Shapiro}}]{Bournaud11}
{Bournaud} F., {Chapon} D., {Teyssier} R., {Powell} L.~C., {Elmegreen} B.~G.,
  {Elmegreen} D.~M., {Duc} P.-A., {Contini} T. {et~al}, 2011, \apj, 730, 4

\bibitem[{{Bournaud} {et~al.}(2009){Bournaud}, {Elmegreen}, \&
  {Martig}}]{Bournaud09}
{Bournaud} F., {Elmegreen} B.~G., {Martig} M., 2009, \apjl, 707, L1

\bibitem[{{Bower} {et~al.}(2006){Bower}, {Benson}, {Malbon}, {Helly}, {Frenk},
  {Baugh}, {Cole}, \& {Lacey}}]{Bower06}
{Bower} R.~G., {Benson} A.~J., {Malbon} R., {Helly} J.~C., {Frenk} C.~S.,
  {Baugh} C.~M., {Cole} S., {Lacey} C.~G., 2006, \mnras, 370, 645

\bibitem[{{Bower} {et~al.}(1992){Bower}, {Lucey}, \& {Ellis}}]{Bower92}
{Bower} R.~G., {Lucey} J.~R., {Ellis} R.~S., 1992, \mnras, 254, 601

\bibitem[{{Bower} {et~al.}(2010){Bower}, {Vernon}, {Goldstein}, {Benson},
  {Lacey}, {Baugh}, {Cole}, \& {Frenk}}]{Bower10}
{Bower} R.~G., {Vernon} I., {Goldstein} M., {Benson} A.~J., {Lacey} C.~G.,
  {Baugh} C.~M., {Cole} S., {Frenk} C.~S., 2010, \mnras, 407, 2017

\bibitem[{{Bruzual} \& {Charlot}(2003)}]{Bruzual03}
{Bruzual} G., {Charlot} S., 2003, \mnras, 344, 1000

\bibitem[{{Cappellari} {et~al.}(2011){Cappellari}, {Emsellem}, {Krajnovi{\'c}},
  {McDermid}, {Scott}, {Verdoes Kleijn}, {Young}, {Alatalo}, {Bacon}, {Blitz},
  {Bois}, {Bournaud}, {Bureau}, {Davies}, {Davis}, {de Zeeuw}, {Duc},
  {Khochfar}, {Kuntschner}, {Lablanche}, {Morganti}, {Naab}, {Oosterloo},
  {Sarzi}, {Serra}, \& {Weijmans}}]{Cappellari11}
{Cappellari} M., {Emsellem} E., {Krajnovi{\'c}} D., {McDermid} R.~M., {Scott}
  N., {Verdoes Kleijn} G.~A., {Young} L.~M., {Alatalo} K. {et~al}, 2011,
  \mnras, 413, 813

\bibitem[{{Cappellari} {et~al.}(2013){Cappellari}, {McDermid}, {Alatalo},
  {Blitz}, {Bois}, {Bournaud}, {Bureau}, {Crocker}, {Davies}, {Davis}, {de
  Zeeuw}, {Duc}, {Emsellem}, {Khochfar}, {Krajnovi{\'c}}, {Kuntschner},
  {Morganti}, {Naab}, {Oosterloo}, {Sarzi}, {Scott}, {Serra}, {Weijmans}, \&
  {Young}}]{Cappellari13}
{Cappellari} M., {McDermid} R.~M., {Alatalo} K., {Blitz} L., {Bois} M.,
  {Bournaud} F., {Bureau} M., {Crocker} A.~F. {et~al}, 2013, \mnras, 432, 1862

\bibitem[{{Catinella} {et~al.}(2010){Catinella}, {Schiminovich}, {Kauffmann},
  {Fabello}, {Wang}, {Hummels}, {Lemonias}, {Moran}, {Wu}, {Giovanelli},
  {Haynes}, {Heckman}, {Basu-Zych}, {Blanton}, {Brinchmann}, {Budav{\'a}ri},
  {Gon{\c c}alves}, {Johnson}, {Kennicutt}, {Madore}, {Martin}, {Rich},
  {Tacconi}, {Thilker}, {Wild}, \& {Wyder}}]{Catinella10}
{Catinella} B., {Schiminovich} D., {Kauffmann} G., {Fabello} S., {Wang} J.,
  {Hummels} C., {Lemonias} J., {Moran} S.~M. {et~al}, 2010, \mnras, 403, 683

\bibitem[{{Cavaliere} \& {Fusco-Femiano}(1976)}]{Cavaliere76}
{Cavaliere} A., {Fusco-Femiano} R., 1976, \aap, 49, 137

\bibitem[{{Cole} {et~al.}(2000){Cole}, {Lacey}, {Baugh}, \& {Frenk}}]{Cole00}
{Cole} S., {Lacey} C.~G., {Baugh} C.~M., {Frenk} C.~S., 2000, \mnras, 319, 168

\bibitem[{{Cortese} {et~al.}(2011){Cortese}, {Catinella}, {Boissier},
  {Boselli}, \& {Heinis}}]{Cortese11}
{Cortese} L., {Catinella} B., {Boissier} S., {Boselli} A., {Heinis} S., 2011,
  \mnras, 415, 1797

\bibitem[{{Davis} {et~al.}(2011){Davis}, {Bureau}, {Young}, {Alatalo}, {Blitz},
  {Cappellari}, {Scott}, {Bois}, {Bournaud}, {Davies}, {de Zeeuw}, {Emsellem},
  {Khochfar}, {Krajnovi{\'c}}, {Kuntschner}, {Lablanche}, {McDermid},
  {Morganti}, {Naab}, {Oosterloo}, {Sarzi}, {Serra}, \& {Weijmans}}]{Davis11}
{Davis} T.~A., {Bureau} M., {Young} L.~M., {Alatalo} K., {Blitz} L.,
  {Cappellari} M., {Scott} N., {Bois} M. {et~al}, 2011, \mnras, 414, 968

\bibitem[{{Davis} {et~al.}(2014){Davis}, {Young}, {Crocker}, {Bureau}, {Blitz},
  {Alatalo}, {Emsellem}, {Naab}, {Bayet}, {Bois}, {Bournaud}, {Cappellari},
  {Davies}, {de Zeeuw}, {Duc}, {Khochfar}, {Krajnovic}, {Kuntschner},
  {McDermid}, {Morganti}, {Oosterloo}, {Sarzi}, {Scott}, {Serra}, \&
  {Weijmans}}]{Davis14}
{Davis} T.~A., {Young} L.~M., {Crocker} A.~F., {Bureau} M., {Blitz} L.,
  {Alatalo} K., {Emsellem} E., {Naab} T. {et~al}, 2014, ArXiv:1403.4850

\bibitem[{{De Lucia} {et~al.}(2006){De Lucia}, {Springel}, {White}, {Croton},
  \& {Kauffmann}}]{DeLucia06}
{De Lucia} G., {Springel} V., {White} S.~D.~M., {Croton} D., {Kauffmann} G.,
  2006, \mnras, 366, 499

\bibitem[{{de Vaucouleurs} {et~al.}(1991){de Vaucouleurs}, {de Vaucouleurs},
  {Corwin}, {Buta}, {Paturel}, \& {Fouque}}]{deVaucouleurs91}
{de Vaucouleurs} G., {de Vaucouleurs} A., {Corwin} Jr. H.~G., {Buta} R.~J.,
  {Paturel} G., {Fouque} P., 1991, \skytel, 82, 621

\bibitem[{{Doyle} {et~al.}(2005){Doyle}, {Drinkwater}, {Rohde}, {Pimbblet},
  {Read}, {Meyer}, {Zwaan}, {Ryan-Weber}, {Stevens}, {Koribalski}, {Webster},
  {Staveley-Smith}, {Barnes}, {Howlett}, {Kilborn}, {Waugh}, {Pierce},
  {Bhathal}, {de Blok}, {Disney}, {Ekers}, {Freeman}, {Garcia}, {Gibson},
  {Harnett}, {Henning}, {Jerjen}, {Kesteven}, {Knezek}, {Mader}, {Marquarding},
  {Minchin}, {O'Brien}, {Oosterloo}, {Price}, {Putman}, {Ryder}, {Sadler},
  {Stewart}, {Stootman}, \& {Wright}}]{Doyle05}
{Doyle} M.~T., {Drinkwater} M.~J., {Rohde} D.~J., {Pimbblet} K.~A., {Read} M.,
  {Meyer} M.~J., {Zwaan} M.~A., {Ryan-Weber} E. {et~al}, 2005, \mnras, 361, 34

\bibitem[{{Efstathiou} {et~al.}(1982){Efstathiou}, {Lake}, \&
  {Negroponte}}]{Efstathiou82}
{Efstathiou} G., {Lake} G., {Negroponte} J., 1982, \mnras, 199, 1069

\bibitem[{{Eke} {et~al.}(1998){Eke}, {Cole}, {Frenk}, \& {Patrick
  Henry}}]{Eke98}
{Eke} V.~R., {Cole} S., {Frenk} C.~S., {Patrick Henry} J., 1998, \mnras, 298,
  1145

\bibitem[{{Elmegreen}(1989)}]{Elmegreen89}
{Elmegreen} B.~G., 1989, \apj, 338, 178

\bibitem[{{Elmegreen} \& {Burkert}(2010)}]{Elmegreen10}
{Elmegreen} B.~G., {Burkert} A., 2010, \apj, 712, 294

\bibitem[{{Fall} \& {Efstathiou}(1980)}]{Fall80}
{Fall} S.~M., {Efstathiou} G., 1980, \mnras, 193, 189

\bibitem[{{Fanidakis} {et~al.}(2012){Fanidakis}, {Baugh}, {Benson}, {Bower},
  {Cole}, {Done}, {Frenk}, {Hickox}, {Lacey}, \& {Del P.~Lagos}}]{Fanidakis10b}
{Fanidakis} N., {Baugh} C.~M., {Benson} A.~J., {Bower} R.~G., {Cole} S., {Done}
  C., {Frenk} C.~S., {Hickox} R.~C. {et~al}, 2012, \mnras, 419, 2797

\bibitem[{{Font} {et~al.}(2008){Font}, {Bower}, {McCarthy}, {Benson}, {Frenk},
  {Helly}, {Lacey}, {Baugh}, \& {Cole}}]{Font08}
{Font} A.~S., {Bower} R.~G., {McCarthy} I.~G., {Benson} A.~J., {Frenk} C.~S.,
  {Helly} J.~C., {Lacey} C.~G., {Baugh} C.~M. {et~al}, 2008, \mnras, 389, 1619

\bibitem[{{Gammie}(2001)}]{Gammie01}
{Gammie} C.~F., 2001, \apj, 553, 174

\bibitem[{{Genzel} {et~al.}(2010){Genzel}, {Tacconi}, {Gracia-Carpio},
  {Sternberg}, {Cooper}, {Shapiro}, {Bolatto}, {Bouch{\'e}}, {Bournaud},
  {Burkert}, {Combes}, {Comerford}, {Cox}, {Davis}, {Schreiber},
  {Garcia-Burillo}, {Lutz}, {Naab}, {Neri}, {Omont}, {Shapley}, \&
  {Weiner}}]{Genzel10}
{Genzel} R., {Tacconi} L.~J., {Gracia-Carpio} J., {Sternberg} A., {Cooper}
  M.~C., {Shapiro} K., {Bolatto} A., {Bouch{\'e}} N. {et~al}, 2010, \mnras,
  407, 2091

\bibitem[{{Giovanelli} {et~al.}(2005){Giovanelli}, {Haynes}, {Kent},
  {Perillat}, {Saintonge}, {Brosch}, {Catinella}, {Hoffman}, {Stierwalt},
  {Spekkens}, {Lerner}, {Masters}, {Momjian}, {Rosenberg}, {Springob},
  {Boselli}, {Charmandaris}, {Darling}, {Davies}, {Garcia Lambas}, {Gavazzi},
  {Giovanardi}, {Hardy}, {Hunt}, {Iovino}, {Karachentsev}, {Karachentseva},
  {Koopmann}, {Marinoni}, {Minchin}, {Muller}, {Putman}, {Pantoja}, {Salzer},
  {Scodeggio}, {Skillman}, {Solanes}, {Valotto}, {van Driel}, \& {van
  Zee}}]{Giovanelli05}
{Giovanelli} R., {Haynes} M.~P., {Kent} B.~R., {Perillat} P., {Saintonge} A.,
  {Brosch} N., {Catinella} B., {Hoffman} G.~L. {et~al}, 2005, \aj, 130, 2598

\bibitem[{{Gonz{\'a}lez} {et~al.}(2009){Gonz{\'a}lez}, {Lacey}, {Baugh},
  {Frenk}, \& {Benson}}]{Gonzalez09}
{Gonz{\'a}lez} J.~E., {Lacey} C.~G., {Baugh} C.~M., {Frenk} C.~S., {Benson}
  A.~J., 2009, \mnras, 397, 1254

\bibitem[{{Gonzalez-Perez} {et~al.}(2014){Gonzalez-Perez}, {Lacey}, {Baugh},
  {Lagos}, {Helly}, {Campbell}, \& {Mitchell}}]{Gonzalez-Perez13}
{Gonzalez-Perez} V., {Lacey} C.~G., {Baugh} C.~M., {Lagos} C.~D.~P., {Helly}
  J., {Campbell} D.~J.~R., {Mitchell} P.~D., 2014, \mnras

\bibitem[{{Granato} {et~al.}(2000){Granato}, {Lacey}, {Silva}, {Bressan},
  {Baugh}, {Cole}, \& {Frenk}}]{Granato00}
{Granato} G.~L., {Lacey} C.~G., {Silva} L., {Bressan} A., {Baugh} C.~M., {Cole}
  S., {Frenk} C.~S., 2000, \apj, 542, 710

\bibitem[{{Hambly} {et~al.}(2001){Hambly}, {MacGillivray}, {Read}, {Tritton},
  {Thomson}, {Kelly}, {Morgan}, {Smith}, {Driver}, {Williamson}, {Parker},
  {Hawkins}, {Williams}, \& {Lawrence}}]{Hambly01}
{Hambly} N.~C., {MacGillivray} H.~T., {Read} M.~A., {Tritton} S.~B., {Thomson}
  E.~B., {Kelly} B.~D., {Morgan} D.~H., {Smith} R.~E. {et~al}, 2001, \mnras,
  326, 1279

\bibitem[{{Hopkins} {et~al.}(2010){Hopkins}, {Bundy}, {Croton}, {Hernquist},
  {Keres}, {Khochfar}, {Stewart}, {Wetzel}, \& {Younger}}]{Hopkins10}
{Hopkins} P.~F., {Bundy} K., {Croton} D., {Hernquist} L., {Keres} D.,
  {Khochfar} S., {Stewart} K., {Wetzel} A. {et~al}, 2010, \apj, 715, 202

\bibitem[{{Huang} {et~al.}(2012){Huang}, {Haynes}, {Giovanelli}, \&
  {Brinchmann}}]{Huang12}
{Huang} S., {Haynes} M.~P., {Giovanelli} R., {Brinchmann} J., 2012, \apj, 756,
  113

\bibitem[{{Jarrett} {et~al.}(2000){Jarrett}, {Chester}, {Cutri}, {Schneider},
  {Skrutskie}, \& {Huchra}}]{Jarrett00}
{Jarrett} T.~H., {Chester} T., {Cutri} R., {Schneider} S., {Skrutskie} M.,
  {Huchra} J.~P., 2000, \aj, 119, 2498

\bibitem[{{Jiang} {et~al.}(2008){Jiang}, {Jing}, {Faltenbacher}, {Lin}, \&
  {Li}}]{Jiang07}
{Jiang} C.~Y., {Jing} Y.~P., {Faltenbacher} A., {Lin} W.~P., {Li} C., 2008,
  \apj, 675, 1095

\bibitem[{{Kauffmann}(1996)}]{Kauffmann96}
{Kauffmann} G., 1996, \mnras, 281, 487

\bibitem[{{Kauffmann} {et~al.}(2012){Kauffmann}, {Li}, {Fu}, {Saintonge},
  {Catinella}, {Tacconi}, {Kramer}, {Genzel}, {Moran}, \&
  {Schiminovich}}]{Kauffmann12}
{Kauffmann} G., {Li} C., {Fu} J., {Saintonge} A., {Catinella} B., {Tacconi}
  L.~J., {Kramer} C., {Genzel} R. {et~al}, 2012, \mnras, 422, 997

\bibitem[{{Kaviraj} {et~al.}(2007){Kaviraj}, {Schawinski}, {Devriendt},
  {Ferreras}, {Khochfar}, {Yoon}, {Yi}, {Deharveng}, {Boselli}, {Barlow},
  {Conrow}, {Forster}, {Friedman}, {Martin}, {Morrissey}, {Neff},
  {Schiminovich}, {Seibert}, {Small}, {Wyder}, {Bianchi}, {Donas}, {Heckman},
  {Lee}, {Madore}, {Milliard}, {Rich}, \& {Szalay}}]{Kaviraj07}
{Kaviraj} S., {Schawinski} K., {Devriendt} J.~E.~G., {Ferreras} I., {Khochfar}
  S., {Yoon} S.-J., {Yi} S.~K., {Deharveng} J.-M. {et~al}, 2007, \apjs, 173,
  619

\bibitem[{{Kennicutt}(1983)}]{Kennicutt83}
{Kennicutt} Jr. R.~C., 1983, \apj, 272, 54

\bibitem[{{Keres} {et~al.}(2003){Keres}, {Yun}, \& {Young}}]{Keres03}
{Keres} D., {Yun} M.~S., {Young} J.~S., 2003, \apj, 582, 659

\bibitem[{{Khochfar} {et~al.}(2011){Khochfar}, {Emsellem}, {Serra}, {Bois},
  {Alatalo}, {Bacon}, {Blitz}, {Bournaud}, {Bureau}, {Cappellari}, {Davies},
  {Davis}, {de Zeeuw}, {Duc}, {Krajnovi{\'c}}, {Kuntschner}, {Lablanche},
  {McDermid}, {Morganti}, {Naab}, {Oosterloo}, {Sarzi}, {Scott}, {Weijmans}, \&
  {Young}}]{Khochfar11}
{Khochfar} S., {Emsellem} E., {Serra} P., {Bois} M., {Alatalo} K., {Bacon} R.,
  {Blitz} L., {Bournaud} F. {et~al}, 2011, \mnras, 417, 845

\bibitem[{{Komatsu} {et~al.}(2011){Komatsu}, {Smith}, {Dunkley}, {Bennett},
  {Gold}, {Hinshaw}, {Jarosik}, {Larson}, {Nolta}, {Page}, {Spergel},
  {Halpern}, {Hill}, {Kogut}, {Limon}, {Meyer}, {Odegard}, {Tucker}, {Weiland},
  {Wollack}, \& {Wright}}]{Komatsu11}
{Komatsu} E., {Smith} K.~M., {Dunkley} J., {Bennett} C.~L., {Gold} B.,
  {Hinshaw} G., {Jarosik} N., {Larson} D. {et~al}, 2011, \apjs, 192, 18

\bibitem[{{Krumholz} \& {Burkert}(2010)}]{Krumholz10}
{Krumholz} M., {Burkert} A., 2010, \apj, 724, 895

\bibitem[{{Lacey} {et~al.}(1993){Lacey}, {Guiderdoni}, {Rocca-Volmerange}, \&
  {Silk}}]{Lacey93}
{Lacey} C., {Guiderdoni} B., {Rocca-Volmerange} B., {Silk} J., 1993, \apj, 402,
  15

\bibitem[{{Lacey} {et~al.}(2008){Lacey}, {Baugh}, {Frenk}, {Silva}, {Granato},
  \& {Bressan}}]{Lacey08}
{Lacey} C.~G., {Baugh} C.~M., {Frenk} C.~S., {Silva} L., {Granato} G.~L.,
  {Bressan} A., 2008, \mnras, 385, 1155

\bibitem[{{Lagos} {et~al.}(2011{\natexlab{a}}){Lagos}, {Baugh}, {Lacey},
  {Benson}, {Kim}, \& {Power}}]{Lagos11}
{Lagos} C.~D.~P., {Baugh} C.~M., {Lacey} C.~G., {Benson} A.~J., {Kim} H.-S.,
  {Power} C., 2011{\natexlab{a}}, \mnras, 418, 1649

\bibitem[{{Lagos} {et~al.}(2014){Lagos}, {Baugh}, {Zwaan}, {Lacey},
  {Gonzalez-Perez}, {Power}, {Swinbank}, \& {van Kampen}}]{Lagos14}
{Lagos} C.~D.~P., {Baugh} C.~M., {Zwaan} M.~A., {Lacey} C.~G., {Gonzalez-Perez}
  V., {Power} C., {Swinbank} A.~M., {van Kampen} E., 2014, \mnras, 440, 920

\bibitem[{{Lagos} {et~al.}(2012){Lagos}, {Bayet}, {Baugh}, {Lacey}, {Bell},
  {Fanidakis}, \& {Geach}}]{Lagos12}
{Lagos} C.~d.~P., {Bayet} E., {Baugh} C.~M., {Lacey} C.~G., {Bell} T.~A.,
  {Fanidakis} N., {Geach} J.~E., 2012, \mnras, 426, 2142

\bibitem[{{Lagos} {et~al.}(2013){Lagos}, {Lacey}, \& {Baugh}}]{Lagos13}
{Lagos} C.~d.~P., {Lacey} C.~G., {Baugh} C.~M., 2013, \mnras, 436, 1787

\bibitem[{{Lagos} {et~al.}(2011{\natexlab{b}}){Lagos}, {Lacey}, {Baugh},
  {Bower}, \& {Benson}}]{Lagos10}
{Lagos} C.~D.~P., {Lacey} C.~G., {Baugh} C.~M., {Bower} R.~G., {Benson} A.~J.,
  2011{\natexlab{b}}, \mnras, 416, 1566

\bibitem[{{Lang} {et~al.}(2014){Lang}, {Wuyts}, {Somerville}, {F{\"o}rster
  Schreiber}, {Genzel}, {Bell}, {Brammer}, {Dekel}, {Faber}, {Ferguson},
  {Grogin}, {Kocevski}, {Koekemoer}, {Lutz}, {McGrath}, {Momcheva}, {Nelson},
  {Primack}, {Rosario}, {Skelton}, {Tacconi}, {van Dokkum}, \&
  {Whitaker}}]{Lang14}
{Lang} P., {Wuyts} S., {Somerville} R.~S., {F{\"o}rster Schreiber} N.~M.,
  {Genzel} R., {Bell} E.~F., {Brammer} G., {Dekel} A. {et~al}, 2014, \apj, 788,
  11

\bibitem[{{Le F{\`e}vre} {et~al.}(2000){Le F{\`e}vre}, {Abraham}, {Lilly},
  {Ellis}, {Brinchmann}, {Schade}, {Tresse}, {Colless}, {Crampton},
  {Glazebrook}, {Hammer}, \& {Broadhurst}}]{LeFevre00}
{Le F{\`e}vre} O., {Abraham} R., {Lilly} S.~J., {Ellis} R.~S., {Brinchmann} J.,
  {Schade} D., {Tresse} L., {Colless} M. {et~al}, 2000, \mnras, 311, 565

\bibitem[{{Lemonias} {et~al.}(2013){Lemonias}, {Schiminovich}, {Catinella},
  {Heckman}, \& {Moran}}]{Lemonias13}
{Lemonias} J.~J., {Schiminovich} D., {Catinella} B., {Heckman} T.~M., {Moran}
  S.~M., 2013, \apj, 776, 74

\bibitem[{{Leroy} {et~al.}(2008){Leroy}, {Walter}, {Brinks}, {Bigiel}, {de
  Blok}, {Madore}, \& {Thornley}}]{Leroy08}
{Leroy} A.~K., {Walter} F., {Brinks} E., {Bigiel} F., {de Blok} W.~J.~G.,
  {Madore} B., {Thornley} M.~D., 2008, \aj, 136, 2782

\bibitem[{{Lintott} {et~al.}(2008){Lintott}, {Schawinski}, {Slosar}, {Land},
  {Bamford}, {Thomas}, {Raddick}, {Nichol}, {Szalay}, {Andreescu}, {Murray}, \&
  {Vandenberg}}]{Lintott08}
{Lintott} C.~J., {Schawinski} K., {Slosar} A., {Land} K., {Bamford} S.,
  {Thomas} D., {Raddick} M.~J., {Nichol} R.~C. {et~al}, 2008, \mnras, 389, 1179

\bibitem[{{Lisenfeld} {et~al.}(2011){Lisenfeld}, {Espada}, {Verdes-Montenegro},
  {Kuno}, {Leon}, {Sabater}, {Sato}, {Sulentic}, {Verley}, \&
  {Yun}}]{Lisenfeld11}
{Lisenfeld} U., {Espada} D., {Verdes-Montenegro} L., {Kuno} N., {Leon} S.,
  {Sabater} J., {Sato} N., {Sulentic} J. {et~al}, 2011, \aap, 534, A102

\bibitem[{{Malbon} {et~al.}(2007){Malbon}, {Baugh}, {Frenk}, \&
  {Lacey}}]{Malbon07}
{Malbon} R.~K., {Baugh} C.~M., {Frenk} C.~S., {Lacey} C.~G., 2007, \mnras, 382,
  1394

\bibitem[{{Maraston}(2005)}]{Maraston05}
{Maraston} C., 2005, \mnras, 362, 799

\bibitem[{{Marigo}(2001)}]{Marigo01}
{Marigo} P., 2001, \aap, 370, 194

\bibitem[{{Martig} {et~al.}(2013){Martig}, {Crocker}, {Bournaud}, {Emsellem},
  {Gabor}, {Alatalo}, {Blitz}, {Bois}, {Bureau}, {Cappellari}, {Davies},
  {Davis}, {Dekel}, {de Zeeuw}, {Duc}, {Falc{\'o}n-Barroso}, {Khochfar},
  {Krajnovi{\'c}}, {Kuntschner}, {Morganti}, {McDermid}, {Naab}, {Oosterloo},
  {Sarzi}, {Scott}, {Serra}, {Griffin}, {Teyssier}, {Weijmans}, \&
  {Young}}]{Martig13}
{Martig} M., {Crocker} A.~F., {Bournaud} F., {Emsellem} E., {Gabor} J.~M.,
  {Alatalo} K., {Blitz} L., {Bois} M. {et~al}, 2013, \mnras, 432, 1914

\bibitem[{{Martin} {et~al.}(2010){Martin}, {Papastergis}, {Giovanelli},
  {Haynes}, {Springob}, \& {Stierwalt}}]{Martin10}
{Martin} A.~M., {Papastergis} E., {Giovanelli} R., {Haynes} M.~P., {Springob}
  C.~M., {Stierwalt} S., 2010, \apj, 723, 1359

\bibitem[{{McCarthy} {et~al.}(2008){McCarthy}, {Frenk}, {Font}, {Lacey},
  {Bower}, {Mitchell}, {Balogh}, \& {Theuns}}]{McCarthy08}
{McCarthy} I.~G., {Frenk} C.~S., {Font} A.~S., {Lacey} C.~G., {Bower} R.~G.,
  {Mitchell} N.~L., {Balogh} M.~L., {Theuns} T., 2008, \mnras, 383, 593

\bibitem[{{Meyer} {et~al.}(2008){Meyer}, {Zwaan}, {Webster}, {Schneider}, \&
  {Staveley-Smith}}]{Meyer08a}
{Meyer} M.~J., {Zwaan} M.~A., {Webster} R.~L., {Schneider} S., {Staveley-Smith}
  L., 2008, \mnras, 391, 1712

\bibitem[{{Meyer} {et~al.}(2004){Meyer}, {Zwaan}, {Webster}, {Staveley-Smith},
  {Ryan-Weber}, {Drinkwater}, {Barnes}, {Howlett}, {Kilborn}, {Stevens},
  {Waugh}, {Pierce}, {Bhathal}, {de Blok}, {Disney}, {Ekers}, {Freeman},
  {Garcia}, {Gibson}, {Harnett}, {Henning}, {Jerjen}, {Kesteven}, {Knezek},
  {Koribalski}, {Mader}, {Marquarding}, {Minchin}, {O'Brien}, {Oosterloo},
  {Price}, {Putman}, {Ryder}, {Sadler}, {Stewart}, {Stootman}, \&
  {Wright}}]{Meyer04}
{Meyer} M.~J., {Zwaan} M.~A., {Webster} R.~L., {Staveley-Smith} L.,
  {Ryan-Weber} E., {Drinkwater} M.~J., {Barnes} D.~G., {Howlett} M. {et~al},
  2004, \mnras, 350, 1195

\bibitem[{{Mitchell} {et~al.}(2013){Mitchell}, {Lacey}, {Baugh}, \&
  {Cole}}]{Mitchell13}
{Mitchell} P.~D., {Lacey} C.~G., {Baugh} C.~M., {Cole} S., 2013, \mnras, 435,
  87

\bibitem[{{Mo} {et~al.}(1998){Mo}, {Mao}, \& {White}}]{Mo98}
{Mo} H.~J., {Mao} S., {White} S.~D.~M., 1998, \mnras, 295, 319

\bibitem[{{Naab} \& {Burkert}(2003)}]{Naab03}
{Naab} T., {Burkert} A., 2003, \apj, 597, 893

\bibitem[{{Naab} {et~al.}(2013){Naab}, {Oser}, {Emsellem}, {Cappellari},
  {Krajnovic}, {McDermid}, {Alatalo}, {Bayet}, {Blitz}, {Bois}, {Bournaud},
  {Bureau}, {Crocker}, {Davies}, {Davis}, {de Zeeuw}, {Duc}, {Hirschmann},
  {Johansson}, {Khochfar}, {Kuntschner}, {Morganti}, {Oosterloo}, {Sarzi},
  {Scott}, {Serra}, {van de Ven}, {Weijmans}, \& {Young}}]{Naab13}
{Naab} T., {Oser} L., {Emsellem} E., {Cappellari} M., {Krajnovic} D.,
  {McDermid} R.~M., {Alatalo} K., {Bayet} E. {et~al}, 2013, ArXiv:1311.0284

\bibitem[{{Navarro} {et~al.}(1997){Navarro}, {Frenk}, \& {White}}]{Navarro97}
{Navarro} J.~F., {Frenk} C.~S., {White} S.~D.~M., 1997, \apj, 490, 493

\bibitem[{{Parry} {et~al.}(2009){Parry}, {Eke}, \& {Frenk}}]{Parry09}
{Parry} O.~H., {Eke} V.~R., {Frenk} C.~S., 2009, \mnras, 396, 1972

\bibitem[{{Portinari} {et~al.}(1998){Portinari}, {Chiosi}, \&
  {Bressan}}]{Portinari98}
{Portinari} L., {Chiosi} C., {Bressan} A., 1998, \aap, 334, 505

\bibitem[{{Ruiz} {et~al.}(2013){Ruiz}, {Cora}, {Padilla}, {Dom{\'{\i}}nguez},
  {Tecce}, {Orsi}, {Yaryura}, {Garc{\'{\i}}a Lambas}, {Gargiulo}, \& {Mu{\~n}oz
  Arancibia}}]{Ruiz13}
{Ruiz} A.~N., {Cora} S.~A., {Padilla} N.~D., {Dom{\'{\i}}nguez} M.~J., {Tecce}
  T.~E., {Orsi} {\'A}., {Yaryura} Y.~C., {Garc{\'{\i}}a Lambas} D. {et~al},
  2013, ArXiv e-prints

\bibitem[{{Saintonge} {et~al.}(2011){Saintonge}, {Kauffmann}, {Kramer},
  {Tacconi}, {Buchbender}, {Catinella}, {Fabello}, {Graci{\'a}-Carpio}, {Wang},
  {Cortese}, {Fu}, {Genzel}, {Giovanelli}, {Guo}, {Haynes}, {Heckman},
  {Krumholz}, {Lemonias}, {Li}, {Moran}, {Rodriguez-Fernandez}, {Schiminovich},
  {Schuster}, \& {Sievers}}]{Saintonge11}
{Saintonge} A., {Kauffmann} G., {Kramer} C., {Tacconi} L.~J., {Buchbender} C.,
  {Catinella} B., {Fabello} S., {Graci{\'a}-Carpio} J. {et~al}, 2011, \mnras,
  415, 32

\bibitem[{{Schiminovich} {et~al.}(2007){Schiminovich}, {Wyder}, {Martin},
  {Johnson}, {Salim}, {Seibert}, {Treyer}, {Budav{\'a}ri}, {Hoopes},
  {Zamojski}, {Barlow}, {Forster}, {Friedman}, {Morrissey}, {Neff}, {Small},
  {Bianchi}, {Donas}, {Heckman}, {Lee}, {Madore}, {Milliard}, {Rich}, {Szalay},
  {Welsh}, \& {Yi}}]{Schiminovich07}
{Schiminovich} D., {Wyder} T.~K., {Martin} D.~C., {Johnson} B.~D., {Salim} S.,
  {Seibert} M., {Treyer} M.~A., {Budav{\'a}ri} T. {et~al}, 2007, \apjs, 173,
  315

\bibitem[{{Schmidt}(1968)}]{Schmidt68}
{Schmidt} M., 1968, \apj, 151, 393

\bibitem[{{Serra} {et~al.}(2012){Serra}, {Oosterloo}, {Morganti}, {Alatalo},
  {Blitz}, {Bois}, {Bournaud}, {Bureau}, {Cappellari}, {Crocker}, {Davies},
  {Davis}, {de Zeeuw}, {Duc}, {Emsellem}, {Khochfar}, {Krajnovi{\'c}},
  {Kuntschner}, {Lablanche}, {McDermid}, {Naab}, {Sarzi}, {Scott}, {Trager},
  {Weijmans}, \& {Young}}]{Serra12}
{Serra} P., {Oosterloo} T., {Morganti} R., {Alatalo} K., {Blitz} L., {Bois} M.,
  {Bournaud} F., {Bureau} M. {et~al}, 2012, \mnras, 2823

\bibitem[{{Serra} {et~al.}(2014){Serra}, {Oser}, {Krajnovic}, {Naab},
  {Oosterloo}, {Morganti}, {Cappellari}, {Emsellem}, {Young}, {Blitz}, {Davis},
  {Duc}, {Hirschmann}, {Weijmans}, {Alatalo}, {Bayet}, {Bois}, {Bournaud},
  {Bureau}, {Davies}, {de Zeeuw}, {Khochfar}, {Kuntschner}, {Lablanche},
  {McDermid}, {Sarzi}, \& {Scott}}]{Serra14}
{Serra} P., {Oser} L., {Krajnovic} D., {Naab} T., {Oosterloo} T., {Morganti}
  R., {Cappellari} M., {Emsellem} E. {et~al}, 2014, ArXiv:1401.3180

\bibitem[{{Simien} \& {de Vaucouleurs}(1986)}]{Simien86}
{Simien} F., {de Vaucouleurs} G., 1986, \apj, 302, 564

\bibitem[{{Smith} {et~al.}(2012){Smith}, {Gomez}, {Eales}, {Ciesla}, {Boselli},
  {Cortese}, {Bendo}, {Baes}, {Bianchi}, {Clemens}, {Clements}, {Cooray},
  {Davies}, {de Looze}, {di Serego Alighieri}, {Fritz}, {Gavazzi}, {Gear},
  {Madden}, {Mentuch}, {Panuzzo}, {Pohlen}, {Spinoglio}, {Verstappen},
  {Vlahakis}, {Wilson}, \& {Xilouris}}]{Smith12}
{Smith} M.~W.~L., {Gomez} H.~L., {Eales} S.~A., {Ciesla} L., {Boselli} A.,
  {Cortese} L., {Bendo} G.~J., {Baes} M. {et~al}, 2012, \apj, 748, 123

\bibitem[{{Spergel} {et~al.}(2003){Spergel}, {Verde}, {Peiris}, {Komatsu},
  {Nolta}, {Bennett}, {Halpern}, {Hinshaw}, {Jarosik}, {Kogut}, {Limon},
  {Meyer}, {Page}, {Tucker}, {Weiland}, {Wollack}, \& {Wright}}]{Spergel03}
{Spergel} D.~N., {Verde} L., {Peiris} H.~V., {Komatsu} E., {Nolta} M.~R.,
  {Bennett} C.~L., {Halpern} M., {Hinshaw} G. {et~al}, 2003, \apjs, 148, 175

\bibitem[{{Springel} {et~al.}(2005){Springel}, {White}, {Jenkins}, {Frenk},
  {Yoshida}, {Gao}, {Navarro}, {Thacker}, {Croton}, {Helly}, {Peacock}, {Cole},
  {Thomas}, {Couchman}, {Evrard}, {Colberg}, \& {Pearce}}]{Springel05}
{Springel} V., {White} S.~D.~M., {Jenkins} A., {Frenk} C.~S., {Yoshida} N.,
  {Gao} L., {Navarro} J., {Thacker} R. {et~al}, 2005, \nat, 435, 629

\bibitem[{{Strateva} {et~al.}(2001){Strateva}, {Ivezi{\'c}}, {Knapp},
  {Narayanan}, {Strauss}, {Gunn}, {Lupton}, {Schlegel}, {Bahcall}, {Brinkmann},
  {Brunner}, {Budav{\'a}ri}, {Csabai}, {Castander}, {Doi}, {Fukugita}, {Gy{\H
  o}ry}, {Hamabe}, {Hennessy}, {Ichikawa}, {Kunszt}, {Lamb}, {McKay},
  {Okamura}, {Racusin}, {Sekiguchi}, {Schneider}, {Shimasaku}, \&
  {York}}]{Strateva01}
{Strateva} I., {Ivezi{\'c}} {\v Z}., {Knapp} G.~R., {Narayanan} V.~K.,
  {Strauss} M.~A., {Gunn} J.~E., {Lupton} R.~H., {Schlegel} D. {et~al}, 2001,
  \aj, 122, 1861

\bibitem[{{Sutherland} \& {Dopita}(1993)}]{Sutherland93}
{Sutherland} R.~S., {Dopita} M.~A., 1993, \apjs, 88, 253

\bibitem[{{Tecce} {et~al.}(2010){Tecce}, {Cora}, {Tissera}, {Abadi}, \&
  {Lagos}}]{Tecce10}
{Tecce} T.~E., {Cora} S.~A., {Tissera} P.~B., {Abadi} M.~G., {Lagos} C.~D.~P.,
  2010, \mnras, 408, 2008

\bibitem[{{Toomre}(1977)}]{Toomre77}
{Toomre} A., 1977, in Evolution of Galaxies and Stellar Populations, {Tinsley}
  B.~M., {Larson} D.~Campbell R.~B.~G., eds., p. 401

\bibitem[{{Wang} {et~al.}(2014){Wang}, {Fu}, {Aumer}, {Kauffmann}, {J{\'o}zsa},
  {Serra}, {Huang}, {Brinchmann}, {van der Hulst}, \& {Bigiel}}]{Wang14}
{Wang} J., {Fu} J., {Aumer} M., {Kauffmann} G., {J{\'o}zsa} G.~I.~G., {Serra}
  P., {Huang} M.-l., {Brinchmann} J. {et~al}, 2014, \mnras, 441, 2159

\bibitem[{{Weinzirl} {et~al.}(2009){Weinzirl}, {Jogee}, {Khochfar}, {Burkert},
  \& {Kormendy}}]{Weinzirl09}
{Weinzirl} T., {Jogee} S., {Khochfar} S., {Burkert} A., {Kormendy} J., 2009,
  \apj, 696, 411

\bibitem[{{Welch} {et~al.}(2010){Welch}, {Sage}, \& {Young}}]{Welch10}
{Welch} G.~A., {Sage} L.~J., {Young} L.~M., 2010, \apj, 725, 100

\bibitem[{{White} \& {Rees}(1978)}]{White78}
{White} S.~D.~M., {Rees} M.~J., 1978, \mnras, 183, 341

\bibitem[{{Wuyts} {et~al.}(2011){Wuyts}, {F{\"o}rster Schreiber}, {van der
  Wel}, {Magnelli}, {Guo}, {Genzel}, {Lutz}, {Aussel}, {Barro}, {Berta},
  {Cava}, {Graci{\'a}-Carpio}, {Hathi}, {Huang}, {Kocevski}, {Koekemoer},
  {Lee}, {Le Floc'h}, {McGrath}, {Nordon}, {Popesso}, {Pozzi}, {Riguccini},
  {Rodighiero}, {Saintonge}, \& {Tacconi}}]{Wuyts11}
{Wuyts} S., {F{\"o}rster Schreiber} N.~M., {van der Wel} A., {Magnelli} B.,
  {Guo} Y., {Genzel} R., {Lutz} D., {Aussel} H. {et~al}, 2011, \apj, 742, 96

\bibitem[{{Young} \& {Knezek}(1989)}]{Young89}
{Young} J.~S., {Knezek} P.~M., 1989, \apjl, 347, L55

\bibitem[{{Young} {et~al.}(2011){Young}, {Bureau}, {Davis}, {Combes},
  {McDermid}, {Alatalo}, {Blitz}, {Bois}, {Bournaud}, {Cappellari}, {Davies},
  {de Zeeuw}, {Emsellem}, {Khochfar}, {Krajnovi{\'c}}, {Kuntschner},
  {Lablanche}, {Morganti}, {Naab}, {Oosterloo}, {Sarzi}, {Scott}, {Serra}, \&
  {Weijmans}}]{Young11}
{Young} L.~M., {Bureau} M., {Davis} T.~A., {Combes} F., {McDermid} R.~M.,
  {Alatalo} K., {Blitz} L., {Bois} M. {et~al}, 2011, \mnras, 414, 940

\bibitem[{{Young} {et~al.}(2013){Young}, {Scott}, {Serra}, {Alatalo}, {Bayet},
  {Blitz}, {Bois}, {Bournaud}, {Bureau}, {Crocker}, {Cappellari}, {Davies},
  {Davis}, {de Zeeuw}, {Duc}, {Emsellem}, {Khochfar}, {Krajnovic},
  {Kuntschner}, {McDermid}, {Morganti}, {Naab}, {Oosterloo}, {Sarzi}, \&
  {Weijmans}}]{Young13}
{Young} L.~M., {Scott} N., {Serra} P., {Alatalo} K., {Bayet} E., {Blitz} L.,
  {Bois} M., {Bournaud} F. {et~al}, 2013, ArXiv:1312.6318

\bibitem[{{Zwaan} {et~al.}(2005){Zwaan}, {Meyer}, {Staveley-Smith}, \&
  {Webster}}]{Zwaan05}
{Zwaan} M.~A., {Meyer} M.~J., {Staveley-Smith} L., {Webster} R.~L., 2005,
  \mnras, 359, L30

\end{thebibliography}

\label{lastpage}
\appendix
\section[]{The effect of varying the partial ram pressure stripping parameters}\label{RPpars}
\begin{figure}
\begin{center}
\includegraphics[trim = 1.5mm 0.5mm 1mm 1mm,clip,width=0.47\textwidth]{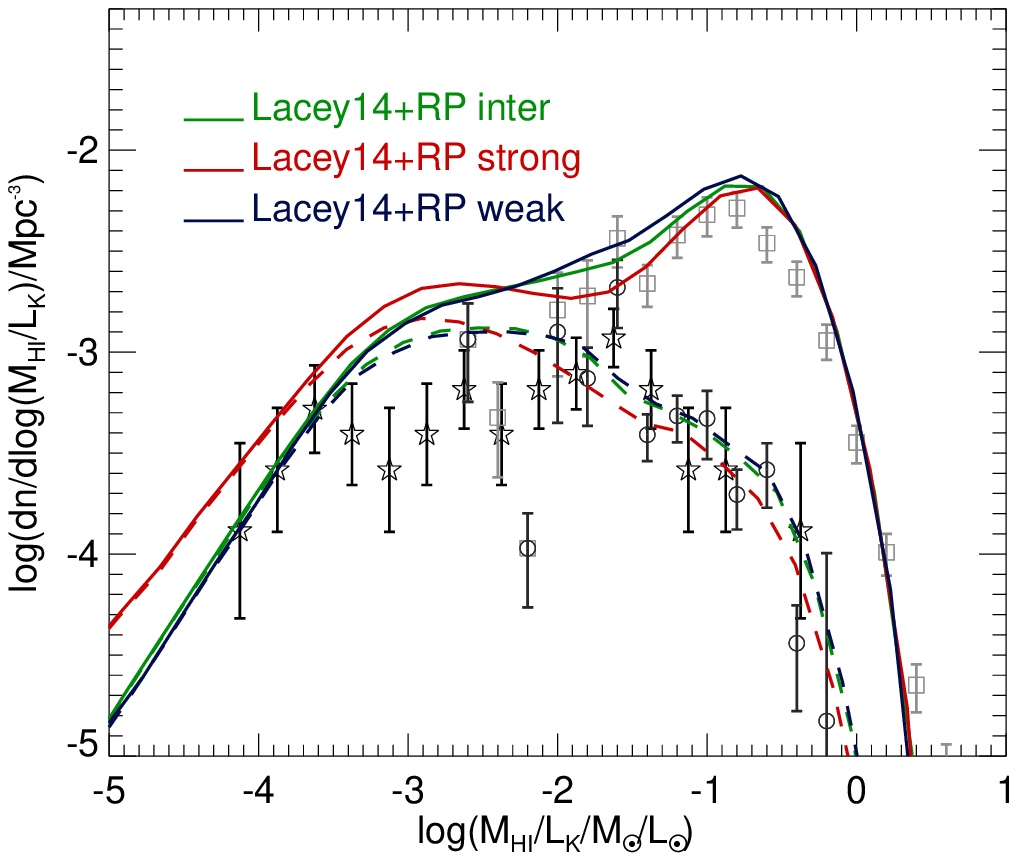}
\caption{The HI gas fraction, $M_{\rm HI}/L_{\rm K}$, distribution 
for all galaxies with $L_{\rm K}>6\times10^9\,L_{\odot}$ (solid lines)
and the sub-sample of ETGs ($B/T>0.5$; dashed lines)
for the Lacey14+RP model. We have adopted three different values for $\epsilon_{\rm strip}$: 0.01 (RP weak), 
0.1 (RP inter) and 1 (RP strong). The observations are described in $\S$~\ref{obssec}.} 
\label{RPparameters}
\end{center}
\end{figure}

We show in Fig.~\ref{RPparameters} the HI gas fractions 
for the Lacey14+RP model using three different values for $\epsilon_{\rm strip}$, which controls the 
fraction of the reheated mass from stellar feedback that was driven out from the galaxy 
after the first passage of the satellite that is affected by stripping. The reheated mass considered here is the 
one that sits outside the stripping radius. 
We remind the reader that the latter is calculated at the pericentre of the 
satellite's orbit (see $\S$~\ref{Sec:rampressure}). 
A value of $\epsilon_{\rm strip}=1$ implies that the reheated gas gets stripped in subsequent steps at the same rate as the hot gas
of the galaxy when the satellite first passed through its pericentre. To adopt this value in the model leads to predictions 
that are very similar
to the Lacey14 model predictions (including strangulation). This shows that $\epsilon_{\rm strip}=1$ drives very efficient 
hot gas stripping close to the fully efficient case.
The model using the values $\epsilon_{\rm strip}=0.01$ and $\epsilon_{\rm strip}=0.1$ produce very similar gas fractions. 
Both cases lead to predict number densities of ETGs with HI and H$_2$ gas fraction $>0.02$ higher than both the standard Lacey14 and
the version including partial ram pressure stripping with $\epsilon_{\rm strip}=1$. 
 This higher number density is due to the higher rates of infalling cold gas, which replenish 
the ISM with newly cooled gas. We conclude that for values of 
$\epsilon_{\rm strip}\lesssim 0.3$ the results presented in the paper do not considerably change, mainly because 
of a self-regulation of outflows and inflows in satellites: if a very small value of $\epsilon_{\rm strip}$ is adopted, 
it will drive higher accretion rates of gas onto the disk, which will lead to higher star formation rates, and therefore 
higher outflows rates. For values $\epsilon_{\rm strip}\gtrsim 0.3$ the satellite's hot gas reservoir is removed too quickly 
driving very little further gas accretion of newly cooled gas onto the galaxy disk. The latter has the effect of quenching 
star formation in the satellite galaxy quickly, driving low HI and H$_2$ gas fractions.

\section[]{Computing the contribution from different gas sources}\label{GasSourcesContribution}

The equations governing the mass exchange between stars (${M}_{\star}$), gas in the disk (${M}_{\rm g}$),  
ejected mass from the disk (${M}_{\rm eject}$) and the hot halo (${M}_{\rm hot}$) are as follows:

\begin{eqnarray}
\dot{M}_{\star}&=&(1-R)\psi,\label{Eqs:SFset1}\\
\dot{M}_{\rm g} &=& \dot{M}_{\rm cool}-(1-R)\psi-\dot{M}_{\rm eject}\\
\dot{M}_{\rm eject} &=& \beta\,\psi\\
\dot{M}_{\rm hot} &=& -\dot{M}_{\rm cool}+\frac{M_{\rm eject}}{\tau_{\rm rein}}.\label{Eqs:SFset2}
\end{eqnarray}

\noindent Here, $\psi$ is the instantaneous SFR described in $\S$~\ref{SFlaw}, 
$\dot{M}_{\rm cool}$ is the cooling rate described in $\S$~\ref{Sec:Cooling}, 
$R$ is the recycled fraction described in $\S$~\ref{Sec:Randp} and 
$\beta$ is the efficiency of supernovae feedback. The latter depends on the circular velocity as 
$\beta=(V/V_{0})^{-\alpha_{\rm hot}}$ (see \citet{Lagos13} for a discussion of the physical motivation of this parametrisation). 
The parameters adopted in each model are 
$\alpha_{\rm hot}=3.2$ in the Lagos12, Gonzalez-Perez14 
and Lacey14 models, $V_{0}=485\,\rm km\,s^{-1}$ in Lagos12, 
$V_{0}=425\,\rm km\,s^{-1}$ in Gonzalez-Perez14 and 
$V_{0}=320\,\rm km\,s^{-1}$  in Lacey14. We define the changes in the quantities above 
in an arbitrary timestep as $\Delta {M}_{\rm cool}$, $\Delta {M}_{\star}$, $\Delta {M}_{\rm g}$ and $\Delta {M}_{\rm eject}$.

In order to follow the three sources of gas in galaxies (galaxy mergers, recycling and gas cooling), we 
define $M_{\rm merg}$, $M_{\rm recycle}$ and $M_{\rm cooling}$, and calculate them as follows. We first 
add the amount of cooled gas in the cooling component and the one following the total gas in the disk, ${M}_{\rm g}$,

\begin{eqnarray}
\Delta M_{\rm cooling} = \Delta {M}_{\rm cool},\label{mcooling1}.
\end{eqnarray}

\noindent We update the quantities $M_{\rm cooling}$ and ${M}_{\rm g}$ by adding $\Delta M_{\rm cooling}$.
From ${M}_{\rm g}$, an amount $\Delta {M}_{\star}$ of stars 
is formed and the amount of gas that is depleted from the ISM 
is $\Delta {M}_{\star}$. This mass is subtracted from the quantities $M_{\rm merg}$, $M_{\rm recycle}$ and $M_{\rm cooling}$, 
preserving their fractional contribution to ${M}_{\rm g}$ before stars formed. We then update ${M}_{\rm g}$ by subtracting 
$\Delta {M}_{\star}$. 

After stars form, a fraction $R$ is returned to the ISM, and we modify 
 $M_{\rm recycle}$ and ${M}_{\rm g}$ by $\Delta M_{\rm recycle}$ defined as:

\begin{eqnarray}
\Delta M_{\rm recycle} = R\,\Delta {M}_{\star},\label{mcooling1}.
\end{eqnarray}

\noindent From the stars formed, an amount $\Delta {M}_{\rm eject}=\beta\,{M}_{\star}$ is ejected from the galaxy, 
and we subtract the amount of gas escaping the disk from $M_{\rm merg}$, $M_{\rm recycle}$ and $M_{\rm cooling}$, 
preserving their fractional contribution to ${M}_{\rm g}$ before the ejection of gas. We then update
${M}_{\rm g}$ by subtracting $\Delta {M}_{\rm eject}$. This procedure ensures that
$M_{\rm merg}+M_{\rm recycle}+M_{\rm cooling}\equiv {M}_{\rm g}$.

During galaxy mergers, we add the amount of gas accreted by the central galaxy from the satellite to $M_{\rm merg}$ before star 
formation takes place, and then we proceed to the set of Eqs.~\ref{Eqs:SFset1}-\ref{Eqs:SFset2} with the formalism described 
above.

\end{document}